\documentclass[fleqn,usenatbib]{mnras}

\usepackage[T1]{fontenc}
\usepackage{ae,aecompl}

\usepackage{graphicx}	
\usepackage{subfig}
\usepackage{amsmath}	
\usepackage{amssymb}	
\usepackage{indentfirst}    
\usepackage{url}
\usepackage{cleveref}   
\usepackage[dvipsnames]{xcolor} 

\usepackage{newtxtext,newtxmath}

\title[\c{C}atmabacak et al. 2020]{Black hole -- galaxy scaling relations in FIRE: the importance of black hole location and mergers}

\author[\c{C}atmabacak et al.]{
Onur \c{C}atmabacak,$^{1}$\thanks{E-mail: onurc@physik.uzh.ch (UZH ICS)}
Robert Feldmann,$^{1}$
Daniel Angl{\'e}s-Alc{\'a}zar,$^{2,3}$
\newauthor Claude-Andr{\'e} Faucher-Gigu{\`e}re,$^{4}$
Philip F. Hopkins,$^{5}$
and Du{\v s}an Kere{\v s},$^{6}$
\\
$^{1}$Institute for Computational Science, University of Zurich, Zurich CH-8057, Switzerland\\
$^{2}$Department of Physics, University of Connecticut, 196 Auditorium Road, U-3046, Storrs, CT 06269-3046, USA\\
$^{3}$Center for Computational Astrophysics, Flatiron Institute, 162 Fifth Avenue, New York, NY 10010, USA\\
$^{4}$Department of Physics and Astronomy and CIERA, Northwestern University, 2145 Sheridan Road, Evanston, IL 60208, USA\\
$^{5}$TAPIR, Mailcode 350-17, California Institute of Technology, Pasadena, CA 91125, USA\\
$^{6}$Department of Physics, Center for Astrophysics and Space Sciences, University of California at San Diego, 9500 Gilman Drive,\\ La Jolla, CA 92093
}

\date{Accepted XXX. Received YYY; in original form ZZZ}

\pubyear{2020}

\crefformat{section}{\S#2#1#3} 
\crefformat{subsection}{\S#2#1#3}
\crefformat{subsubsection}{\S#2#1#3}
\creflabelformat{equation}{#2#1#3}

\begin{document}
\label{firstpage}
\pagerange{\pageref{firstpage}--\pageref{lastpage}}
\maketitle

\begin{abstract}

The concurrent growth of supermassive black holes (SMBHs) and
their host galaxies remains to be fully explored, especially at high redshift. While often understood as a consequence of self-regulation via AGN feedback, it can also be explained by alternative SMBH accretion models. Here, we expand on previous work by studying the growth of SMBHs with the help of a large suite of cosmological zoom-in simulations (\textsc{\small{MassiveFIRE}}) that are part of the Feedback in Realistic Environments (FIRE) project. The growth of SMBHs is modelled in post-processing with different black hole accretion models, placements, and merger treatments, and validated by comparing to on-the-fly calculations. Scaling relations predicted by the gravitational torque driven accretion (GTDA) model agree with observations at low redshift \emph{without} the need for AGN feedback, in contrast to models in which the accretion rate depends strongly on SMBH mass. At high redshift, we find deviations from the local scaling relations in line with previous theoretical results. In particular, SMBHs are under-massive, presumably due to stellar feedback, but start to grow efficiently once their host galaxies reach $M_* \sim 10^{10} M_{\odot}$. We analyse and explain these findings in the context of a simple analytic model. Finally, we show that the predicted scaling relations depend sensitively on the SMBH location and the efficiency of SMBH merging, particularly in low-mass systems. These findings highlight the relevance of understanding the evolution of SMBH-galaxy scaling relations to predict the rate of gravitational wave signals from SMBH mergers across cosmic history.
    
\end{abstract}




\section{Introduction}\label{sec:Introduction}

Roughly $50$ years ago, \cite{Lynden-Bell1969} suggested that the high mass-to-light ratio objects (active galactic nuclei, AGN) observed at the centre of galaxies are supermassive black holes (SMBHs). Today it is commonly accepted that the luminosity of AGN is powered by accretion onto SMBHs \citep{Soltan1982} and almost all massive galaxies host SMBHs of millions to billions of solar masses at their centres \citep{Lynden-BellRees1971, Rees1984, KormendyRichstone1995, Salucci::etal::1999, Shankar::etal::2004}. The universal existence of SMBHs in the centres of galaxies has raised questions about their nature, formation mechanisms, and relation with the environment.

Observational studies in the past two decades have shown strong correlations between the SMBH mass and various properties of their host galaxies such as the bulge mass \citep{KormendyRichstone1995, MarconiHunt2003, HaringRix2004, McConnellMa2013, KormendyHo2013, GrahamScott2015}, the stellar velocity dispersion of the bulge \citep{Magorrian+1998, FerrareseMerritt2000, Gebhardt::etal::2000, Gultekin+2009, Bennert::etal::2015}, and the stellar mass of the host galaxy \citep{ReinesVolonteri2015, Beifiori::etal::2012, Savorgnan::etal::2016} in the local Universe. A good understanding of the scaling relations is essential, as they may reflect a co-evolution of SMBHs and galaxies. While the scaling relations at low redshift are empirically well constrained, their observational status at higher redshift is less clear, with different authors suggesting both redshift-dependent \citep{Treu::etal::2004, Walter::etal::2004, Merloni::etal::2010, Targett::etal::2012, NetzerBenny2014, Bongiorno::etal::2014} and redshift-independent relations \citep{Shields::etal::2003, Jahnke::etal::2009, Cisternas::etal::2011, Ding::etal::2017}. 

Even though SMBHs and galaxies follow relatively tight scaling relations in the local Universe, it is currently unknown whether such tight relations hold in the early Universe \citep{Huang::etal::2018, Trakhtenbrot::etal::2017, Delvecchio::etal::2019, Shirakata::etal::2016, Izumi::etal::2018}. In particular, SMBHs at high redshift may be over- or under-massive compared to their host galaxies or could grow in lock-step with each other \citep{Volonteri2012}.
The redshift evolution and the scatter of various SMBH-galaxy scaling relations may provide critical insights into the physics of black hole and galaxy growth.
Which physical processes might be responsible for reproducing the local scaling relations? Is it possible to produce the local scaling relations without self-regulating black hole feedback? If so, how do the SMBH-galaxy scaling relations evolve at high redshift? These are the questions we would like to address in this paper.

The standard approach to model the growth of SMBHs is via the spherical accretion approximation \citep{HoyleLyttleton1939, BondiHoyle1944, Bondi1952}. In its basic form, Bondi-like models assume radial accretion of non-self-gravitating gas onto a point-like source to estimate the accretion rate from large scales to black holes at the centre of galaxies. However, there are caveats to these prescriptions. \cite{Springel::etal::2005} and \cite{BoothSchaye2009} introduced an ad hoc boost factor of the Bondi model to avoid underestimating the accretion rate. Furthermore, the Bondi ansatz does not account for the angular momentum of the inflowing gas \citep{HopkinsQuataert2010, HopkinsQuataert2011, AnglesAlcazar::etal::2020}. However, modifications of the Bondi model that include rotation have been proposed by, e.g. \cite{Hobbs:etal:2012, Tremmel+2017}.

On their own, Bondi-like models tend to overpredict the SMBH mass since they scale with $M_{\rm BH} ^2$. Black hole feedback is thus critical as it avoids overly massive SMBHs relative to the local scaling relations by regulating both black hole growth and star formation \citep{Springel::etal::2005, DiMatteo::etal::2005, Sijacki::etal::2007}. This idea has been widely used to investigate the evolution of galaxies and SMBHs in cosmological simulations such as $\mathrm{\textsc{\small{Illustris}}}$ \citep{Vogelsberger::etal::2014, Genel::etal::2014}, $\mathrm{\textsc{\small{Horizon-AGN}}}$ \citep{Dubois::etal::2014, Volonteri::etal::2016, Kaviraj::etal::2017}, $\mathrm{\textsc{\small{Eagle}}}$ \citep{Schaye::etal::2015}, $\mathrm{\textsc{\small{MassiveBlack}}}$ \citep{Khandai::etal::2015}, $\mathrm{\textsc{\small{BlueTides}}}$ \citep{Feng::etal::2016}, $\mathrm{\textsc{\small{Romulus}}}$ \citep{Tremmel+2017}, and $\mathrm{\textsc{\small{Illustris-TNG}}}$ \citep{Springel::etal::2018}. 

On the other hand, alternative models for the gas accretion onto SMBHs have been proposed. \cite{HopkinsQuataert2010} performed nested simulations of star-forming galaxies to understand how gas can accrete from galactic scales ($\sim 10-100$ kpc) to smaller scales ($<1$ pc). Non-axisymmetric features that result in gravitational torques caused by galaxy mergers, spiral instabilities and eccentric disc modes efficiently remove angular momentum of the gas and drive it further down to the sub-pc scales \citep{HopkinsQuataert2011}. This model has been successfully used in galaxy simulations to reproduce the local scaling relations without the need for self-regulatory black hole feedback \citep{AnglesAlcazar::etal::2013, AnglesAlcazar::etal::2015, Angles-Alcazar+17a, Angles-Alcazar+17c, Dave::etal::2019, Thomas::etal::2019}. 

The present work studies the SMBH growth in a fully cosmological context with the help of high-resolution, zoom-in, hydrodynamical simulations. The simulations used in this paper (\textsc{\small{MassiveFIRE}}) resolve scales down to tens of pc in a cosmological environment. High resolution is essential to properly trace the flow of gas into the centres of galaxies. Furthermore, the relatively large number of well-resolved galaxies in our sample ($43$ different galaxies at $z=6$ and $132$ at $z=2$) compared to full cosmological simulations allows us to address the questions listed above with a statistically significant set of simulated galaxies over a wide range of redshifts ($2\leq z\leq 12$) and halo masses ($10 < \log(M_{\rm h} / M_\odot) < 13.5$).

The outline of the paper is as follows; section \Cref{sec:Simulations} introduces the simulation properties. Section \Cref{sec:Methodology} lays out the details of our post-processing analysis. We present our main results in the following section \Cref{sec:Results}. Specifically, section \Cref{sec:Toy Model} presents a toy model to explain the physical origin of the $M_* - M_{\rm BH}$ scaling relation. We discuss the caveats of our post-processing analysis in \Cref{sec:Discussion} and give our summary in \Cref{sec:Summary}.

\section{Simulations}\label{sec:Simulations}

\begin{table}
\centering
\begin{tabular}{l|c p{1.3cm} l c c }

Name & Physics & Box Size ($\mathrm{h^{-1}}$ Mpc) & Final $\rm z$ & $\log(M_{\rm halo}/M_\odot)$ & Ref. \\
\hline
A1  & FIRE-1 & 100 & 0     & 12.75 & 3 \\
A2  & FIRE-1 & 100 & 1.7   & 12.48 & 3 \\
A3  & FIRE-1 & 100 & 1.7   & 12.38 & 3 \\
A4  & FIRE-1 & 100 & 0.27  & 12.84 & 3 \\
A5  & FIRE-1 & 100 & 1.7   & 12.37 & 3 \\
A6  & FIRE-1 & 100 & 1.7   & 12.45 & 3 \\
A7  & FIRE-1 & 100 & 1.7   & 12.41 & 3 \\
A8  & FIRE-1 & 100 & 1.7   & 12.57 & 3 \\
A9  & FIRE-1 & 100 & 1.7   & 12.48 & 3 \\
A10 & FIRE-1 & 100 & 1.7   & 12.53 & 3 \\
B1  & FIRE-1 & 100 & 1.7   & 12.94 & 3 \\
B2  & FIRE-1 & 100 & 1.7   & 12.97 & 3 \\
B3  & FIRE-1 & 100 & 1.7   & 13.00  & 3 \\
B4  & FIRE-1 & 100 & 1.7   & 12.94 & 3 \\
B5  & FIRE-1 & 100 & 1.7   & 12.97 & 3 \\
C1  & FIRE-1 & 100 & 2     & 13.39  & 1 \\
C2  & FIRE-1 & 100 & 2     & 13.37  & 2 \\
C3  & FIRE-1 & 100 & 2     & 13.35  & 1 \\
D1  & FIRE-1 & 400 & 6     & 12.29  & 4 \\
D2  & FIRE-1 & 400 & 6     & 12.38  & 4 \\
D3  & FIRE-1 & 400 & 6     & 12.67  & 4 \\
D4  & FIRE-1 & 400 & 6     & 12.50  & 4 \\
D5  & FIRE-1 & 400 & 6     & 12.39 & 4 \\
D6  & FIRE-1 & 400 & 6     & 12.57  & 4 \\
D7  & FIRE-1 & 400 & 6     & 12.29  & 4 \\
D8  & FIRE-1 & 400 & 6     & 12.36  & 4 \\
D9  & FIRE-1 & 400 & 6     & 12.01 & 4 \\
E1  & FIRE-1 & 762 & 6     & 12.81 & 1 \\
E2  & FIRE-1 & 762 & 6     & 12.80 & 1 \\
E3  & FIRE-1 & 762 & 6     & 12.77 & 1 \\
A1  & FIRE-2 & 100 & 1     & 12.60 & 5 \\
A2  & FIRE-2 & 100 & 1     & 12.89 & 5 \\
A4  & FIRE-2 & 100 & 1     & 12.66 & 5 \\
A8  & FIRE-2 & 100 & 1     & 13.10 & 5 \\
    
\end{tabular}
\caption{List of simulations used in this work. Column 1 refers to the name of the simulation, see \protect\cite{Feldmann+17}. Column 2 lists whether simulations were run with FIRE-1 or FIRE-2 physics. Column 3 provides the box sizes from which the zoom-in simulations were selected (in comoving units). The final redshift reached by each simulation is listed in column 4. Column 5 lists the halo mass at the final redshift given in column 4. References to the papers that first present or use the simulations are provided in the last column as follows: 1-This work 2-\protect\cite{Feldmann+17} 3-\protect\cite{Feldmann+16} 4-\protect\cite{Feldmann17} 5-\protect\cite{Angles-Alcazar+17c}.}
\label{tab:simulation details}
\end{table}

We use $34$ high-resolution, cosmological zoom-in simulations from the $\textsc{\small{MassiveFIRE}}$ suite \citep{Feldmann+16, Feldmann+17, Feldmann17, Angles-Alcazar+17c} that is part of the Feedback in Realistic Environments (FIRE\footnote{See the FIRE project website:  \url{http://fire.northwestern.edu}}) project \citep{Hopkins:etal:2014, Hopkins+18}. Simulations were run with the gravity-hydrodynamics solver GIZMO\footnote{A public version of GIZMO is available at \url{http://www.tapir. caltech.edu/~phopkins/Site/GIZMO.html}} in Pressure-Energy Smoothed Particle Hydrodynamics (P-SPH, FIRE-1) and Meshless Finite Mass (MFM, FIRE-2) mode. \Cref{tab:simulation details} provides an overview of the simulations used in the present work.

The selection of the zoom-in regions for runs from series A, B, and C is described in detail in \cite{Feldmann+17}. In brief, isolated halos are selected from a low-resolution DM-only run of an $L=100$ Mpc h$^{-1}$ comoving cosmological volume. The halos are selected based on their $z=2$ masses (we consider 3 narrow mass bins corresponding to $2.5-3.6 \times 10^{12} M_\odot$, $0.9-1.1 \times 10^{13} M_\odot$, and $2.5-3.6 \times 10^{13} M_\odot$) and local environmental densities (based on the enclosed mass within a $1.8$ pMpc radius). In total, $18$ haloes are selected with a range of masses (10, 5, and 3 haloes from the low, intermediate, and high mass bins) and environmental densities. Initial conditions for the zoom-in runs were generated using the multi-scale initial condition tool MUSIC \citep{Hahn&Abel2011} using a convex hull for all particles within $3 \times R_{\rm vir}$ at $z=2$.

Additional zoom-in simulations (series D and E) are created in a similar fashion from low-resolution DM-only simulations of $L=400$ Mpc h$^{-1}$ and $L=762$ Mpc h$^{-1}$ comoving cosmological volumes. The 3 most massive halos at $z=6$ are selected from each volume. In addition, 5 zoom-in regions are created based on the $L=400$ Mpc h$^{-1}$ volume by selecting halos with $M_{\rm halo}(z=6)>10^{12.5}$ $M_\odot$ from a range of local environmental densities. Another zoom-in region is selected from the same volume based on having a halo mass at $z=2$ of approximately $10^{14}$ $M_\odot$.

FIRE-1 simulations use a quintic spline kernel with 60-62 neighbors for gravitational softening \citep{Morris1996, DehnenAly2012}, see \cite{Hopkins:etal:2014}. The gravitational softening lengths of dark matter and star particles are fixed at $143$ and $21$ pc (physical) respectively, while the softening length of gas particles is adaptive and reaches a minimum value of $9$ pc (physical) in the dense interstellar medium. FIRE-2 simulations use a cubic spline kernel with 32 neighbors \citep{Morris1996, DehnenAly2012}, see \cite{Hopkins+18}. The gravitational softening lengths of dark matter and star particles are $57$ and $7$ pc. The minimum softening length of gas particles is $0.7$ pc. In all simulations, the gas softening lengths are chosen sufficiently small to capture gas densities well above the star formation threshold. All simulations have a mass resolution of $1.7 \times 10^5 M_\odot$ for dark matter particles and $3.3 \times 10^4 M_\odot$ for gas and star particles.
 
Star formation takes place only in self-gravitating, dense, molecular gas with a density above 5 and 1000 atoms per cm$^3$ for FIRE-1 and FIRE-2 simulations, respectively. The simulations include various stellar feedback channels such as energy, momentum, and mass injection from stellar winds and supernovae, local and long-range momentum flux from radiative pressure, a uniform UV background using the model from \citep{Faucher-Giguere::etal::2009} and photo-ionization and photo-electric heating \citep{Hopkins:etal:2014, Hopkins+18}. 

The growth of black holes is modelled fully in post-processing. Our FIRE-1 simulations do not directly account for black hole physics, while the FIRE-2 runs include live black hole sink particles but do not model AGN feedback. However, we adopt the same post-processing approach for FIRE-1 and FIRE-2 simulations in the present study. A comparison between the prediction of our post-processing model and the on-the-fly calculation is shown in appendix \ref{Appendix:Additional analysis}.

We refer the reader to \cite{Feldmann+16, Feldmann+17, Angles-Alcazar+17c, Hopkins:etal:2014, Hopkins+18} for more detailed information about the simulations and the properties of the simulated galaxies.

\section{Methodology}\label{sec:Methodology}

In this section, we will introduce our post-processing approach and describe the various SMBH accretion models studied in this work. By considering different models of black hole accretion, we can analyze how the resultant SMBH masses change in the absence of self-regulating AGN feedback. As we will show in subsequent sections, the choice of the accretion model affects both the mass evolution of SMBHs as well as the resulting $M_*-M_{\rm BH}$ scaling relation. We follow the exact same post-processing approach in both our FIRE-1 and FIRE-2 simulations.

\subsection{Post-processing}
\label{sect:postprocessing}

\begin{figure}
    \centering
    \includegraphics[width=\linewidth]{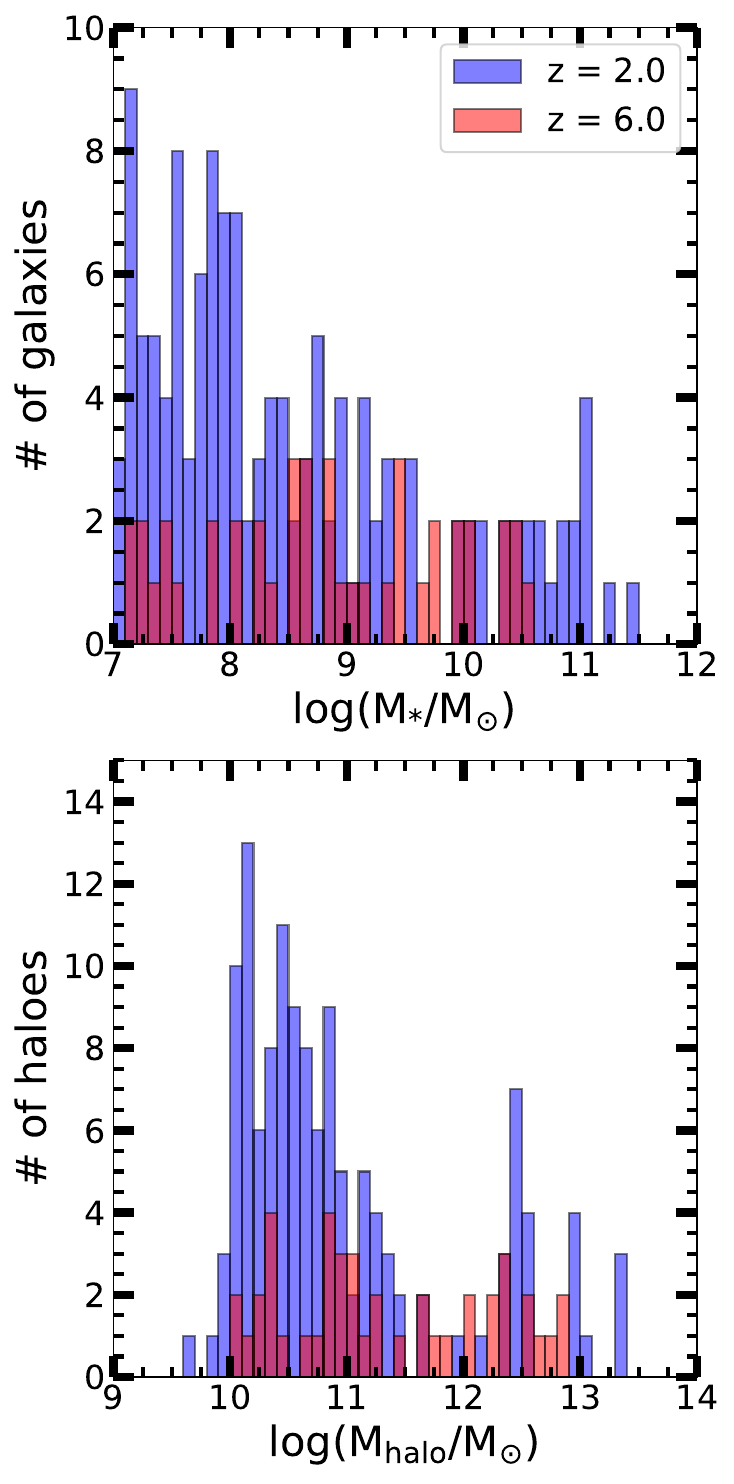}
    \caption{Stellar mass and halo mass distributions of the simulated galaxies listed in \Cref{tab:simulation details} at $z=2$ and $z=6$. Our sample includes the central galaxy of the primary halo from 34 different zoom-in simulations and all galaxies in halos above $M_{\rm halo} = 10^{10} M\odot$ from 6 zoom-in simulations evolved to $z \lesssim  1$. The total number of distinct galaxies followed across cosmic time varies with redshift (132 at $z=2$, 43 at $z=6$). Our sample includes a broad range of low to moderately massive galaxies and halos.}
    \label{fig:galaxy properties}
\end{figure}

We use the publicly available Amiga Halo Finder\footnote{AMIGA Halo Finder: \url{http://popia.ft.uam.es/AHF/Download.html}} (AHF, \citealt{KnollmannKnebe2009}) to identify dark matter haloes and to find their centres in the $\textsc{\small{MassiveFIRE}}$ simulations. The identified virialized structures contain at least $100$ particles ($M_{\rm halo} \sim 10^7 \; M_\odot h^{-1}$).

SMBHs are placed either at the centre of mass (COM) or the maximum density centre (MAX) of each halo as provided by AHF. The former is defined as the centre of mass of the gas, star, and dark matter particles on the finest level of refinement of the host halo. Therefore, the COM often represents a typical environment in the central region of the host halo. The MAX is calculated as the position of the highest-density cell within the halo by AHF. These positions typically correspond to dense star clusters or gas clouds. Clearly, the SMBH placement can have a substantial impact on the early SMBH growth, in addition to the role played by stellar feedback \citep{Angles-Alcazar+17c}.

The difference between the two centering approaches is shown in physical units in \Cref{fig:halo center diff for MAX and COM} for simulation A1 with FIRE-2 physics. The center positions differ by  $0.5-1.0$ kpc at relatively high redshift ($z\gtrsim5$) and the difference becomes much smaller ($<100$ pc) at later times. This behavior is potentially linked to the transition in galactic structure from an irregular morphology at high redshift to well-settled disc galaxies at later times \citep{Sparre::etal::2017, Stern::etal::2020}.


Our sample consists of the most massive halo in each simulation and of all the haloes above $M_{\rm halo}=10^{10} M_\odot$ at the final simulation snapshot in the $6$ simulations that run to $z \lesssim  1$ (FIRE-1 A1 and A4, and FIRE-2 A1, A2, A4, and A8). Our selection is subject to the following contamination condition. The mass fraction of high-resolution dark matter particles is required to be larger than $98\%$ so that haloes are not significantly polluted with low-resolution dark matter particles. Progenitors of selected halos are traced back in time with the AHF \textsc{MergerTree} tool. We only consider progenitors with a stellar mass of at least $10^7 M_\odot$ within $10\%$ of their virial radii \citep{Moore::etal::1998, Devriendt::etal::2010}. Progenitors of lower stellar mass do not host a SMBH in our model and are thus neglected in this study.


\Cref{fig:galaxy properties} shows the distribution of stellar masses in our sample at $z = 2$ and $z = 6$. The stellar mass range of \textsc{\small{MassiveFIRE}} galaxies is $7.0<\log{M_*/M_\odot}<11.5$ at $z\sim2$ and $7.0<\log{M_*/M_\odot}<10.6$ at $z\sim6$ while the halo mass range is $8.3<\log{M_*/M_\odot}<12.1$ at $z\sim2$ and $8.4<\log{M_*/M_\odot}<11.6$ at $z\sim6$. In total, we have $4681$ galaxy data sets corresponding to $\sim 100$ distinct galaxies traced across cosmic time ($z<12$). We refer the reader to \cite{Feldmann+16, Price::etal::2017, Feldmann+17, Cochrane::etal::2019, Wellons::etal::2020, Parsotan::etal::2021} for the general properties of galaxies in \textsc{\small{MassiveFIRE}} simulations.


The radii of central galaxies, $R_{\rm gal}$, are defined as $\mathrm{10\%}$ of the virial radii of their parent halos, e.g., \cite{Price::etal::2017}. The total stellar mass of a given central galaxy is subsequently calculated as the mass of stellar particles within $R_{\rm gal}$. We separate the stellar mass of the galaxy into a bulge and a disc component following \cite{Angles-Alcazar+14}. Specifically, the bulge mass in an enclosed region with a given radius $r$ is calculated as twice the mass of all star particles that are counter-rotating ($v_{\rm \phi} < 0$) according to the unit stellar angular momentum vector of the galaxy. The stellar disc mass within $r$ is then defined as the difference between the stellar mass within $r$ and the bulge mass within $r$. The total disc mass is defined as the sum of the stellar disc mass and the gas mass within $r$. We calculate the velocity dispersion as the velocity dispersions of all star particles that belong to the stellar bulge. The comparison of velocity dispersion within different radial apertures ($R_0$) is shown in the appendix \ref{Appendix:Additional analysis}.

The evolution and growth of SMBHs are treated in post-processing. A (virtual) SMBH seed is placed at the centre of each progenitor halo. This choice establishes early SMBHs on the local $M_* - M_{\rm BH}$ relation when $10^4 \; M_\odot$ SMBH seed masses are adopted (our fiducial option). Lower or higher seed masses result in high-redshift SMBHs that start either below or above the local $M_* - M_{\rm BH}$ scaling relation. Also, the velocity of each SMBH is defined as the average velocity of the $100$ youngest star particles in its vicinity (within $R_0$, see \cref{sect:AccretionRatesGTDA}). The post-processing analysis is performed in the rest-frame of the SMBH of the given galaxy.

We define the stellar growth rate ($\rm SGR = d M_* / dt$) as the past change in stellar mass over a time interval $t_{\rm avg}$ in a three-dimensional sphere of fixed physical size divided by $t_{\rm avg}$. In practice, we measure the difference in stellar mass in the considered physical volume between the current snapshot at time $t$ and an earlier snapshot near time $t-t_{\rm avg}$. We thus do not calculate SGRs when $t_{\rm avg}$ is much shorter than the time interval between snapshots ($10-25$ Myr).
In contrast, the star formation rate (SFR) is defined as the stellar mass belonging to recently (within time $t_{\rm avg}$) formed stars currently present in a three-dimensional sphere divided by $t_{\rm avg}$. It can be thus be calculated for both long and short $t_{\rm avg}$. Both SFRs and SGRs are subject to stellar mass loss.

We need to differentiate between mergers of SMBHs and mergers of the dark matter halos in which they reside as not every halo merger results in the merger of their central galaxies and not every galaxy merger results in a prompt merger of their SMBHs. We thus analyse three different SMBH merger scenarios for the remainder of this paper.

Our first model (``\textsc{all mergers}'') assumes that SMBHs merge as soon as their parent dark matter haloes merge, regardless of the mass ratio of the two haloes. This scenario results in the maximum possible number of SMBH mergers. Hence, this extreme scenario explores the most optimistic contribution of BH merging to SMBH growth.

Our second model (``\textsc{few mergers}'') poses more stringent constraints on SMBH mergers and is based on the distance $d_{\rm sep}$ of the two parent halos when they are identified as separate (sub-)halos for the last time ($z_{\rm merge}$). Specifically, the SMBHs at the centres of these halos are assumed to have merged by the next snapshot if (i) their $d_{\rm sep}$ is smaller than $10\%$ $R_{\rm vir}$ of the more massive of the two halos or (ii) their dynamical friction timescale $t_{\rm df}$ is smaller than the Hubble time at $z_{\rm merge}$. We adopt the following analytic estimate of the dynamical friction time \citep{BinneyTremaine1987}:

\begin{equation}
    \mathrm{t_{df}= \frac{M_{h1} \, R_{vir}}{\ln{\Lambda} \, M_{h2} \, V_{vir}}}
\end{equation}

Here, $M_{\rm h1}$ and $M_{\rm h2}$ are the virial masses of the more massive and the less massive halo, respectively. Furthermore, $\Lambda$ is Coulomb logarithm where $\ln(\Lambda)=\ln(1+M_{\rm h1}/M_{\rm h2})$. $R_{\rm vir}$ is the virial radius and $V_{\rm vir}$ is the circular velocity of the more massive of the merging haloes. The less massive halo is often affected by tidal stripping at the late stages of the halo merger. To mitigate the effect of stripping -which may extend the inferred merging timescale- the maximum mass of the progenitors of the less massive halo is used instead of its current halo mass. A halo typically reaches its maximum mass $60-100$ Myrs before the halo merger.

Our third option (``\textsc{no mergers}'') is to neglect SMBH mergers altogether so that SMBHs only grow via mass accretion throughout cosmic time.  

\subsection{SMBH Accretion Rates}
\label{sect:AccretionRates}
    
\subsubsection{Gravitational Torque-Driven Accretion} \label{sect:AccretionRatesGTDA}
\cite{HopkinsQuataert2011} introduced a model of gas accretion from kpc to sub-pc scales driven by gravitational torques. The functional form of the black hole accretion rate (BHAR) in the gravitational torque-driven accretion (GTDA) model is

\begin{equation}
\begin{aligned}
    & \dot{\mathrm{M}} \mathrm{_{BH,GTDA}= \epsilon_m \, \alpha_T \times f_d(<R_0)^{5/2} \times \left( \frac{M_{BH}}{10^8 M_\odot} \right)^{1/6}}  \\ 
    \\
    & \mathrm{\times \left( \frac{M_{tot}(<R_0)}{10^9 M_\odot} \right) \times \left( \frac{R_0}{100pc} \right)^{-3/2}}
    \\
    \\
    & \mathrm{ \times \left( 1 + \frac{f_0(<R_0)}{f_{gas}(<R_0)} \right)^{-1} M_\odot yr^{-1} }
    \\
    \\
    & = \mathrm{\theta(\epsilon_m, \, \alpha_T, \, f_0, \, f_{gas} \, R_0, \, M_{BH}, \, M_*, \, M_{tot} ) \times \frac{M_*(<R_0)}{10^9 M_\odot}}
    \label{eq:GTDA equation}
\end{aligned}
\end{equation}

In the GTDA model, the accretion rate is calculated based on the properties of gas particles inside of a sphere with a radial aperture ($R_0$). All the terms in \Cref{eq:GTDA equation} have a radial aperture ($R_0$) dependency except the SMBH mass, $\epsilon_{\rm m}$, and $\alpha_{\rm T}$. Here, $\alpha_{\rm T}$ is a function of nuclear star formation law (see equations 39 and 65 in \cite{HopkinsQuataert2011}), however for practical reasons we set $\alpha_{\rm T}$ to $5$ (see Figure 10 in \cite{HopkinsQuataert2011}). The mass retention rate $\epsilon_{\rm m}$ is a fudge factor between 0 and 1 which reduces the analytically-derived accretion rate of the GTDA model. Physically, $\epsilon_{\rm m}$ captures the reduction in accretion rate due to unresolved winds. The product of $\epsilon_{\rm m}$ and $\alpha_{\rm T}$ is an overall normalisation that covers the effects on gas dynamics (i.e. stellar and BH feedback) at unresolved scales \citep{Angles-Alcazar+17a}.

The disc fraction $f_{\rm d}$ is the ratio of the disc mass (stellar disc mass + gas mass) to total baryonic mass (stellar mass + gas mass). The accretion rate scales super linearly with the disc fraction, $\dot{M} \propto f_{\rm d} ^{5/2}$. $M_{\rm BH}$ is the mass of the SMBH and $M_{\rm tot}$ is the total mass within the radial aperture ($R_0$). The total mass is the sum of dark matter and baryonic matter inside $R_0$. However, especially at the central regions, baryonic matter dominates over dark matter. Although the accretion rate scales only linearly with the total mass inside $R_0$, the total mass is the determining factor for the accretion rate onto SMBH. Amongst the remaining parameters, $f_{\rm gas}$ is the ratio of the gas mass to the total baryonic mass inside $R_0$ and $f_{\rm 0} \approx 0.31 \, f_{\rm d}^{\rm 2} \, (M_{\rm tot} / 10^9 M_\odot)^{\rm -1/3}$.

We update black hole masses in our post-processing analysis iteratively for each accretion model, as briefly explained in \Cref{sect:AccretionRates}, $M_{\rm BH,i+1} = M_{\rm BH,i} \, + \, \Delta t \, \times \, \dot{M}_{\rm BH,i}$. We repeat the same analysis for the densest (MAX) and average density (COM) centres, different SMBH merger treatments (with or without SMBH mergers) and radial apertures ($R_0$) changing from $1$ kpc to $100$ pc. The BHAR is limited to the Eddington rate for the Bondi-like models (see below) and to ten times the Eddington limit for all other accretion models.

\subsubsection{Bondi accretion and variants}

Spherical accretion onto a point object has a solution known as Bondi accretion, \citep{BondiHoyle1944, Bondi1952} and the corresponding accretion rate is
\begin{equation}
    \dot{\mathrm{M}} \mathrm{_{BH,Bondi} = \alpha_B \times \frac{4 \pi \left( G \, M_{BH} \right)^2 \rho(<R_0)}{\left( v_{bulk}^2 + c_s^2 \right)^{3/2}} .}
    \label{eq:bondi accretion equation}
\end{equation}

In the equation above, $\alpha_{\rm B}$ is the boost factor introduced by \cite{Springel::etal::2005}, $G$ is the gravitational constant, and $\rho$ is the volume density of gas particles within $R_0$. The bulk velocity of gas and the sound speed are denoted by $v_{\rm bulk}$ and $c_{\rm s}$, respectively.

The Bondi model is valid for the case of hot virialised gas with negligible angular momentum and radiative cooling. It does not account for SMBH growth via accretion of high density gas which tends to cool efficiently. \cite{Hobbs:etal:2012} proposed a modification to the Bondi model to account for the contribution of the halo to the gas dynamics. They replaced the relative velocity with the velocity dispersion for the external potential, $\sigma \sim \sqrt{G M_{\rm enc}(r) / r}$, and the SMBH mass with the enclosed mass of the external potential.

\begin{equation}
    \dot{\mathrm{M}}\mathrm{_{BH,Hobbs} = \alpha_H \times \frac{4 \pi \left( G M_{enc}(<R_0) \right)^2 \rho(<R_0)}{\left( \sigma^2 + c_s^2 \right)^{3/2}} }
    \label{eq:hobbs accretion equation}
\end{equation}

\subsubsection{Empirical parametrizations}

Observational and theoretical evidence for a roughly constant ratio between BHAR and SFR rate has led to the idea that the growth of SMBHs and galaxies are coupled, especially at the nuclear scales \citep{HopkinsQuataert2010, Volonteri::etal::2015b, Dai::etal::2018, Yang::etal::2017}. A simple ansatz is to model the SMBH accretion rate as a linear function of the SFR inside $\mathrm{R_0}$:

\begin{equation}
    \dot{\mathrm{M}} \mathrm{_{BH,SFR} = \frac{SFR(<R_0)}{500} }.
    \label{eq:sfr accretion equation}
\end{equation}

Finally, we also use a simple accretion model for comparison where the BHAR scales with the free-falling gas inside $\mathrm{R_0}$ \citep{Angles-Alcazar+17c}:

\begin{equation}
\begin{split}
    \dot{\mathrm{M}} \mathrm{_{BH,dyn} = \gamma \times \frac{M_{gas}(<R_0)}{t_{dyn}} }
    \label{eq:dyn accretion equation}
\end{split}
\end{equation}

Here, $\gamma$ is a scaling factor that controls the percentage of free-falling gas accreted onto SMBHs. The typical values of $\gamma$ changes from $0$ to $100\%$. The free-fall timescale of the gas is $t_{\rm dyn} = R_0^3 / (2GM_{\rm tot}(<R_0))$.

\section{Results}\label{sec:Results}

\subsection{Parameter Study}\label{sec:Parameter Study}

\begin{table}
\centering
\begin{tabular}{|l|l|}

Parameter Name & Fiducial Setting \\
\hline
Accretion model     & GTDA Model \\
Accretion cap       & $\mathrm{10 \; \times \;} \dot{\mathrm{M}} \mathrm{_{Edd}}$ \\
Mass retention rate      & $\mathrm{10\%}$ \\
Radial aperture     & $\mathrm{100 \; pc}$ \\
SMBH seed mass      & $\mathrm{10^4 \; M_\odot}$

\end{tabular}
\caption{The fiducial model parameters used in our post-processing analysis (see text). These parameters result in SMBH masses that are consistent with the local scaling relations, for all considered variants of BH merging and positioning.}

\label{tab:standard values}
\end{table}

In this section, we analyse how the choice of our model parameters affects the predicted SMBH growth. Our post-processing analysis includes several key parameters that influence the growth of SMBHs such as the SMBH seed mass $(M_{\rm seed})$, the black hole merger treatment and, most importantly, the accretion model (\Cref{eq:GTDA equation} - \ref{eq:dyn accretion equation}) and its free parameters, e.g. the mass retention rate ($\epsilon_{\rm m}$) and the radial aperture ($R_0$) for the GTDA model. We will use the fiducial values for our model parameters in \Cref{tab:standard values} together with the MAX centring method and the \textsc{few mergers} model to see the precise impact of our model parameters. Throughout this paper, these settings (the fiducial parameters from \Cref{tab:standard values}, the MAX centring method and the \textsc{few mergers} model) will be our primary choice unless stated otherwise.

\begin{figure}
  \centering
    \includegraphics[width=\linewidth]{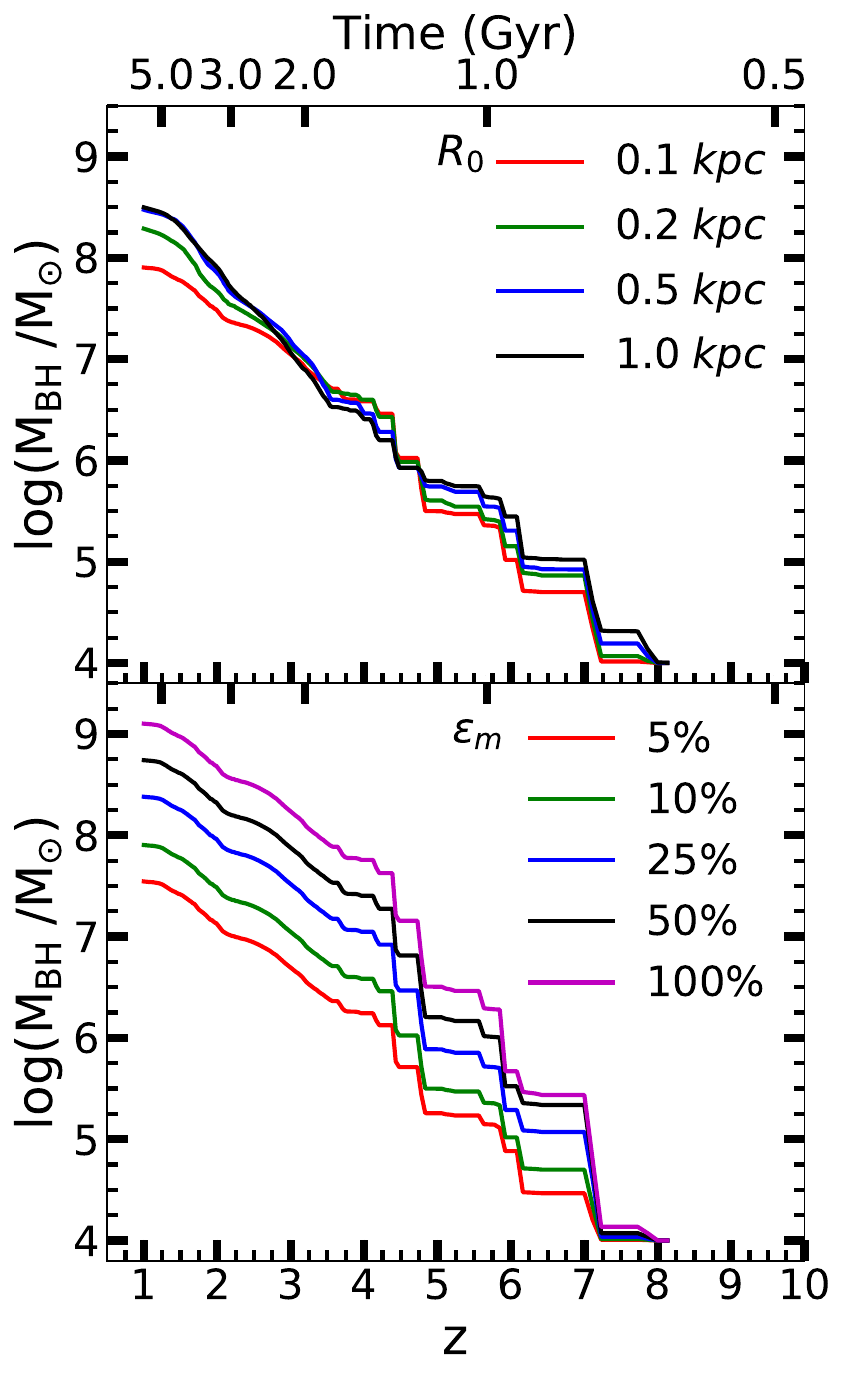} 
    \caption{Effect of varying the free parameters of the GTDA model on the SMBH mass for simulation A1 (FIRE-2). Top panel shows how SMBH mass changes with radial aperture ($R_0$) while bottom panel shows the dependence of SMBH mass on mass retention rate, $\epsilon_{\rm m}$. All other model parameters are at their fiducial values, see \Cref{tab:standard values}. SMBHs are placed in the densest centres (MAX), and \textsc{few mergers} model is used to model SMBH growth. A larger $R_0$ slightly increases the SMBH mass, but overall the choice of $R_0$ has little impact on the growth history of the SMBH. In contrast, the SMBH scales approximately linearly with $\epsilon_{\rm m}$.}
    \label{fig:BH growth for different radii and mass retention rates}
\end{figure}

We tested various values for the radial aperture ($R_0$) in the absence of self-regulating AGN feedback. As the top panel of \Cref{fig:BH growth for different radii and mass retention rates} shows, the growth history of SMBHs is only mildly affected if $R_0$ is varied between $100$ pc and $1$ kpc. This finding is perhaps surprising, given that the GTDA model has a strong dependence on the radial aperture ($R_0$), $\dot{M} \propto R_0^{-3/2}$. However, various other terms in \Cref{eq:GTDA equation} also depend on the radial aperture ($R_0$), largely cancelling the overall dependence on $R_0$ in agreement with similar tests in \cite{AnglesAlcazar::etal::2015}.

As shown in \cite{HopkinsQuataert2010}, a smaller $R_0$ results in a more precise prediction (lower scatter) of the instantaneous accretion rate on small scales. Hence, to mimic the accretion from galactic scales to sub-pc scales as accurately as we can, we adopt $R_0 = 100$ pc as the fiducial value.

The mass retention rate $\epsilon_m$ is a normalisation of the overall gas accretion rate. As the bottom panel of \Cref{fig:BH growth for different radii and mass retention rates} highlights, varying the mass retention rate creates a noticeable shift in the normalisation of the SMBH mass, at any redshift. Hence, we can adjust the normalisation of the predicted $\mathrm{M_* - M_{BH}}$ scaling relation by choosing an appropriate value of the mass retention rate \citep{AnglesAlcazar::etal::2013}. We adopt a mass retention rate of 10\% as our fiducial value. The fiducial values in \Cref{tab:standard values} are chosen so that SMBH masses in our post-processing analysis are in approximate agreement with the local $M_*-M_{\rm BH}$ scaling relation.

\begin{figure}
  \centering
    \includegraphics[width=\linewidth]{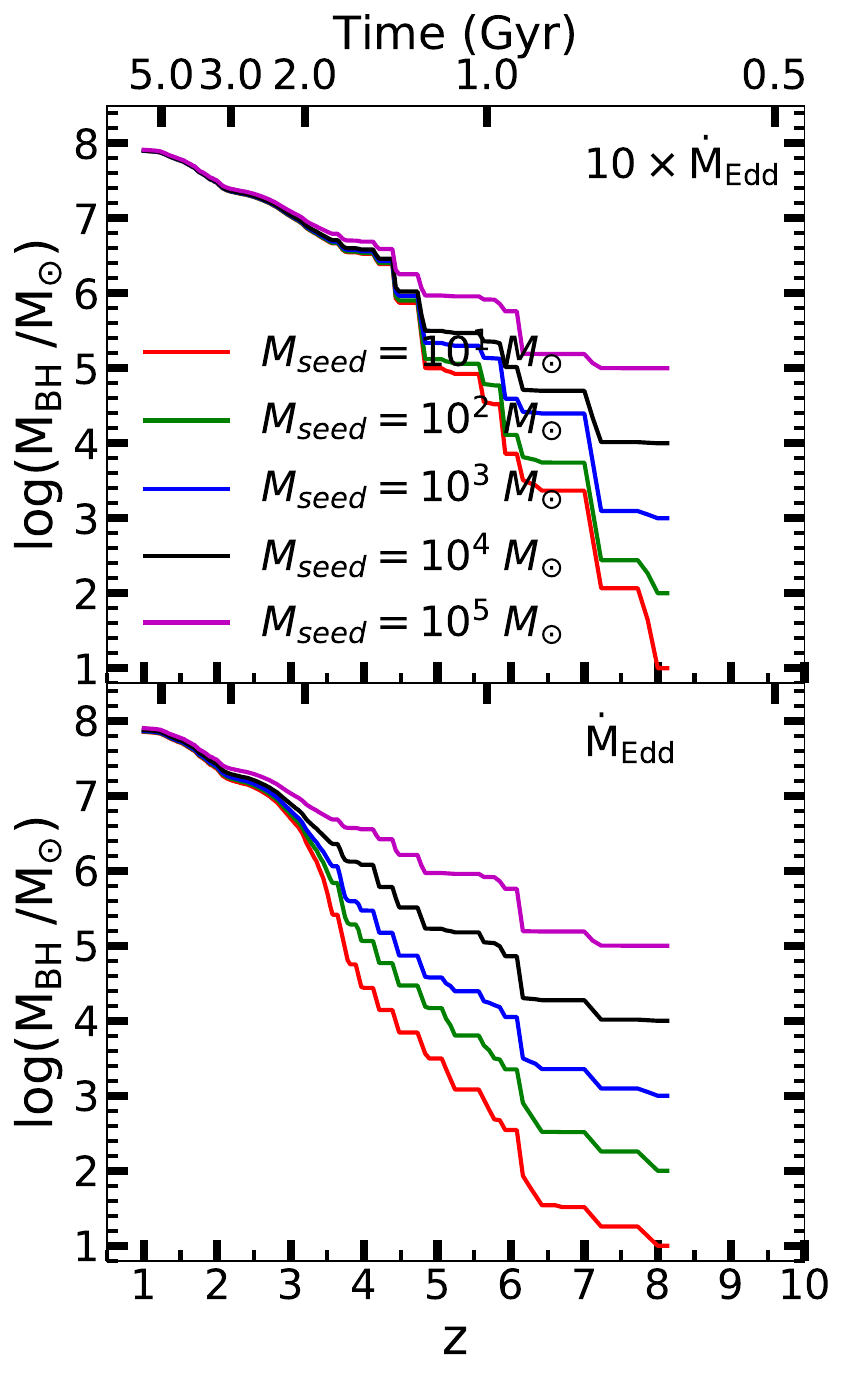} 
  
    \caption{Effect of the SMBH seed mass choice and a limit on the growth rate on the mass evolution of SMBHs for our fiducial settings with the \textsc{few mergers} model and SMBHs located in the densest centres (MAX). The top (bottom) panel shows the prediction if the growth rates are limited to ten times the Eddington rate (to the Eddington rate). The masses of SMBHs converge to $\sim 10^7 M_\odot$ after $1.5$-$2$ Gyr of cosmic time (by $\rm z \sim 3-4$) independent of the initial seed mass. The influence of the seed mass on the growth of SMBHs is thus limited to high redshift in the progenitors of massive galaxies, and it is smaller if super-Eddington accretion rates are possible.} 
  \label{fig:BH growth for different seed mass choice}
\end{figure}

\Cref{fig:BH growth for different seed mass choice} shows the insensitivity of the GTDA model predictions to the black hole seed mass choice \citep{AnglesAlcazar::etal::2013, AnglesAlcazar::etal::2015, Angles-Alcazar+17a}. Specifically, SMBH seeds with masses $10 - 10^5$ $M_\odot$ result in a similar SMBH mass by $z\sim4.5$ if accretion is limited to less than ten times Eddington and by $z\sim 3$ in the case of Eddington-limited accretion. This convergence is a consequence of the SMBH accretion rate being only a weak function of black hole mass in the GTDA model. This figure suggests that observations of SMBH masses in the progenitors of massive galaxies at $z>3-5$ may provide useful constraints on the masses of the first black hole seeds.

\begin{figure}
    \includegraphics[width=\linewidth]{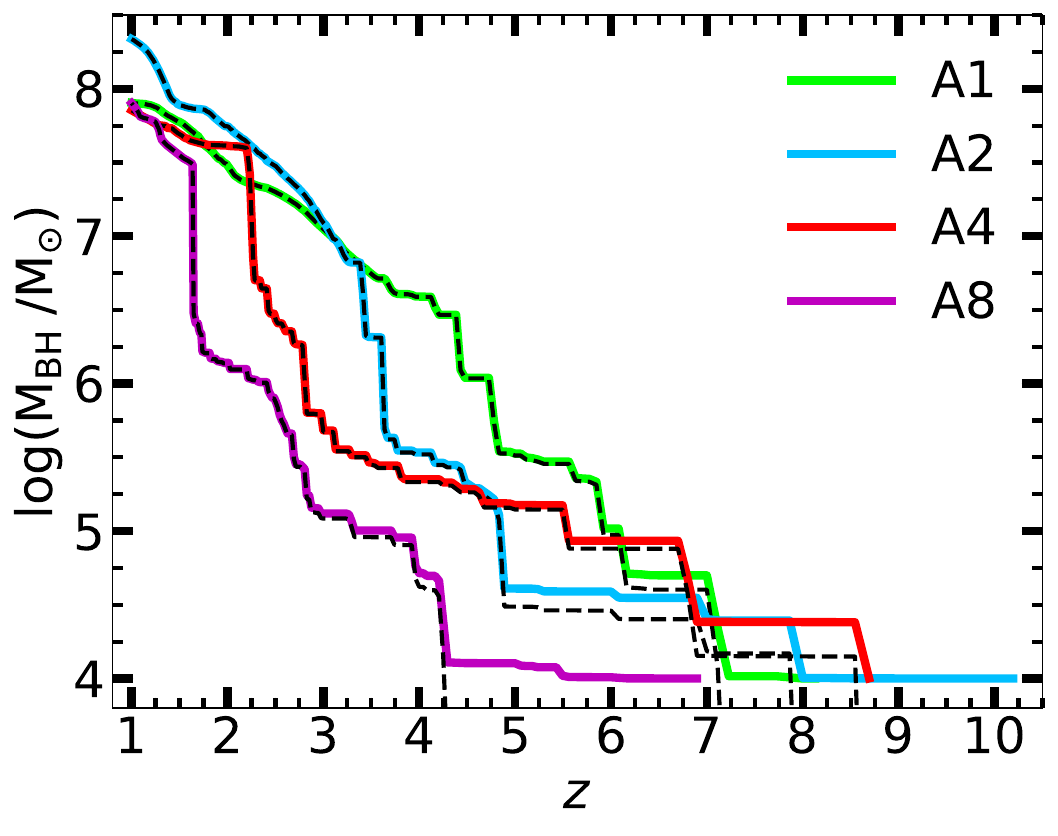}
    \caption{Contribution of gas accretion to the total SMBH mass for different simulations with FIRE-2 physics. Solid lines show the total SMBH mass $M_{\rm BH}$, while dashed lines show only the contribution from gas accretion computed via $M_{\rm BH, acc}(\rm t) = M_{\rm BH}(\rm t) - n \times M_{\rm seed}$, where n is the number of mergers up to the relevant redshift. The mass contribution from SMBH seeds is typically significant only at $3>z>6$ and for intermediate-mass black holes even when adopting a relatively heavy seed mass ($10^4 M_\odot$). Overall, gas accretion is the driving force for black hole growth throughout much of cosmic history.}
    \label{fig:BH growth for different accretion contribution}
\end{figure}

\Cref{fig:BH growth for different accretion contribution} compares the SMBH growth via gas accretion with the contribution from black hole seed masses for the A-series simulations run with FIRE-2 physics \citep{Angles-Alcazar+17c}. For low-mass SMBHs ($M_{\rm BH} < 10^6 M_\odot$) at $3<z<6$, the seed mass contribution can be significant if relatively heavy seeds (here $10^4$ $M_\odot$) are chosen. However, most of the SMBH mass is acquired via gas accretion either in-situ or by merging at lower redshifts. We refer the reader to \Cref{fig:Accretion contribution for different seed masses} for the same analysis with different seed masses.

\begin{figure}
    \includegraphics[width=\linewidth]{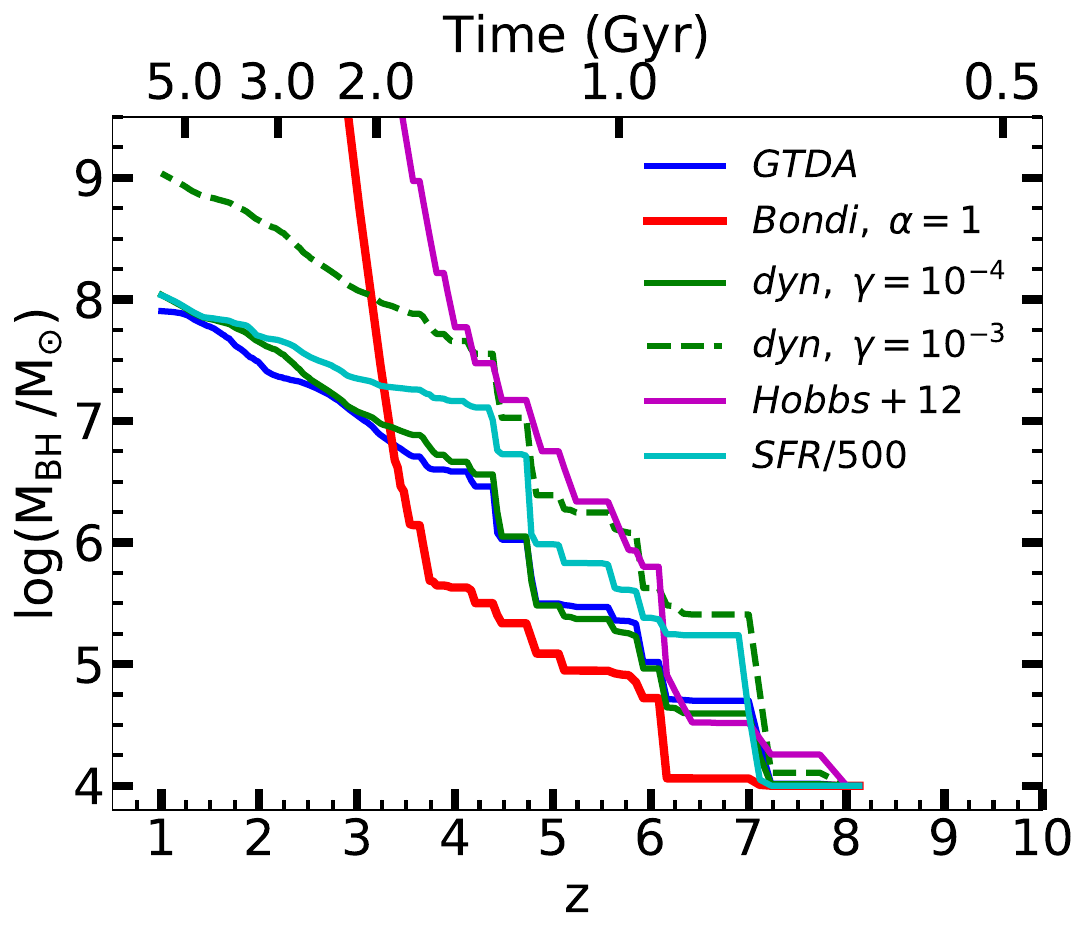}
    \caption{Growth histories of SMBHs for different accretion models (see section \ref{sect:AccretionRates}). In the absence of self-regulating AGN feedback, the Bondi accretion model and its modification by \protect\cite{Hobbs:etal:2012} result in overly massive black holes at $z<4$. In contrast, SMBHs grow more steadily in GTDA, SFR-based, and dynamical free-fall models, and, for appropriate choices of the overall normalisation, they lead to similar mass growth histories.
    }
    \label{fig:BH growth for different accretion models}
\end{figure}

\Cref{fig:BH growth for different accretion models} compares the results for SMBH growth via GTDA with the predictions of various other accretion models employed in the literature (see section \ref{sect:AccretionRates}).
These models can be divided into two types. The first group includes models in which the accretion rate depends strongly on the black hole mass, such as in Bondi accretion. In this case, black holes may grow extremely fast in the absence of AGN feedback. As a consequence, the predicted black holes are overly massive compared to the local scaling relations. 
 
In the second group of models, accretion rates scale weakly with SMBH mass. These models use the properties of the host galaxy such as stellar mass, disc fraction, free-fall timescale of gas and SFR. Such models result in a much more steady black hole growth in massive galaxies from an initial seed mass of $10^4$ $M_\odot$ at $z>7$ to SMBHs of $10^8$ $M_\odot$ or more by $z=1$. The most significant difference between these two groups of accretion models is the necessity of self-regulating AGN feedback to reproduce the local scaling relations. The Bondi-like accretion models have a high dependence on the SMBH mass, thus require strong AGN feedback to regulate SMBH growth to provide reasonable SMBH masses \citep{AnglesAlcazar::etal::2015}. Other models can produce matching results without the need for expelling material from the centre of the host galaxy.

\begin{figure}        \includegraphics[width=\linewidth]{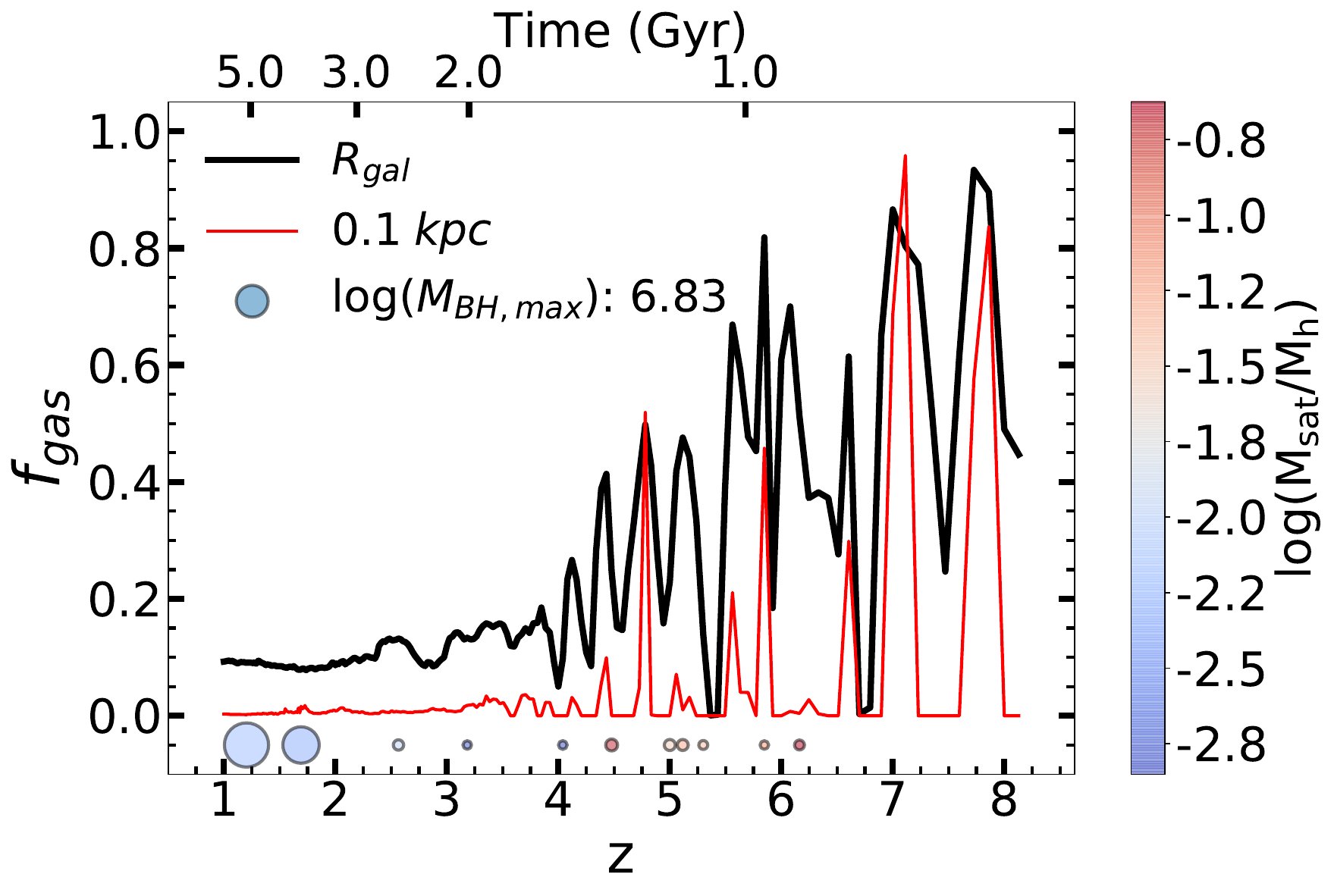}
    \caption{Comparison of redshift evolution of gas fraction within the $100$ pc and the entire galaxy of the most massive halo in the simulation A1 (FIRE-2). The solid black line is the gas fraction for the central galaxy, and the solid red line is for the central $100$ pc region. Colorbar shows the mass ratio of merging haloes. Dots indicate the merging haloes coloured according to the mass ratio of the haloes. The area of the dots is proportional to the SMBH mass of the second halo. We only include the merging haloes with a mass ratio above $0.001$. 
    }
    \label{fig:fgas for R_0 and R_gal}
\end{figure}

\Cref{fig:fgas for R_0 and R_gal} compares the gas fraction within the central $100$ pc region and the whole galaxy for the simulation A1 with FIRE-2 physics. At high redshift, gas fraction demonstrates a bursty behaviour until stellar feedback becomes inefficient to remove gas from the central region of galaxy \citep{Muratov::etal::2015, Angles-Alcazar+17c}. Gas fraction in the centre region follows the increase in the galactic gas fraction. This suggests that gas inflow at high redshift can reach to the central regions more easily compared to the galaxies in the local Universe. The change in the galaxy size could play an important role for the bursty behaviour of gas fraction \citep{Torrey::etal::2017, Faucher-Giguere2018}. Furthermore, peaks in gas fraction decrease rapidly at early times. The replenished gas reservoir triggers expeditious SF and feedback from the newly formed stars evacuates gas from the star forming regions. We refer the reader to \Cref{fig:fgas for MAX and COM} for a comparison of the gas fraction for different centring methods.

\subsection{Scaling Relations}\label{sec:Scaling Relations}

\begin{figure*}
    \includegraphics[width=\textwidth]{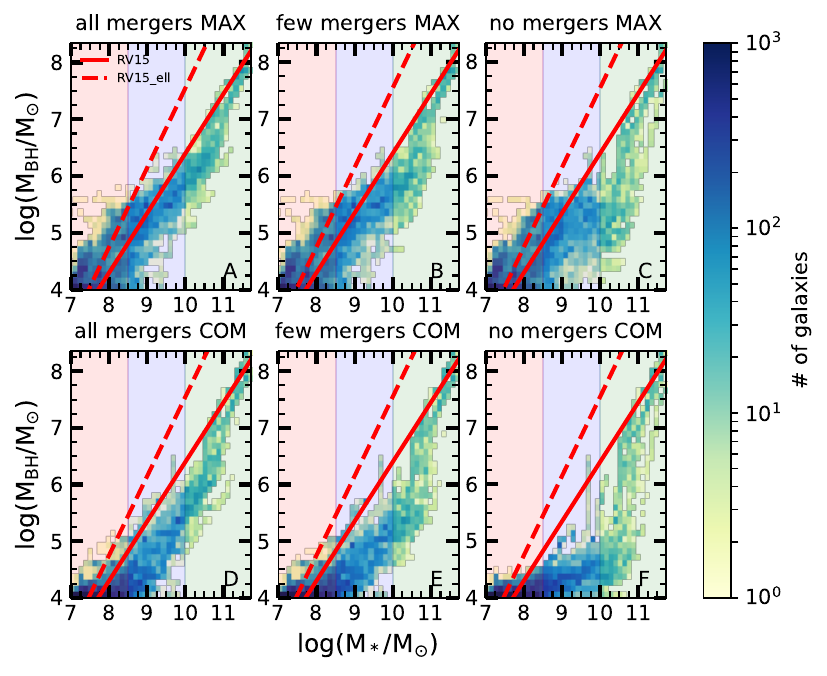} 
    \caption{Different evolutionary phases in the $M_* - M_{\rm BH}$ scaling relation for the fiducial case ($M_{\rm seed} = 10^4 M_\odot$, $\;R_0=100$ pc, and $\epsilon_{\rm m} = 10\%$), for different SMBH locations and merger treatments using the complete \textsc{\small{MassiveFIRE}} sample. Colour reflects the number of galaxies in each pixel for all simulation snapshots. The panels show SMBHs from their time of seeding ($\rm z \sim 6-12$) until the final redshift of each simulation, see \Cref{tab:simulation details}. In each panel, the solid red line shows the $M_*-M_{\rm BH}$ scaling relation for disc galaxies in \protect\cite{ReinesVolonteri2015}. The first row of panels use maximum-density centre and the second row of panels use centre-of-mass as the halo centre. Columns indicate $M_*-M_{\rm BH}$ scaling relation for different merger treatments. Early growth of SMBHs is suppressed when SMBHs are placed at the centre of mass of the halo (bottom panels), which is more strongly affected by stellar feedback compared to the maximum-density centre. The maximum-density location (top panels) result in SMBH growth at high redshift, in line with the local scaling relation. SMBH mergers make a considerable contribution to the total SMBH mass in intermediate-mass galaxies ($8.5<\log(M_*/M_\odot)<10$). Efficient SMBH growth starts when the stellar mass reaches $\sim 10^{10} M_\odot$. SMBHs grow at a similar rate as their hosts in massive galaxies ($M_*>10^{11} M_\odot$). In all six cases, SMBHs end up with similar masses. Panel A is the most optimistic scenario for SMBH growth and results from Panel E and F are in line with the findings of \protect\cite{Angles-Alcazar+17c}, in which early SMBH growth is suppressed due to strong stellar feedback. We refer the reader to \Cref{fig:mstar vs mbh scaling relation for different seed masses} to see the effect of different seed mass choice on the $M_*-M_{\rm BH}$ scaling relation.}
    \label{fig:mstar vs mbh scaling relation}
\end{figure*}

\begin{figure*}
    \centering
    \begin{tabular}{cc}
    \includegraphics[width=0.5\textwidth, height=0.3\textheight]{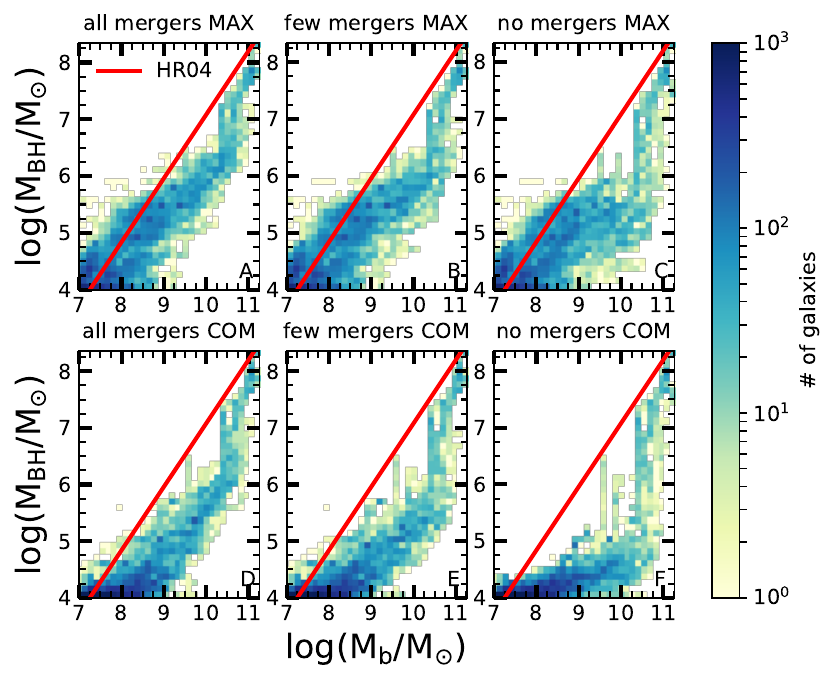}
       &
    \includegraphics[width=0.5\textwidth, height=0.3\textheight]{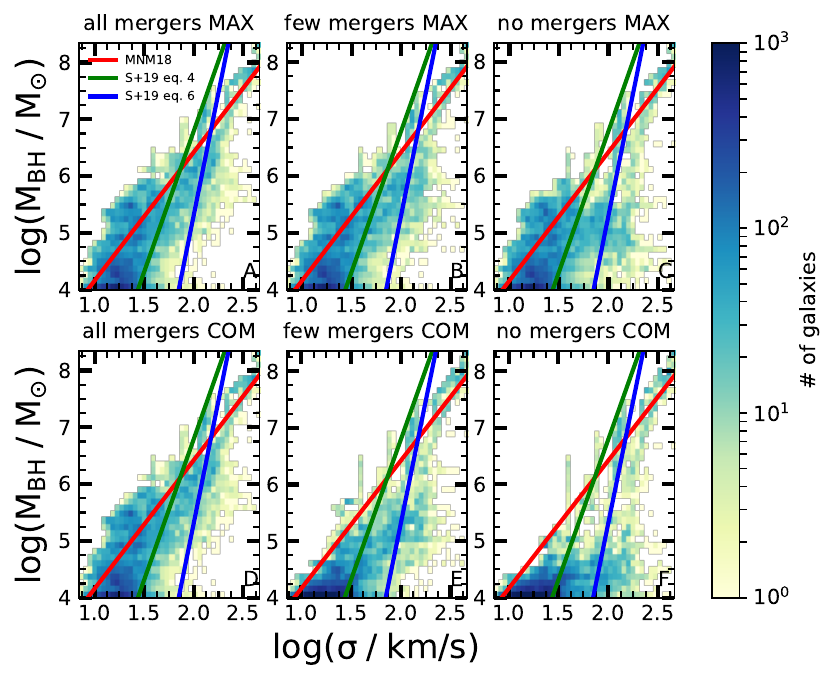}
    \end{tabular}
    
    \caption{Same as \Cref{fig:mstar vs mbh scaling relation} for the local $M_{\rm b} - M_{\rm BH}$ scaling relation of \protect\cite{HaringRix2004} (red solid line in the left-hand panels) and the local $\sigma-M_{\rm BH}$ scaling relation of \protect\cite{Martin-NavarroMezcua2018} (red solid line in the right-hand panels) for the fiducial model ($M_{\rm seed} = 10^4 M_\odot$, $R_0=100$ pc, and $\epsilon_{\rm m} = 10\%$). The solid blue and green lines in the right panel are for local S\'{e}rsic and Core-S\'{e}rsic ETGs \protect\cite{Sahu::etal::2019b} which are steeper than the \textsc{\small{MassiveFIRE}} prediction. The redshift range is same as in \Cref{fig:mstar vs mbh scaling relation}}.
    \label{fig:mbulge vs mbh scaling relation}
\end{figure*}

\subsubsection{$\mathrm{M_* ,\; M_{\rm b} ,\; and \; \sigma \; vs \; M_{\rm BH}}$}\label{sec:SMBH vs Gal scaling relations}

\Cref{fig:mstar vs mbh scaling relation} shows the relation between the stellar mass of the host galaxies and SMBH masses for our fiducial set of parameters as well as for different assumptions regarding the SMBH location and merger treatment. In all cases, SMBHs reach similar masses ($M_{\rm BH}\gtrsim 10^{7.5}$ $M_\odot$) in massive galaxies ($M_* \gtrsim 10^{11}$ $M_\odot$). Different combinations of SMBH location and merger treatment result in different tracks ( see \citealt{Volonteri2012} for details) in the $M_* - M_{\rm BH}$ plane, notably at high redshift. For each case, efficient SMBH growth starts around a similar stellar mass threshold $\sim 10^{10}$ $M_\odot$.

The growth trend of SMBHs shown in \Cref{fig:mstar vs mbh scaling relation} can be divided into three phases. During phase I ($M_* < 10^{8.5} M_\odot$), the contribution of black hole mergers to the total SMBH mass is negligible. Instead, early SMBH growth is driven by accretion from the densest central region (upper panels in \Cref{fig:mstar vs mbh scaling relation}). In this scenario, the growth of SMBHs and their host galaxies follow the local scaling relation for AGN galaxies in \cite{ReinesVolonteri2015}. In contrast, the early growth of SMBHs is shifted if they reside in average density central regions, lower panels in \Cref{fig:mstar vs mbh scaling relation}.

During phase II ($8.5 < \log(M_*/M_\odot) < 10$), SMBH mergers play an important role. In particular, SMBH mergers contribute more to the growth of the SMBH mass than gas accretion. In contrast, SMBH growth stalls during this phase if SMBH mergers are not considered. In our sample, the contribution from mergers peaks at $2 < z < 6$ (see \Cref{fig:BH growth for different accretion contribution}). The impact of mergers is more pronounced when the seed mass is heavier. We refer the reader to \Cref{Appendix:Additional analysis} for the comparison of different seed masses.

The SMBH merger treatment and the seed mass choice appear not to have a major impact on the SMBH mass at low redshift in massive galaxies. In each case, we find $\sim 10^8 M_\odot$ SMBHs in $\sim 10^{11} M_\odot$ galaxies, see \Cref{fig:mstar vs mbh scaling relation}. While this result disagrees with other theoretical studies \cite{Shirakata::etal::2016, Park::etal::2016}, we suspect the origin of this difference is in the modelling of the SMBH accretion and stellar feedback. For any accretion model that has a black hole mass dependency in the form of BHAR $\propto M_{\rm BH}^{\rm p}$ for $p > 0$ (e.g. $p = 1/6$ for the GTDA model), the specific BHAR ($\rm sBHAR \equiv \rm{BHAR} / M_{\rm BH}$) scales with $M_{\rm BH}^{\rm p - 1}$. In the case of Bondi-like models, p is equal to $2$ and sBHAR scales linearly with the black hole mass. This means that heavier black holes grow faster than smaller black holes in Bondi-like models for the same environmental conditions.

SMBH masses are asymptotically insensitive to the seed mass if BH accretion is sublinear ($p < 1$) \citep{AnglesAlcazar::etal::2015, Angles-Alcazar+17a}. In contrast, BH growth histories are highly sensitive to the initial seed mass in the case of super-linear accretion ($p>1$), especially in the absence of AGN feedback and/or at early times when $M_{\rm BH}\sim M_{\rm seed}$. We note that the presence of feedback may \citep{Taylor::etal::2014} or may not \citep{Dubois::etal::2015} affect the sensitivity of SMBH growth on the seed mass.

Phase III of the SMBH growth starts with accelerated SMBH growth when the stellar mass of the host galaxy reaches $\sim 10^{10} M_{\odot}$, roughly coinciding with the time when the escape velocity of the central region becomes comparable to the velocity of galactic winds \citep{Angles-Alcazar+17b}. The transition around the threshold mass is an empirical result from the simulations \citep{Dubois::etal::2015, Habouzit::etal::2017, McAlpine::etal::2018}. Interestingly, the threshold stellar mass for efficient SMBH growth is similar to the divider between early and late type galaxies seen in both observations \citep{Kauffmann::etal::2003} and simulations \citep{Bower::etal::2017, Taylor::etal::2017}. We refer the reader to Byrne et al. (in prep; see also the discussion in \citealt{Stern::etal::2020}) for a more detailed analysis of the physical drivers of delayed vs. efficient SMBH growth. Finally, once galaxies reach a stellar mass of $M_* \gtrsim 10^{11} M_\odot$; this accelerated growth comes to an end. After this time, galaxies and SMBHs grow again at a similar rate.

The relation between the stellar bulge and SMBH mass shown in \Cref{fig:mbulge vs mbh scaling relation} is qualitatively similar to the relation between galaxy stellar mass and SMBHS mass (\Cref{fig:mstar vs mbh scaling relation}). Typically, the $M_{\rm b} - M_{\rm BH}$ relation predicted by our sample at high redshift falls below the local $M_{\rm b}-M_{\rm BH}$ scaling relation \cite{HaringRix2004}. Our post-processing analysis thus predicts that black hole masses are lower (or bulge masses and galaxy masses are larger) than expected from the local scaling relations.

We refer the reader to \Cref{fig:DAA vs our PP} for a comparison of the $M_{\rm b}$--$M_{\rm BH}$ scaling relation predicted by our post-processing analysis with the corresponding on-the-fly and post-processing predictions by \cite{Angles-Alcazar+17c}. The early growth of SMBHs is suppressed if SMBHs are placed in typical regions in the centre of galaxies instead of the densest regions (COM model vs MAX model). In either case, the resulting SMBH masses at low redshift are not affected by this SMBH placement choice. This result can be understood from \Cref{fig:halo center diff for MAX and COM} which shows that at late times ($z\ll4$) the centre of the galaxy becomes well defined. Both choices lead to virtually identical SMBH placements.

Our standard approach to determine the velocity dispersion assumes non-rotating bulges (see \Cref{sec:Methodology} for details). We also show the effect of replacing the velocity dispersion of the bulge with the velocity dispersion of all star particles in $1$ kpc, within the half stellar mass radius, and within the galactic radius in \Cref{fig:vel disp comparison}, finding little difference.

\subsubsection{SFR vs BHAR and Inferred Luminosities}\label{sec:SFR vs accretion scaling relations}

\begin{figure*}
  \centering
  \begin{tabular}{cc}
    \includegraphics[width=0.5\textwidth]{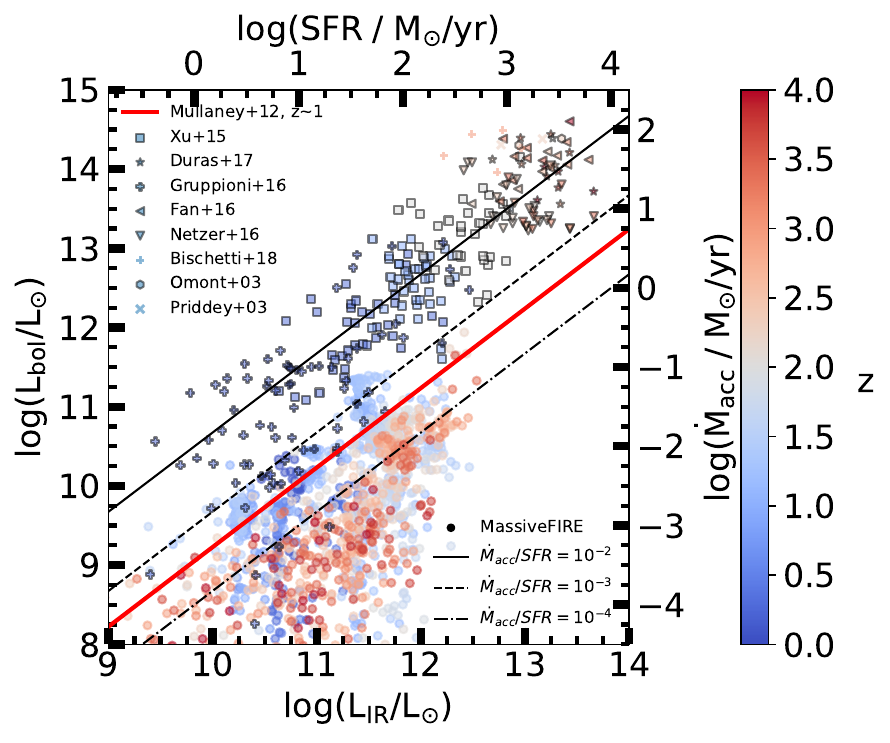}
       &
    \includegraphics[width=0.5\textwidth]{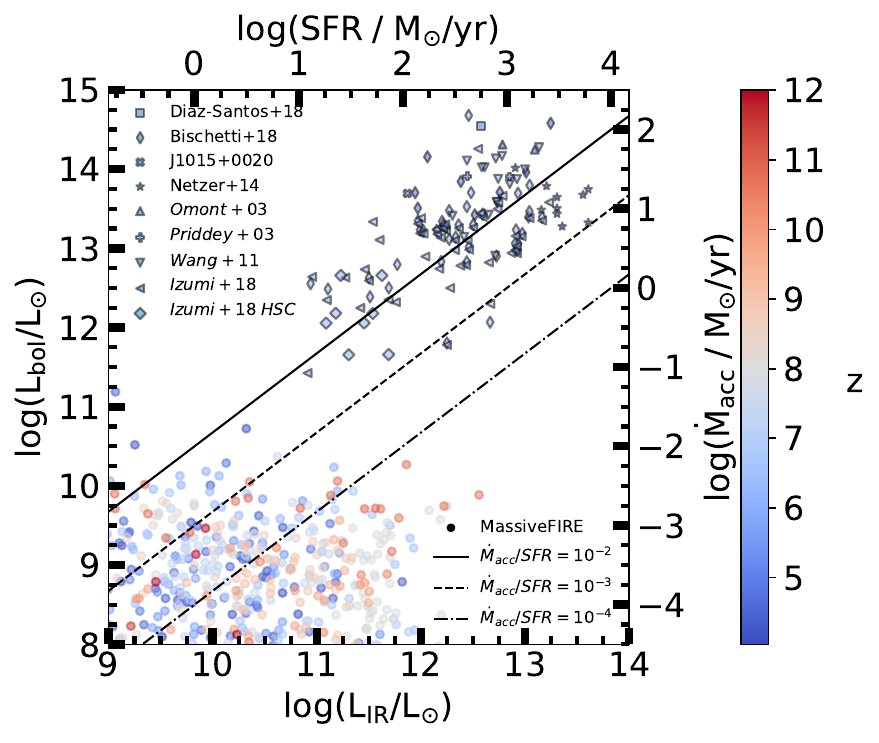}
  \end{tabular}
  \caption{Comparison of IR and bolometric luminosities derived from star formation and mass accretion rates with the observational data from literature \citep{Omont::etal::2003, Priddey::etal::2003, Wang::etal::2011, Lei::etal::2015, Fan::etal::2016, Gruppioni::etal::2016, Netzer::etal::2016, Duras::etal::2017, Bischetti::etal::2018, Diaz-Santos::etal::2018, Izumi::etal::2018}. Data from observations are shown with black edge colours, coloured dots without edge colours are from post-processing analysis. Our predictions are consistent with the IR luminosity of low-redshift sources. However, our model is unable to reproduce the bolometric luminosities of the most luminous observed AGNs, falling short by several orders of magnitude. This suggests that most of the black holes at low luminosities in the early Universe are below the detection limit of current observational surveys.}
  \label{fig:bolometric luminosity vs IR luminosity}
\end{figure*}

We convert our SFRs and BHARs into IR and bolometric luminosities via the following conversions: $L_{\rm IR}= {\rm SFR} \times 1.49 \times 10^{10} \; L_{\odot}$ \citep{Kennicutt1998} and $L_{\rm bol} = \eta / (1-\eta) \times \dot{M} c^2$, see \Cref{fig:bolometric luminosity vs IR luminosity}. Here $\eta$ is the radiative efficiency of the accreting black hole which is generally taken as $10\%$. In general, we find good agreement between our model predictions and observations at $z<2$ and large differences at high redshift. 

Specifically, \Cref{fig:bolometric luminosity vs IR luminosity} compares SFR and BHAR of our sample with the ones in galaxies hosting luminous AGN for $z \leq 4$ (left-hand panel) and $z \geq 4$ (right panel). The BHAR-SFR ratio scatters around a mean value of $1/10^3$, which is consistent with what \cite{Mullaney::etal::2012b} predict for main-sequence AGN at $z=1$. High SFRs and BHARs seem to be  characteristic features of observed AGN beyond $z=2$, while \textsc{\small{MassiveFIRE}} simulations produce relatively moderate SFRs and BHARs. Assuming our sample is a good representation of less luminous AGN and galaxies at high redshift, we predict a large number of yet unobserved low-luminosity AGN at high redshift. 

Observation and theory show that the average BHAR and the average SFR correlate well \citep{Mullaney::etal::2012b, Calhau::etal::2017, Dai::etal::2015, Volonteri::etal::2015b}. While the cause of this correlation is not yet fully understood, several explanations have been proposed. The first explanation refers to a common cause. In particular, gas reservoirs in the galaxy are both the primary source for BH feeding and SF \citep{AnglesAlcazar::etal::2015}. An alternative explanation is the idea that AGN activity may drive SF. For instance, the high-velocity outflows from AGN can sweep the gas away and pierce a cavity along its way but also trigger SF by induced pressure of the edges \citep{Cresci::etal::2015}. Hence, AGN feedback may enhance SF and be responsible for its suppression at the same time \citep{Maiolino::etal::2017, BestHeckman2012, Ivison::etal::2012, Norris::etal::2012}.

\begin{figure*}
    \includegraphics[width=\textwidth]{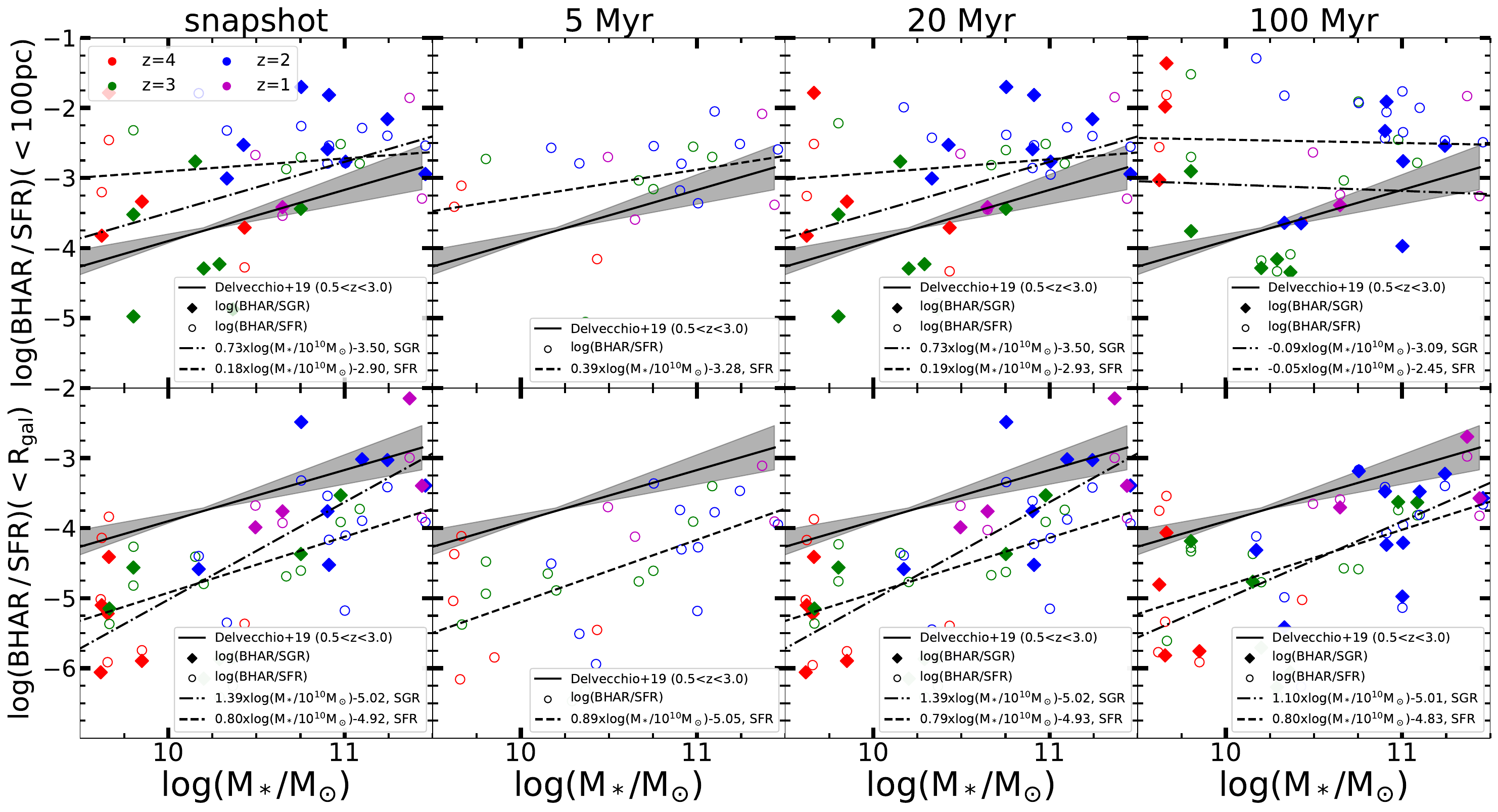}
    \caption{The relations between the galaxy stellar mass and the black hole accretion rate (BHAR) in the GTDA model normalized to the star formation rate (SFR) or the stellar growth rate (SGR). The upper panels show $\rm{BHAR/SFR}$ (circles) and $\rm{BHAR/SGR}$ (diamonds) within the central $100$ pc region of galaxies in the A-C series of \textsc{\small{MassiveFIRE}}, while the lower panels show the same relations based on the total SFR and SGR. The columns refer to different averaging times used to compute the SFR and SGR: (1st column) time between snapshots, (2nd column) 5 Myr averaging time, (3rd column) 20 Myr averaging time, (4th column) 100 Myr averaging time. The time difference between snapshots varies from $10$ Myr at high redshift to $25$ Myr at $z=2$, thus preventing us from calculating SGRs in the 2nd column. Dashed and dot-dashed lines show best fits of the $M_*-\rm{BHAR/SFR}$ and $M_*-\rm{BHAR/SGR}$ relations from our simulated sample at $z=1-4$. Data points are color-coded according to redshift: $z=4$ (red), $z=3$ (green), $z=2$ (blue), and $z=1$ (purple). The relation between BHAR and SFR (or SGR) shows less scatter when galaxy-integrated SFRs (or SGRs) are used.
    Solid black lines show the fit from \protect\cite{Delvecchio::etal::2019} based on SFRs of $0.5<z<3$ galaxies. It should be compared to the model predictions (dashed lines) in the lower panels. The BHAR normalized to SFR (or SGR) increases with the stellar mass of galaxies in \textsc{\small{MassiveFIRE}} in qualitative agreement with observations. However, we predict a large number of galaxies with BHARs that are an order of magnitude lower than expected from the fit by \protect\cite{Delvecchio::etal::2019}. Overall, slope, normalization, and amount of scatter can vary widely depending on how SFRs and SGRs are calculated.}
    \label{fig:bhar vs sfr}
\end{figure*}

\Cref{fig:bhar vs sfr} shows how the ratios of BHAR/SFR and BHAR/SGR scale with the total stellar mass of host galaxies in the A, B, and C-series of \textsc{\small{MassiveFIRE}}. We also investigate how the relation between BHAR and SFR (or SGR) depends on spatial scale ($<R_{\rm gal}$ and $<100$ pc) as well as on the averaging timescale $t_{\rm avg}$ of the SFR and SGR (ranging from 5 Myr to 100 Myr). The description of how SFRs and SGRs are calculated is provided in section \ref{sect:postprocessing}. We caution that our results are inferred by combining galaxies over a range of redshifts with lower masses being probed at higher redshift. These scaling relations may thus look different if measured for a sample of galaxies of different mass at a fixed redshift.

When employing galaxy-wide SFRs or SGRs (bottom row of \Cref{fig:bhar vs sfr}), we find that the slope of the $M_*$--BHAR/SFR relation predicted by our study ($\sim 0.80 \pm 0.22$) is in good agreement with the slope ($\sim 0.73[+0.22,-0.29]$) inferred based on a compilation of observations of star-forming galaxies by \cite{Delvecchio::etal::2019}. However, we predict a normalization of the relation that is approximately an order of magnitude smaller than observed. Interestingly, the highest BHAR/SFR ratios that we obtain at a given stellar mass match observed values well suggesting that observational selection biases against low-luminosity AGN could be an explanation for the difference in normalization. If we aim to match the normalization by, e.g., boosting the mass retention rate from $5\%$ to $25\%$, our predicted SMBHs become too massive at low redshift and thus inconsistent with the local $M_*-M_{\rm BH}$ scaling relation.

The analogous relations measured within the central 100 pc (top row of \Cref{fig:bhar vs sfr}) are significantly different. They have significantly larger normalization and a shallower slope. Furthermore, we find that the scatter of the BHAR/SFR and BHAR/SGR relations remains very substantial even when SFRs and SGRs are measured within the central regions of galaxies. In fact, we find that using central SFRs and SGRs result in a \emph{larger} scatter than using galaxy-wide SFRs and SGRs.

This latter result is perhaps somewhat unexpected. The BHAR should better correlate with the nuclear SFR rather than the total SFR, except during galaxy mergers, since the timescales of the nuclear star formation and the accretion onto SMBHs are close to the dynamical timescale of matter in the nuclear region ($\sim 100$ pc) \cite{HopkinsQuataert2010}. A similar conclusion was reached by \cite{Yang::etal::2019}. However, during mergers the total SFR of the host galaxy correlates well with BHAR because global dynamics becomes more important than the local processes in terms of angular momentum loss \citep{Volonteri::etal::2015b}. \Cref{fig:bhar vs sfr} shows that the difference in scatter between the central and the whole-galaxy  BHAR/SFR relations is most noticeable at high redshift, when mergers are expected to be much more frequent \citep{Netzer::etal::2016, Silva::etal::2021}. Hence, high merger rates at high $z$ provide a possible explanation of our result \citep{Duncan::etal::2019}.

The precise slope, normalization, and scatter of the relation between $M_*$ and BHAR/SFR (or BHAR/SGR) in \Cref{fig:bhar vs sfr} varies depending on the averaging timescale (the different columns of the figure show averaging times ranging from 5 to 100 Myr). First, the slope decreases when the averaging timescales are increased. Secondly, the scatter increases especially when comparing the $M_*$-BHAR/SFR relation on 100 pc scales for short and long averaging times. Overall, however, the choice of the averaging timescale plays a much smaller role than the choice of the spatial scale over which SFRs (or SGRs) are measured.

The figure also reveals that the BHAR/SGR relation at $z=1-4$ has a steeper slope ($\sim 1.40 \pm 0.25$) than the BHAR/SFR relation. This finding implies that the SFR increases more strongly with increasing stellar mass than the SGR, while the opposite behavior would have been expected based on the increased galaxy merging activity in massive galaxies \citep{Ferreras::etal::2014, Ferreras::etal::2016, Zahid::etal::2019}. However, the SGR differs from the SFR not only by the additional merger contribution, but also by the decrease in stellar mass of stars already present at time $t-t_{\rm avg}$, i.e., those formed before $t-t_{\rm avg}$. This reduction in stellar mass can arise in multiple ways. First, supernovae and stellar winds  return mass from the stellar component to the gas component. The contribution of this stellar mass loss to the SGR (when averaged over $t_{\rm avg}$) becomes less severe with increasing $t_{\rm avg}$, in agreement with the results shown in \Cref{fig:bhar vs sfr}, as a single stellar population loses much of its mass early on \citep{Chabrier2003}. Secondly, stellar mass may also be lost when stellar particles migrate outside the fixed physical radius used to calculate the SGRs. We thus conclude that the steeper scaling of the $M_*$ - BHAR/SGR relation reflects the higher importance of stellar mass loss compared with galaxy mergers in our sample.

\subsubsection{Redshift evolution of $\mathrm{M_*-M_{BH}}$ }\label{sec:Redshift Evolution of Mstar vs Mbh Scaling Relation}

\begin{figure}
  \centering

  \includegraphics[width=\linewidth]{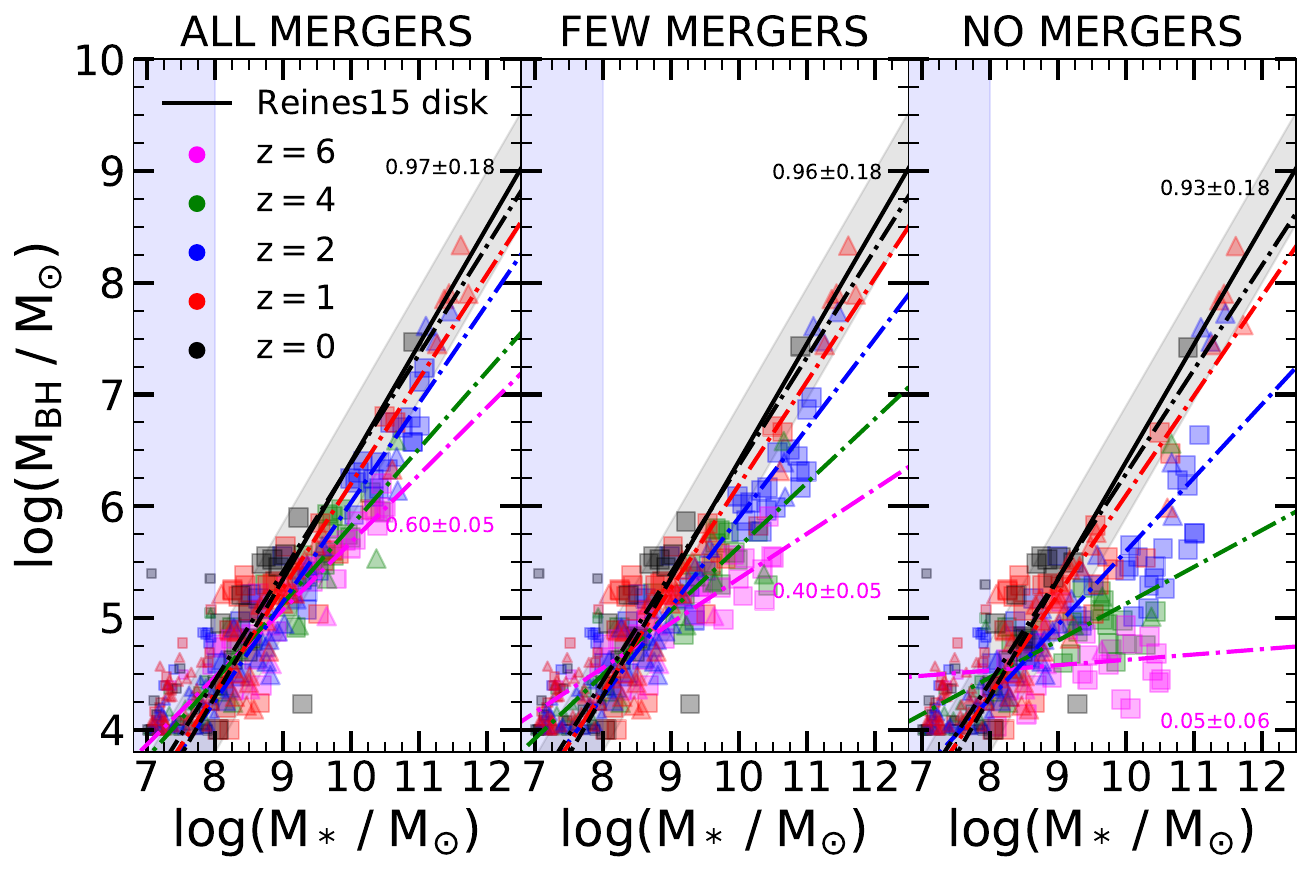}

  \caption{Effect of SMBH mergers on $M_*-M_{\rm BH}$ scaling relation for the full sample of 34 \textsc{\small{FIRE}} simulations from the \textsc{\small{MassiveFIRE}} suite for our fiducial model parameters and the densest central region (MAX). Triangles and squares show data belonging to simulations that were run with  \textsc{\small{FIRE-1}} and \textsc{\small{FIRE-2}} physics, respectively. Different colours represent different redshifts. The solid line stands for the fit function from \protect\cite{ReinesVolonteri2015}, while the dashed line is its extrapolation. Dotted dashed lines show linear fits to the $\log M_*-\log M_{\rm BH}$ at $z=0-6$. The fit only uses galaxies with $M_* \geq  10^8 M_\odot$ to minimize the impact of the SMBH seed mass choice on the slope. The slope of the $M_*-M_{\rm BH}$ scaling relation is close to linear at $z=0$ ($0.93 \pm 0.18, 0.96 \pm 0.18, 0.97 \pm 0.18$) while the slope of \protect\cite{ReinesVolonteri2015} for disk galaxies is $1.05 \pm 0.11$, but decreases with increasing redshift (e.g., $0.05-0.6$ at $z=6$ depending on the SMBH merger model). A model with a higher number of SMBH mergers typically results in a steeper slope. The importance of mergers is particularly evident at high redshift, while the slope at the late times is not affected much, in agreement with our previous finding (\Cref{fig:BH growth for different accretion contribution}) that SMBH mass at low redshift is primarily set by gas accretion and not SMBH mergers.}
  \label{fig:scaling relations MDC}
\end{figure}

\Cref{fig:scaling relations MDC} presents how the $M_*-M_{\rm BH}$ scaling relation changes with redshift. To arrive at quantitative estimates of the slope of the scaling relation, we fit log stellar masses of all galaxies above $10^8 M_\odot$ in our sample as well as their SMBHs at redshifts $z=0, 1, 2, 4, 6$ with a linear function. We also estimate the slopes for galaxies with stellar masses between $10^8$ $M_\odot$ and $10^{10.5}$ $M_\odot$ to mitigate biases due to the increasing stellar mass with redshift for the galaxies in our sample.

\begin{table}
\centering
\begin{tabular}{c c c}
&& \, $8 \geq  \log{M_*} < 10.5$
\end{tabular}
\begin{tabular}{l|l l l}

Merger Model & z & Slope & Norm. \\
\hline
all mergers      & 6.0   & 0.603 $\pm$ 0.057 & 5.680 $\pm$ 0.062 \\
all mergers      & 4.0   & 0.656 $\pm$ 0.073 & 5.789 $\pm$ 0.070 \\
all mergers      & 2.0   & 0.819 $\pm$ 0.058 & 5.965 $\pm$ 0.066 \\
all mergers      & 1.0   & 0.745 $\pm$ 0.092 & 5.960 $\pm$ 0.120 \\
all mergers      & 0.0   & 0.790 $\pm$ 0.324 & 6.117 $\pm$ 0.480 \\
\hline
few mergers  & 6.0   & 0.401 $\pm$ 0.055 & 5.354 $\pm$ 0.0600 \\
few mergers  & 4.0   & 0.508 $\pm$ 0.076 & 5.570 $\pm$ 0.073 \\
few mergers  & 2.0   & 0.696 $\pm$ 0.056 & 5.788 $\pm$ 0.064 \\
few mergers  & 1.0   & 0.717 $\pm$ 0.092 & 5.920 $\pm$ 0.120 \\
few mergers  & 0.0   & 0.788 $\pm$ 0.322 & 6.099 $\pm$ 0.477 \\
\hline
no mergers  & 6.0   & 0.063 $\pm$ 0.065 & 4.644 $\pm$ 0.070 \\
no mergers  & 4.0   & 0.194 $\pm$ 0.067 & 4.991 $\pm$ 0.064 \\
no mergers  & 2.0   & 0.421 $\pm$ 0.058 & 5.345 $\pm$ 0.066 \\
no mergers  & 1.0   & 0.653 $\pm$ 0.094 & 5.799 $\pm$ 0.123 \\
no mergers  & 0.0   & 0.676 $\pm$ 0.310 & 5.909 $\pm$ 0.460 \\
\\
&& \, $\log{M_*} \geq  8$ \\
\hline
all mergers      & 6.0   & 0.603 $\pm$ 0.054 & 5.680 $\pm$ 0.057 \\
all mergers      & 4.0   & 0.691 $\pm$ 0.070 & 5.824 $\pm$ 0.066 \\
all mergers      & 2.0   & 0.887 $\pm$ 0.038 & 6.035 $\pm$ 0.041 \\
all mergers      & 1.0   & 0.936 $\pm$ 0.053 & 6.204 $\pm$ 0.069 \\
all mergers      & 0.0   & 0.972 $\pm$ 0.183 & 6.390 $\pm$ 0.266 \\
\hline
few mergers  & 6.0   & 0.400 $\pm$ 0.052 & 5.352 $\pm$ 0.055 \\
few mergers  & 4.0   & 0.572 $\pm$ 0.078 & 5.635 $\pm$ 0.074 \\
few mergers  & 2.0   & 0.800 $\pm$ 0.043 & 5.895 $\pm$ 0.047 \\
few mergers  & 1.0   & 0.926 $\pm$ 0.054 & 6.186 $\pm$ 0.070 \\
few mergers  & 0.0   & 0.965 $\pm$ 0.182 & 6.363 $\pm$ 0.264 \\
\hline
no mergers  & 6.0   & 0.048 $\pm$ 0.061 & 4.626 $\pm$ 0.065 \\
no mergers  & 4.0   & 0.329 $\pm$ 0.096 & 5.129 $\pm$ 0.091 \\
no mergers  & 2.0   & 0.657 $\pm$ 0.062 & 5.600 $\pm$ 0.067 \\
no mergers  & 1.0   & 0.890 $\pm$ 0.058 & 6.100 $\pm$ 0.075 \\
no mergers  & 0.0   & 0.932 $\pm$ 0.179 & 6.292 $\pm$ 0.260 \\

\end{tabular}
\caption{Linear regression of the $M_*-M_{\rm BH}$ scaling relation shown in \Cref{fig:scaling relations MDC} as $\log M_{\rm BH} = \rm{Slope} \times \log(M_*/10^{10} M_\odot) + \rm{Norm}$. The first and second columns shows the mergers model used in the post-processing analysis and the redshift. The third and the fourth columns show slope and normalization of the linear fit when we include only galaxies with $8 \geq  \log{M_*}/M_\odot < 10.5$ from our sample. The lower table lists the slope and normalization that we obtain when we include all galaxies with  $M_* \geq  10^8 M_\odot$. SMBH mergers have a significant impact on the slope and normalization of the $M_*-M_{\rm BH}$ scaling relation, especially at $z\geq 2$. In the \textsc{all mergers} and \textsc{few mergers} models the slope and normalization evolve much more gradually with redshift (with only mild changes since $z=4$) than in the \textsc{no mergers} model.}
\label{tab:slope of different merger treatments}
\end{table}

We fit the slope and normalization of the $M_*-M_{\rm BH}$ relation shown in \Cref{fig:scaling relations MDC} via linear regression, see \Cref{tab:slope of different merger treatments}. In the \textsc{all mergers} and \textsc{few mergers} models, the slope and normalization evolve only mildly with redshift, especially at $z\leq 4$. In contrast, the \textsc{no mergers} model shows a significant evolution of the $M_*-M_{\rm BH}$ relation with shallower slopes and lower normalizations at higher redshift. We perform the linear regression both on our complete galaxy sample and on galaxies with stellar masses between $10^8$ $M_\odot$ and $10^{10.5}$ $M_\odot$, finding very similar results in either case. Clearly, the frequency of SMBH mergers play a critical role in shaping the $M_*-M_{\rm BH}$ relation at higher redshift.

The left panel shows the $M_*-M_{\rm BH}$ scaling relation in the \textsc{all mergers} model. Overall, the predictions of this model are in line with the local $M_*-M_{\rm BH}$ scaling relation from \cite{ReinesVolonteri2015}. The dotted dashed lines are the best fit lines for \textsc{MassiveFIRE} data selected at certain redshifts. The slope of these fit lines increases from $z=6$ to $z=0$. The middle panel of \Cref{fig:scaling relations MDC} show the predictions when we consider a more realistic model of SMBH merging (the \textsc{few mergers} model). Here, the $M_*-M_{\rm BH}$ scaling relation has a noticeably shallower slope, especially at $z>2$, compared to the \textsc{all mergers} ansatz. When SMBH mergers are not considered (\textsc{no mergers}), SMBH masses at $z=2,4,6$ are well below the local scaling relation but they still catch up to the local relation at $z<1$. In summary, SMBH mergers have a significant effect on the $M_*-M_{\rm BH}$ scaling relation, especially at high redshift. Hence, by accurately measuring the redshift evolution of the galaxy-SMBH scaling relation, it may be possible to constrain the rates of SMBH merging.

\begin{figure*}
    \centering
    \includegraphics[width=\linewidth]{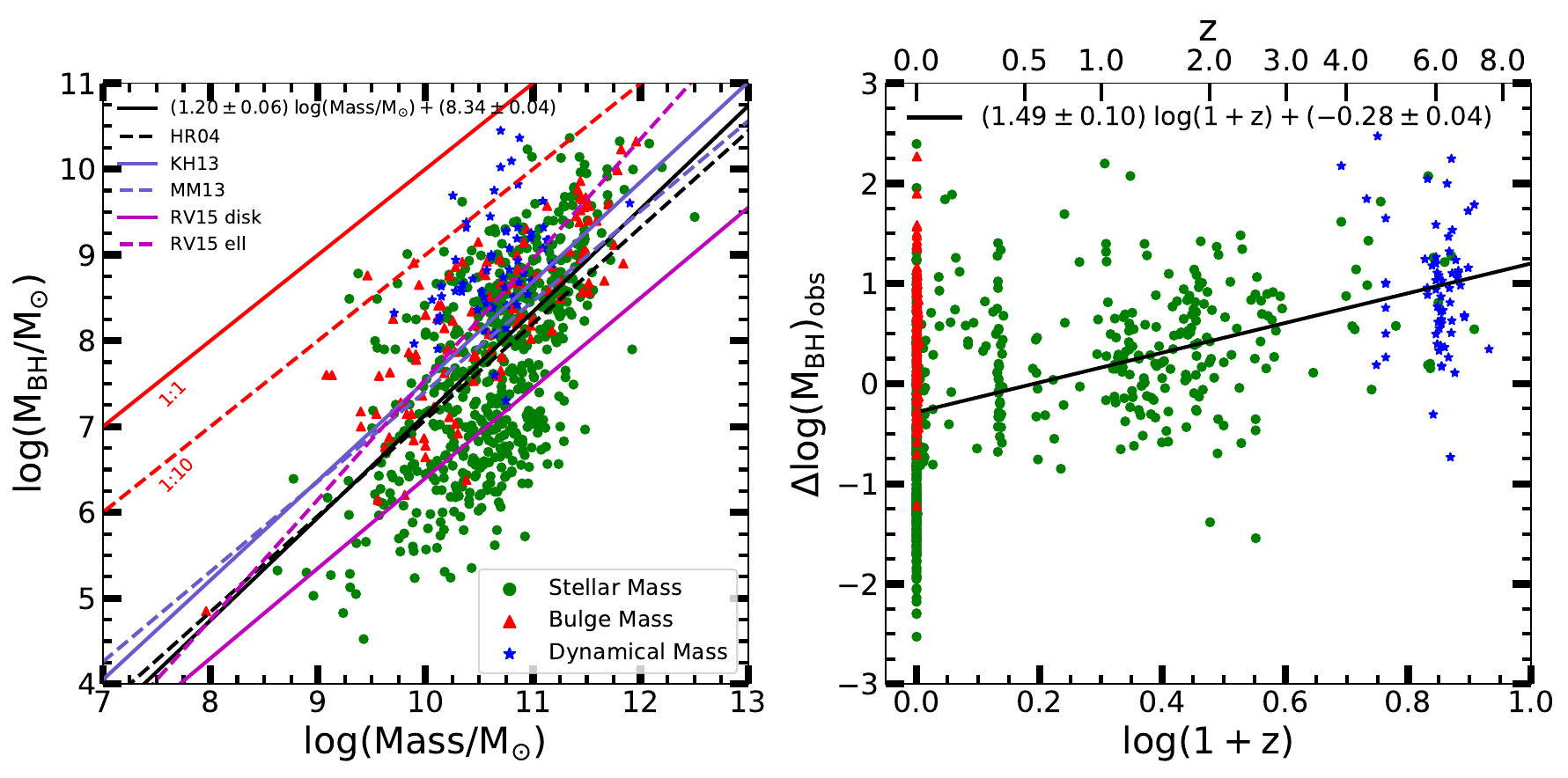}
  
    \caption{\textbf{Left Panel:} Observed $M_*,M_b,M_{\rm dyn} - M_{\rm BH}$ scaling relation for $7.5 \lesssim z \lesssim 0$. Red, green and blue data points indicate, respectively, bulge mass, stellar mass , and dynamical mass for both panels. Bulge masses put a lower limit to the stellar mass of the galaxies while dynamical mass is the upper limit for the galaxy stellar mass. Different lines correspond to the best fit lines taken from literature, HR04 \citep{HaringRix2004}, KH13 \citep{KormendyHo2013}, MM13 \citep{McConnellMa2013}, RV15 \citep{ReinesVolonteri2015}. The black solid line is our best fit line to the whole data. Red solid and dashed lines show $1:1$ and $1:10$ mass ratios. \textbf{Right Panel:} Observed redshift evolution of $\Delta \log \left( M_{\rm BH} \right)_{\rm obs}$ calculated using a best fit line (black solid line) of left panel. The black solid line shows the best fit line for all data in the form of $\Delta \log(M_{\rm BH})_{\rm obs} = \delta_1 + \delta_2 \log(1+z)$, which suggests a relatively steep redshift evolution ($\propto 1.49 \pm 0.10$) of the $M_* - M_{\rm BH}$ scaling relation from $z \sim 0$ to $z=7.54$, compared to what \protect\cite{Merloni::etal::2010}  and \protect\cite{Ding::etal::2020} found: $\delta_2 = 0.68 \pm 0.12$ for $1.1<z<2.2$ and $\delta_2 = 1.03 \pm 0.25$ for $0<z<1.7$, respectively.} 
    \label{fig:redshift evolution of mstar vs mbh scaling relation}
\end{figure*}

\Cref{fig:redshift evolution of mstar vs mbh scaling relation} contains a literature compilation of the correlation between the properties of host galaxies such as stellar \citep{Merloni::etal::2010, KormendyHo2013, ReinesVolonteri2015, Bentz::etal::2018}, bulge \citep{Savorgnan::etal::2016, Sahu::etal::2019b}, and dynamical mass \citep{Maiolino::etal::2005, Riechers::etal::2009, Venemans::etal::2012, Wang::etal::2013, Venemans::etal::2013, Kimball::etal::2015, Banados::etal::2015, Willott::etal::2015, Bischetti::etal::2016, Wang::etal::2016, Venemans::etal::2016, Trakhtenbrot::etal::2017, Tsai::etal::2018, Decarli::etal::2018, Eilers::etal::2018, Izumi::etal::2018, Feruglio::etal::2018} and SMBH mass for $0 \lesssim z \lesssim 7.5$. Here, bulge mass represents a lower limit for the stellar mass of the host galaxy while dynamical mass reflects an upper limit for the stellar mass. SMBH masses in this sample are determined via spectral lines for the AGN host galaxies and dynamics for the elliptical galaxies in the local Universe. Combining the various observational data we find that $\log M_{\rm BH}\sim (1.20\pm 0.06) \log({ M_{\rm gal} / 10^{11} \, M_\odot})+ (8.34 \pm 0.04)$ where $M_{\rm gal}$ is the stellar mass for most observations, and bulge or dynamical mass for the remaining observations.

We also study how the offset of SMBH mass from this average relation, $\Delta \log(M_{\rm BH})_{\rm obs}$, evolves with redshift. We fit the observational data in \Cref{fig:redshift evolution of mstar vs mbh scaling relation} via linear regression finding $\Delta \log (M_{\rm BH})_{\rm obs} = (1.49 \pm 0.10) \times \log(1+z) - (0.28 \pm 0.04)$ pointing to a super linear correlation. Measuring the offset with respect to the observed  $M_b-M_{\rm BH}$ or $M_{\rm dyn}-M_{\rm BH}$ results in slopes which vary between $1.38 \pm 0.11$ to $1.57 \pm 0.10$. \cite{Merloni::etal::2010} report a positive slope of $\sim 0.68 \pm 0.12$ for their sample between $1<z<2.2$, as well as \cite{Ding::etal::2020} reporting $1.03 \pm 0.25$ for $0<z<1.7$. 

The slope we obtain with the observational sample is $\approx 2.5$ times larger than the slope reported by \cite{Merloni::etal::2010}. Also, the offset of SMBH mass appears to steadily increase with redshift. The difference between our re-analysis of observational data and the results by \cite{Merloni::etal::2010} are thus likely attributable to selection effect as many new sources at high redshift were discovered in the last decade.

\begin{figure*}
    \centering
    \includegraphics[width=\textwidth]{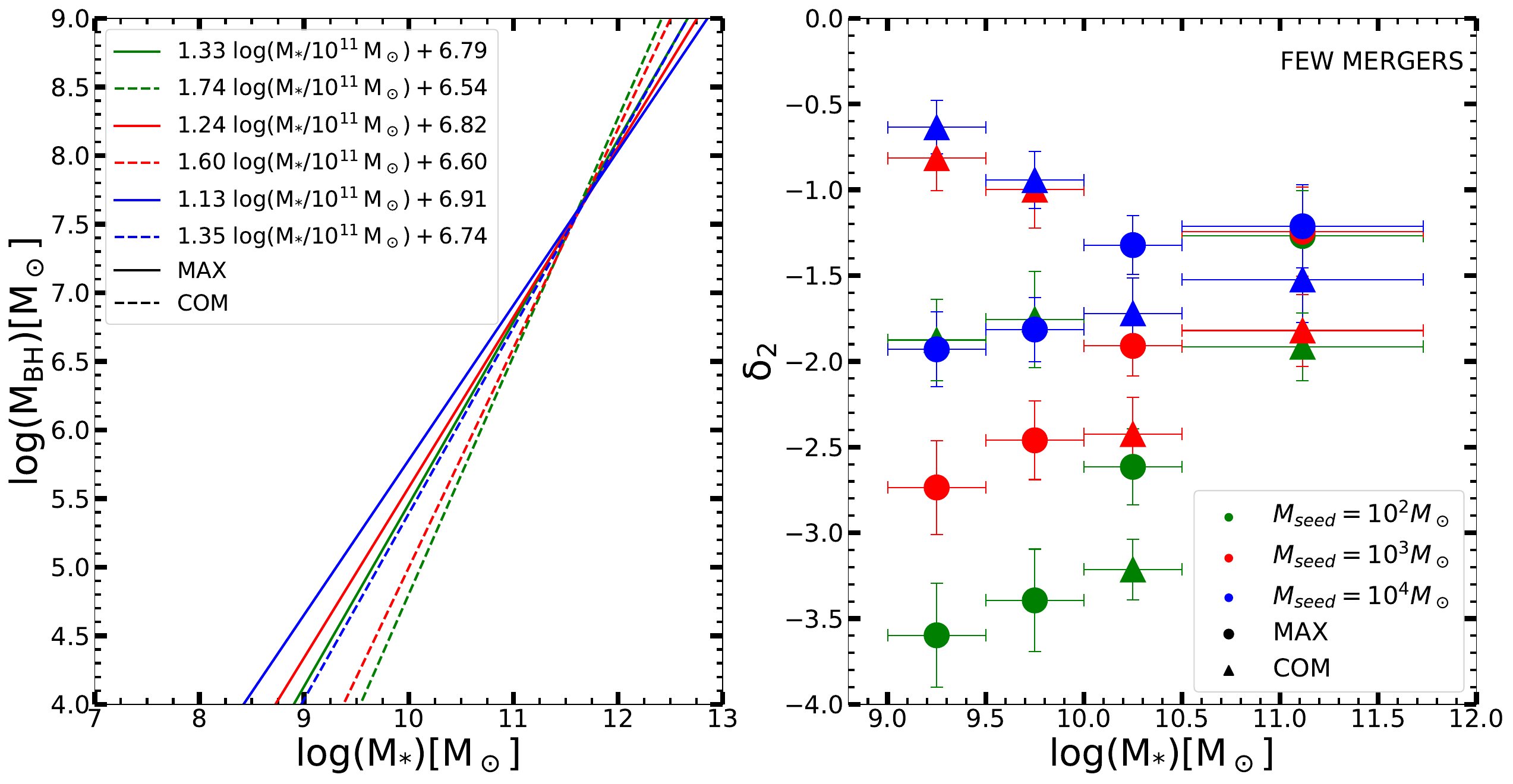}
    
    \caption{\textbf{Left Panel}: Best fit lines of the $M_*-M_{\rm BH}$ relation in the \textsc{few mergers} model for various seed masses ($10^2$ $M_\odot$ in green, $10^3$ $M_\odot$ in red, and $10^4$ $M_\odot$ in blue) and centring methods (solid lines for MAX, dashed lines for COM). We calculate $\rm \Delta \log(M_{BH})_{pp}$ (see text) as the offset from these best fit lines. \textbf{Right Panel}: Change in slope ($\delta_2$) for different stellar mass bins in the "\textsc{few mergers}" model. Here, $\delta_2$ represents the slope of the best fit lines in $\mathrm{\log(1+z)}$ vs $\mathrm{\Delta \log(M_{BH})_{pp}}$, see \Cref{fig:redshift evolution of scaling relation}. Different colors refer to different seed masses (see legend). Dots and triangles show results for MAX and COM centring methods, respectively.  
    The $x$ error bars indicate half the size of the stellar mass bins, while $y$ error bars show the fit error of the slope, $\delta_2$. The slope is generally negative and tends to increase with increasing stellar mass, except in the COM centring model with $M_{\rm seed}\geq 10^3$ $M_\odot$.}
     
    \label{fig:delta2 vs mstar}
\end{figure*} 

The result of a similar analysis for the \textsc{\small{MassiveFIRE}} sample is shown in \Cref{fig:delta2 vs mstar} for a variety of seed masses, centring methods and stellar mass bins. A detailed version of \Cref{fig:delta2 vs mstar} is shown in \Cref{fig:redshift evolution of scaling relation}. We only include SMBHs with host galaxies that have stellar masses larger than $10^9 \; M_\odot$, to stay consistent with the observational data from the literature. The left panel of \Cref{fig:delta2 vs mstar} contains the best fit lines for different cases. The offset of $M_{\rm BH}$ from the best fit line ($\Delta \log(M_{\rm BH})_{\rm pp}$) is computed for post-processing data following the same method in \Cref{fig:redshift evolution of mstar vs mbh scaling relation}. However, our simulations show that the offset generally decreases with redshift on the right panel of \Cref{fig:delta2 vs mstar}, i.e., SMBHs at high redshifts tend to be undermassive compared to the local $M_* - M_{\rm BH}$ scaling relation, see \Cref{fig:redshift evolution of scaling relation}. Consequently, our simulations predict a large number of low-luminosity AGN in high redshift galaxies. The result holds for all different post-processing models. Furthermore, the most negative slopes are obtained in low mass galaxies ($M_*<10^{10} \, M_\odot$) when SMBHs are placed at the densest regions (MAX) and in massive galaxies ($M_*>10^{10} \, M_\odot$) when a COM placing is used. The difference between our simulation results and observational data may be explained by selection effects that bias the samples of observed high redshift AGN. Efforts in the search for the low-luminosity AGNs in the early Universe (i.e. Subaru High-z Exploration of Low-luminosity Quasars Project, \citealt{Matsuoka::etal::2016}) are therefore paramount to better constrain the redshift evolution of $\Delta \log(M_{\rm BH})_{obs}$.

\subsection{The slope of the \protect{$\mathrm{M_*-M_{BH}}$} scaling relation} \label{sec:The drive of the slope of the Mstar MBH scaling relation}

The slope of the $M_*-M_{\rm BH}$ scaling relation contains essential information about the growth trends of SMBHs and their host galaxies. In particular, it is set by the interplay of the stellar mass, the black hole mass, the black hole accretion rate, and the stellar growth rate, see \Cref{eq:slope of scaling relation in toy model}. In this section we will discuss the slope of the trajectory of black holes evolving in the $M_*-M_{\rm BH}$ plane. We note that this slope is not strictly identical to the slope of the $M_*-M_{\rm BH}$ relation of a population of galaxies with different stellar masses at a fixed redshift. However, for the latter we find only a small amount of redshift dependence at $z<4$ implying that the slope of trajectories in the $M_*-M_{\rm BH}$ plane will be similar to the slope of a population of galaxies and SMBHs at fixed redshift for $z<4$.

\begin{equation}
    \mathrm{slope = \frac{d log M_{BH}}{d log M_*} \;} \mathrm{=} \frac{\mathrm{sBHAR}}{\mathrm{sSGR}} \mathrm{=} \frac{\mathrm{M_*}} {\mathrm{M_{BH}}} \mathrm{\times } \frac{\mathrm{BHAR}}{\mathrm{SGR}}
    \label{eq:slope of scaling relation in toy model}
\end{equation}

According to \Cref{eq:slope of scaling relation in toy model}, the slope is larger (smaller) than unity when the sBHAR exceeds (is less than) the sSGR. The slope goes to zero when the black hole ceases to grow, and it reaches infinity for a growing SMBH in a non-growing galaxy. When the specific growth rates of the SMBH equals that of its host galaxy, the slope is unity.

\begin{figure*}
    \centering
    \includegraphics[width=\textwidth]{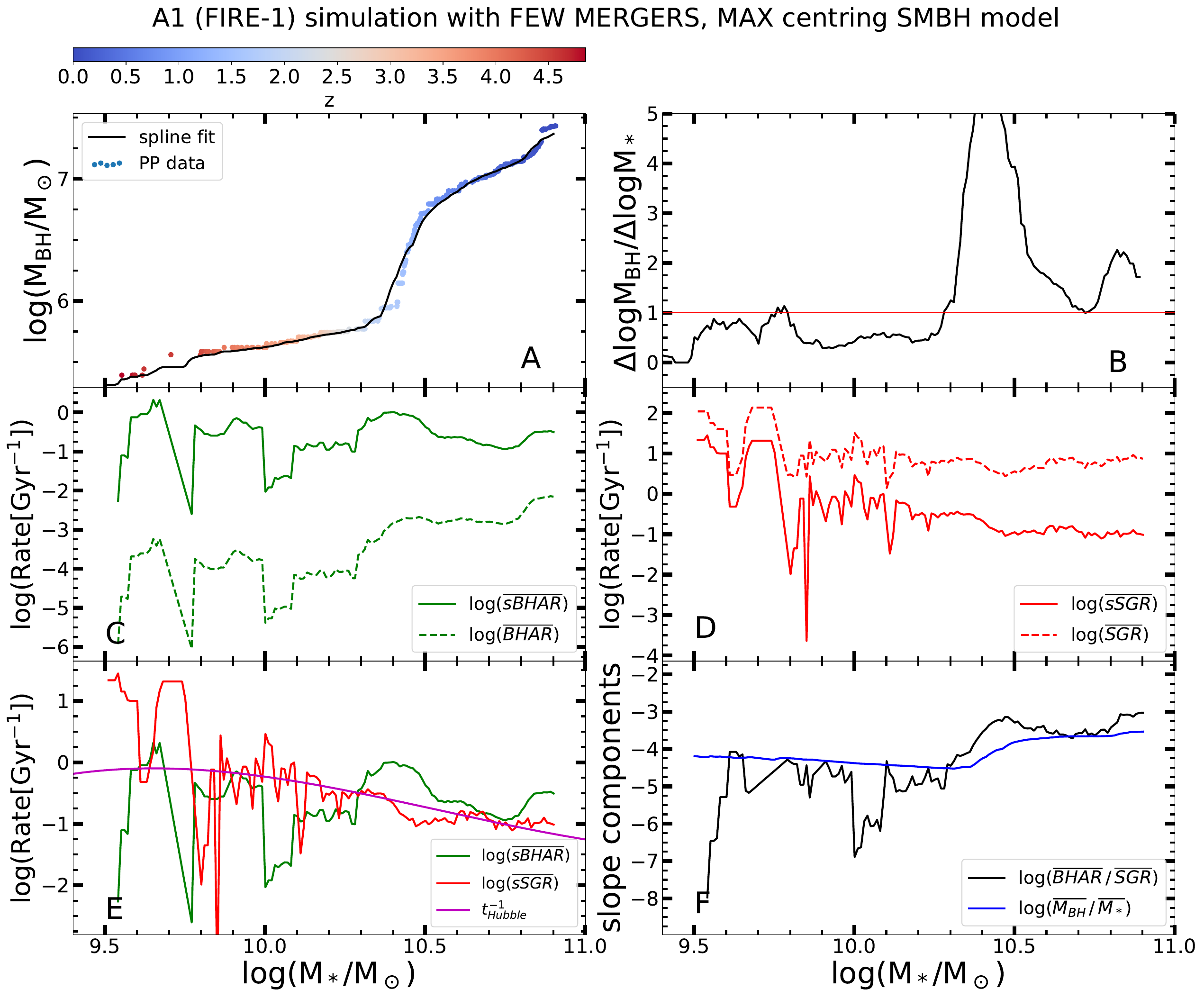}
    \caption{Factors that affect the slope of the $M_*-M_{\rm BH}$ scaling relation for simulation the A1 run with FIRE-1 physics. Panel A shows the predicted $M_*-M_{\rm BH}$ scaling relation for the \textsc{few mergers} model and MAX centring method. The coloured data points are from our post-processing analysis, and the solid black line is the smoothed and averaged $M_*-M_{\rm BH}$ relation with a bin size of $\Delta \log(M_*) = 0.1 \, M_\odot$ and a step size of $0.01\, M_\odot$. The colorbar indicates the redshift of the coloured data points in Panel A. The instantaneous slope ($\equiv \Delta \log M_{\rm BH} / \Delta \log M_*$) of $M_*-M_{\rm BH}$ is shown with a solid black line in Panel B. Panel C and D show how the average specific growth rates of both the SMBH and the host galaxy change over stellar mass. For $\log(M_*) \sim 10.4$, the $\overline{sBHAR}$ reaches its maximum while the $\overline{sSGR}$ reaches its minimum. Panel E includes the comparison of the inverse Hubble time ($t_{\rm Hubble}^{-1}$, magenta line), average specific growth rate history of the SMBH ($\overline{sBHAR}$, green line) and the host galaxy ($\overline{sSGR}$, blue line). The difference between the green and red lines determines the slope of the $M_*-M_{\rm BH}$ scaling relation. The final panel shows the components of \Cref{eq:slope of scaling relation in toy model} that are responsible for the shape and the normalisation of the slope in Panel B. The difference between the black and blue lines gives the slope where BHAR seems responsible for its trend.}
    \label{fig:slope of Mstar vs MBH}
\end{figure*}

Panels A and B of \Cref{fig:slope of Mstar vs MBH} show how the trajectory and the slope of the most massive galaxy in the simulation A1 evolves in the \textsc{few mergers} model and with MAX centring. Here, we calculate the slope based on the (smoothed) trajectory of the galaxy and its SMBH in the  $M_*-M_{\rm BH}$ plane. More precisely, the smoothed trajectory is calculated with the help of sliding bins in log stellar mass of width 0.1 and with subsequent shifts of 0.01. The slope is then calculated from the smoothed trajectory using a 0.1 dex in stellar mass. 

Subsequently, we analyze the various terms in \Cref{eq:slope of scaling relation in toy model} to see how they affect the slope. According to the panel C of \Cref{fig:slope of Mstar vs MBH}, the sBHAR shows significant variations but not a strong evolutionary trend, except for a moderate decrease with increasing stellar mass when $M_*\gtrsim 10^{10.3}$ $M_\odot$. Overall, the sBHAR mostly lies between $10^{-1}$ and 1 Gyr$^{-1}$. In contrast, the sSGR in panel D decreases steadily with time (and stellar mass). It starts at $10^{1.5}$ Gyr$^{-1}$ and reaches $10^{-1}$ Gyr$^{-1}$ when the galaxy becomes massive. Hence, at early times (when the galaxy mass is low), the sBHAR is often much lower than the sSGR resulting in a sub-unity slope. By the time the galaxy reaches a stellar mass of $M_*\sim 10^{10.3}$ $M_\odot$, the sSGR has decreased sufficiently such that the sBHAR is now larger than the sSGR and the slope becomes very large. Subsequently, the sBHAR decreases to a similar level as the sSGR and the slope reaches unity.

Another perspective can be gained by comparing the sSGR and sBHAR to the inverse Hubble time $t_{\rm Hubble}^{-1}$ in panel E. When the galaxy has a relatively low stellar mass, the sSGR exceeds the inverse Hubble time indicating a quickly growing galaxy. In contrast, the sBHAR typically falls below $t_{\rm Hubble}^{-1}$ during this time, indicating slow SMBH growth. However, as soon as the galaxy reaches $M_*\sim 10^{10.3}$ $M_\odot$, the sBHAR approaches and then exceeds $t_{\rm Hubble}^{-1}$. Hence, when the galaxy becomes massive, the SMBH grows as fast as or even faster than its host galaxy on a inverse timescale similar to $t_{\rm Hubble}^{-1}$.

\Cref{eq:slope of scaling relation in toy model} allows us to further understand the slope of the $M_*-M_{\rm BH}$ relation via the BHAR, SGR, and the $M_*/M_{\rm BH}$ ratio. In particular, the panel F of \Cref{fig:slope of Mstar vs MBH} shows that the $M_*/M_{\rm BH}$ ratio does not change enough to affect the trend of the slope over much of the history of this galaxy. A change in $M_*/M_{\rm BH}$ is thus clearly not driving the slope of the $M_*-M_{\rm BH}$ relation. Instead, the slope is set by the BHAR to SGR ratio. Whenever this ratio exceeds the $M_*/M_{\rm BH}$ ratio, the slope becomes large, while for small values of the BHAR to SGR ratio (i.e., when ${\rm BHAR}<10^{-4}\,{\rm SGR}$), the slope is below unity.

Finally, \Cref{fig:slope of Mstar vs MBH} also offers insights into which of the two terms, BHAR and SGR, plays a more important role in setting the slope of the $M_*-M_{\rm BH}$ relation. The slope is low at early times because the BHAR is there much lower than the SGR. A higher slope would require either faster SMBH growth or slower galaxy growth. Subsequently, when the galaxy grows its stellar mass from $M_*\sim 10^{10.1}$ $M_\odot$ to $M_*\sim 10^{10.4}$ $M_\odot$, the BHAR increases and the SGR decreases. This combination results in a BHAR to SGR ratio that finally exceeds $\sim 10^{-4}$ and thus a high value of the slope. Subsequently, the SGR slightly increases again while the BHAR remains nearly constant resulting in  ${\rm BHAR}\approx {\rm SGR}$ and thus an approximately linear slope. Furthermore, a slope change at a later time is well correlated with a change in the BHAR while the SGR is approximately constant. We therefore conclude that both the BHAR and the SGR contribute in a significant manner to the slope evolution of the $M_*-M_{\rm BH}$ relation.

\subsection{Understanding the slope of the $M_*-M_{\rm BH}$ relation}\label{sec:Toy Model}

In this section, we introduce two simple models, a one- and a two- zone model, for the growth of SMBHs based on the stellar growth history of the host galaxy to offer additional insight into the origin of the $M_*-M_{\rm BH}$ scaling relation in the context of the GTDA model.
In both models, the stellar mass of the galaxy is given by the integral of their star formation rates. However, the models differ in how they calculate the black hole accretion rate. In the one-zone model, the central mass which determines the accretion rate is set to a fixed fraction (5\%) of the total stellar mass of the galaxy. In contrast, in the two-zone model the central to total stellar mass ratio is allowed to vary.
We start by making the following assumptions:

\begin{itemize}
    
    \item First, we assume that the total mass within $R_0$ is dominated by the stellar mass, $M_{\rm tot}(<R_0) \sim M_*(<R_0)$. This assumption holds in our simulations, see \Cref{fig:fgas for MAX and COM}, as the stellar-to-total ratio is close to unity for $z \lesssim 6$. At higher redshift, this basic assumption may break down. We refer the reader to \cref{sec:Discussion} for the caveats when modelling SMBH growth via the GTDA model in the early Universe.

    \item Secondly, we split \Cref{eq:GTDA equation} into the term  $M_{\rm tot}(<R_0) \sim M_*(<R_0)$ and collect all other dependencies into a time-dependent function $\theta$. In this approximation, the BHAR becomes proportional to the total stellar mass inside $R_0$. This ansatz allows us to directly tie the slope to the stellar growth history of the host galaxy.
    
    \item Thirdly, we ignore the merger contribution to SMBH growth.
    
\end{itemize}

With the assumptions listed above, we can link the growth of SMBHs to the growth of their host galaxies. The BHAR scales pseudo-linearly with the stellar mass of a galaxy,

\begin{equation}
	\mathrm{BHAR = \theta(t) \times \frac{M_ *}{10^9 M_\odot},}
    \label{eq:accretion rate in toy model}
\end{equation}

where $\theta(t)$ encapsulates the rather complex dependencies of \Cref{eq:GTDA equation}. Finally, we assume that $\theta$ can be modeled by a parametric function of the form

\begin{equation}
    \mathrm{\theta(t) = x_1 \, \exp{ \left( - \frac{t - t_*}{\tau} \right) } + x_2}
\end{equation}

to simplify our analytic calculations. The values $\mathrm{x_1 = 0.052 \,} \mathrm{M_\odot \, yr^{-1}}$, $\mathrm{x_2 = 0.002 \, } \mathrm{M_\odot \, yr^{-1}}$, $\mathrm{t_* = 0.35}$ Gyr, and $\mathrm{\tau = 0.23}$ Gyr result in $\theta(t)$ that is (at $z\geq 1$) in good agreement with our full post-processing analysis, see \Cref{fig:toy model sgr 1-Zone model}. 

The SGR is simply expressed as the change in the stellar mass within $R$ ($\Delta M_*(<R) / \Delta t$) between two adjacent snapshots in the post-processing analysis. Here, $R$ refers to either the radius of the central region $R_0$ or to the size of the galaxy $R_{\rm gal}$. The stellar mass ($M_*(<R)$) is obtained by integrating a function fitted (see \Cref{eq:SGR fit function}) to the central SGR data from the post-processing analysis.

\begin{equation}
    \mathrm{SGR(<R) = a_1 \, \tanh{\left( \frac{t-t_*}{\tau} \right)} + a_2}
    \label{eq:SGR fit function}
\end{equation}

The stellar mass in fixed radius $R$ is connected to the stellar growth rate within the same radius via the integral:

\begin{equation}
    \mathrm{M_*(<R)} = \int_{t_0}^t \rm SGR(<R) \, \mathrm{dt'.} 
    \label{eq:stellar mass in toy model}
\end{equation}

Finally, the black hole mass is given as
\begin{equation}
    \mathrm{M_{BH} = M_{seed} + \int_{t_0}^t BHAR \, dt',}
    \label{eq:mbh in toy model}      
\end{equation}

with $M_{\rm seed}=10^4\,M_\odot$ placed at $z=20$ ($t_0 \sim 180$ Myr) as the fiducial case.

\subsubsection{Predictions of the one-zone model} \label{sec:1-Zone Model}

The one-zone model makes the simplifying assumption that $M_*(<R_0)\propto{}M_*(<R_{\rm gal})$. As a result, the BHAR is directly linked to the stellar mass of galaxies. We adopt a pre-factor of 5\% to approximately match the observed normalization for the $M_*-M_{\rm BH}$ relation and to bring the predictions of the one-zone model in better agreement with the results of the two-zone model discussed in the next section.

First, we demonstrate how different analytical SGR histories affect the evolution of the slope of the $M_*-M_{BH}$ relation. For instance, let us assume a constant SGR evolution in the form of $\rm SGR = c_1$, where the unit of $c_1$ is mass over time. Then, the host galaxy stellar mass becomes $\rm M_* = c_1 \, t$, and

\begin{equation}
    \mathrm{BHAR = 0.05 \, \frac{\theta(t) \, c_1 \, t}{10^9 M_\odot}}.  
    \label{eq:slope calculation BHAR}
\end{equation} 

By inserting \Cref{eq:slope calculation BHAR} into \Cref{eq:mbh in toy model}, we arrive at the following estimate for the SMBH mass

\begin{equation}
\begin{aligned}
    & \mathrm{M_{BH}} = \mathrm{M_{seed} + 0.05 \frac{ x_1 \, c_1}{10^9 \, M_\odot} \, \tau^2 \, \Omega_1\left(t\right) + 0.05 \, \frac{ x_2 \, c_1}{10^9 \, M_\odot} \, \frac{t^2}{2}}
    \\
    & \mathrm{\Omega_1(t) = - \left( \frac{t}{\tau} + 1 \right) \, \exp{\left( -\frac{t-t_*}{\tau} \right)} + \exp{\left( \frac{t_*}{\tau} \right) }}.
\end{aligned}
\end{equation}

A constant SGR thus yields a slope of

\begin{equation}
    \begin{aligned}
        \mathrm{slope} &= \frac{\mathrm{M_*}} {\mathrm{SGR}} \mathrm{\times } \frac{\mathrm{BHAR}}{\mathrm{M_{BH}}}
        \\
        &= \mathrm{ \frac{c_1 \, t}{c_1} \times \frac{ x_1 \, t \, \exp{\left( - \frac{t-t_*}{\tau} \right)} +  x_2 \, t}{\frac{M_{seed} \, 10^9 \, M_\odot}{0.05 \, c_1} +  x_1 \, \tau^2 \, \Omega_1(t) +  x_2 \, \frac{t^2}{2}} } 
        \\
        \\
        &= 
        \left\{ \begin{array}{cc}
            \mathrm{0} & \mathrm{
            as \,\, t \xrightarrow{}  0}  
            \\
            \\
            \mathrm{2} & \mathrm{
            as \,\, t \xrightarrow{}  \infty}.
        \end{array} \right.
    \end{aligned}
    \label{eq:constant SGR 1-zone model}
\end{equation}

In particular, the slope is $0$ when the black hole seed mass is much larger than the accretion contribution ($M_{\rm seed} \gg \rm \int BHAR \, dt'$). At early times the slope is thus rather shallow unless the seed mass is sufficiently small, see \Cref{fig:mstar vs mbh scaling relation}. Instead, when the accretion contribution matches surpasses the seed mass, the slope approaches $\sim 2$ for a constant SGR. The slope takes various values for $0\lesssim t < \infty$, and sometimes the same value more than once. Therefore
\cref{eq:constant SGR 1-zone model} explains the slow-to-fast transition of SMBH growth seen, e.g., in \Cref{fig:mstar vs mbh scaling relation}. At early times the SMBH mass is dominated by its seed mass. Hence, the slope is shallow. At late times, the seed mass is small compared to the accreted mass and the slope is thus steep.

Another simple scenario is a galaxy in which the SGR increases linearly with time, ${\rm SGR}=c_2 \,t$ where the unit of $c_2$ is mass over time squared, resulting in

\begin{equation}
    \begin{aligned}
        \mathrm{slope} 
        &= \mathrm{\frac{c_2 \, \frac{t^2}{2}}{c_2 \, t} \times \frac{ x_1 \, \frac{t^2}{2} \, \exp{\left( - \frac{t-t*}{\tau} \right)} +  x_2 \, \frac{t^2}{2}}{ \frac{M_{seed} \, 10^9 \, M_\odot}{0.05 \, c_2} +  x_1 \, \tau^3 \, \Omega_2(t) +  x_2 \, \frac{t^3}{6}} }
        \\
        \\
        &=
        \left \{ 
        \begin{array}{cc}
            \mathrm{0} & \mathrm{
            as \,\, t \xrightarrow{}  0}
            \\
            \\
            \mathrm{3/2} & \mathrm{
            as \,\, t \xrightarrow{}  \infty}
        \end{array} \right. 
        \\
        \\
        \textrm{with }\,\mathrm{\Omega_2(t)}
        &= \mathrm{\left[ -\frac{t^2}{2 \, \tau^2} - \frac{t}{\tau} - 1\right] \, \exp{\left( - \frac{t-t_*}{\tau} \right)} + \exp{\left( \frac{t_*}{\tau} \right)}}.
    \end{aligned}
    \label{eq:increasing SGR 1-zone model}
\end{equation}

An increasing SGR lowers the slope compared with the case of a constant SGR. However, the SMBH can still grow quickly compared to its host galaxy whenever the accretion contribution exceeds the black hole seed mass.

Finally, a galaxy may experience an epoch in which the SGR decreases with time, e.g., ${\rm SGR}(t) = {\rm SGR}(t_1)\,e^{-c_3(t-t_1)}$ for $t\geq 0$, where the unit of $c_3$ is inverse time. In this case,

\begin{equation}
    \begin{aligned}
        \mathrm{slope}
        & \mathrm{\, = \frac{1 - e^{-c_3 \, t}}{c_3 \, e^{-c_3 \, t}} } \times \mathrm{ \frac{ x_1 \left( 1 - e^{-c_3 \, t} \right) \, \exp{\left( -\frac{t-t_*}{\tau} \right)} +  x_2 \left( 1 - e^{-c_3 \, t} \right) }{ \frac{M_{seed} \, c_3 \, 10^9 \, M_\odot}{0.05 \, SGR(t_1) \, e^{c_3 \, t_1}} +  x_1 \, \tau \, \xi_1 +  x_2 \, \tau \, \xi_2}}
        \\
        \\
        &=
        \left\{ \begin{array}{cc}
            \mathrm{0} & \mathrm{
            as \,\, t \xrightarrow{}  0}
            \\
            \\
            \mathrm{\infty} & \mathrm{
            as \,\, t \xrightarrow{}  \infty}
        \end{array} \right.
        \\
        \\
        \mathrm{\xi_1} & \mathrm{ = \left( -1 + \frac{e^{-c_3 \, t}}{c_3 \, \tau + 1} \right) \, \exp{\left( - \frac{t-t_*}{\tau} \right)} + \left( 1 - \frac{1}{c_3 \, \tau + 1} \right) \exp{\left( \frac{t_*}{\tau} \right)} }
        \\
        \\
        \mathrm{\xi_2} & \mathrm{ = \frac{t}{\tau} - \frac{1-e^{-c_3 \, t}}{c_3 \, \tau}}.
    \end{aligned}
    \label{eq:late time SGR 1-zone model}
\end{equation}

\begin{figure*}

    \includegraphics[width=\textwidth]{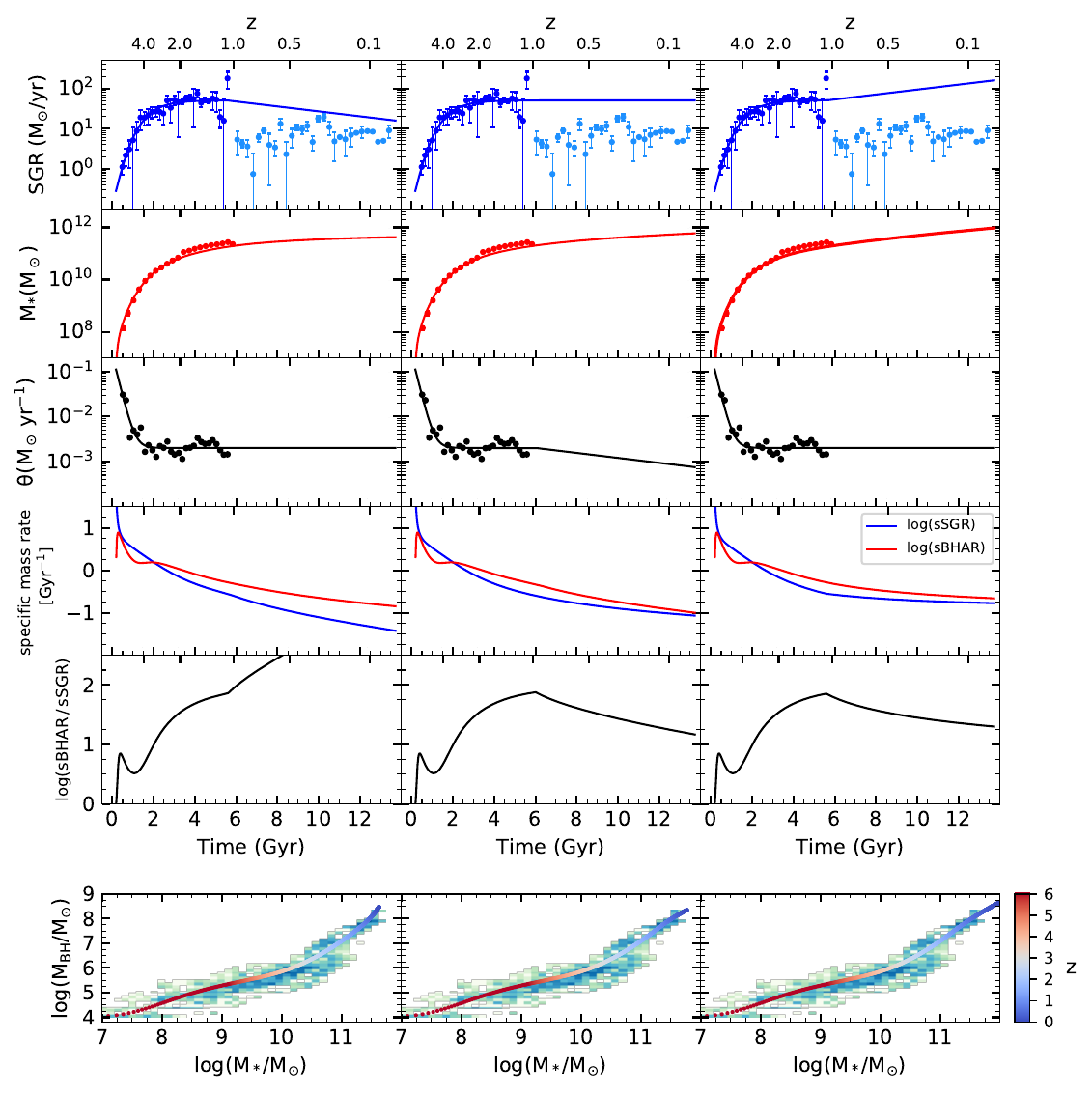}
    \caption{Results from the one-zone model. The first row show the average SGR measured from the A, B, and C-series of \textsc{MassiveFIRE} at $z\geq 1$ (dark blue) and for the two simulations that continue below $z=1$ (light blue). The solid line shows 3 scenarios of how the $z>1$ SGR can be extended to low z. (From left-to-right) the SGR decreases by half a dex between $z=1$ and $z=0$, remains constant at the value at $z=1$, or increases by half a dex. The second row shows the average stellar mass (red) while the third row shows the average $\theta(t)$ parameter, see \Cref{eq:GTDA equation}, which normalizes the pseudo-linear dependence of the BHAR on $M_*$. The $\theta(t)$ parameter is assumed to remain constant at the $z=1$ value in the left and right columns but it decreases by half a dex between $z=1$ and $z=0$ in the middle column. The fourth row show the sSGR, the sBHAR while the fifth panel shows the slope of the trajectory in $M_*-M_{\rm BH}$ space derived from \Cref{eq:slope of scaling relation in toy model}. The trajectory in $M_*-M_{\rm BH}$ space is shown in the bottom row as a colored line indicating the redshift. These panels also show the $M_*-M_{\rm BH}$ scaling relation from the post-processing analysis as an histogram. The shaded region shows the result from post-processing analysis using fiducial settings, see \Cref{fig:mstar vs mbh scaling relation}. The one-zone model describes the overall evolution of simulated galaxies in the $M_*-M_{\rm BH}$ space well, despite its high degree of simplification compared to the full GTDA model.}
    \label{fig:toy model sgr 1-Zone model}
\end{figure*}

\Cref{fig:toy model sgr 1-Zone model} shows the predictions for the SMBH growth when applying the one-zone model to the aggregated stellar growth histories of the A, B, and C-series of \textsc{MassiveFIRE}, see \Cref{tab:simulation details}. In agreement with the simple analytical examples discussed above, the one-zone model predicts a shallow-to-steep transition of the slope. The transition takes place between $z=4$ and $z=2$, when the galaxy transitions from being low mass to becoming a massive galaxy (near $M_*<10^{10}$ $M_\odot$). Interestingly, at this time the SMBH mass already exceeds the seed mass by an order of magnitude, i.e., we are already in the $M_{\rm seed}\ll  \int_0^t BHAR \, dt'$ regime discussed in, e.g., {eq:constant SGR one-zone model}. We also confirmed that reducing the seed mass to, e.g., $M_{\rm seed}<10^{2}$ $M_\odot$ does not change the location of the slope transition, again indicating that this increase in slope is not a seed mass effect. Instead, \Cref{fig:toy model sgr 1-Zone model} shows that during the time of the transition ($z=2-4$) the sBHAR declines much slower than the sSGR, resulting in an increasing slope. Hence, the increase in the slope of the $M_*-M_{\rm BH}$ relation in $\sim M_*<10^{10}$ $M_\odot$ galaxies is a natural consequence of the specific evolutionary history of the SGRs of such galaxies. Ultimately, their SGRs are set by the complex interplay of gas inflows, feedback, and galaxy mergers.

\Cref{fig:mstar vs mbh scaling relation} shows that the $M_*-M_{\rm BH}$ relation flattens at late times in very massive galaxies. While this finding is only based on a small number of simulated galaxies at relatively low redshift, and thus tentative, we would like to explore the physical origin of this finding with the help of the one-zone model. To this end, we study 3 possible scenarios. In our first scenario, we assume that the SGR decreases linearly with increasing log stellar mass by half a dex between $z=1$ and $z=0$ while $\theta$ is kept constant at its $z=1$ value, $\sim 10^{-3} \, yr^{-1}$. In the second scenario, the SGR stays constant until $z=0$ but the $\theta(t)$ parameter decreases by half a dex. The third scenario is identical to the first, except that the SGR \emph{increases} by half a dex between $z=1$ and $z=0$.

For each of the three scenarios, we first fit the average behaviour of the galaxy-wide SGR and of $\theta(t)$ until $z=1$ and then extrapolate to $z=0$. Next, we integrate the SGR to obtain the stellar mass and the SMBH mass as function of time as described in \Cref{sec:1-Zone Model}, see in particular \Cref{eq:accretion rate in toy model}.

In the first scenario, the SGR decreases at low $z$ which results in a $M_*$ that stays nearly constant after $z=1$. However, the average BHAR continues to grow since the BHAR is assumed to be proportional to the stellar mass. Consequently, the slope of the trajectory in $M_*-M_{\rm BH}$ space gradually steepens between $z=1$ and $z=0$ in line with \Cref{eq:late time SGR 1-zone model}. This trend disagrees however with the results shown in \Cref{fig:mstar vs mbh scaling relation} leading us to exclude this scenario.

In second scenario the SGR is kept constant between $z=1$ and $z=0$, while $\theta(t)$ decreases. In this case, the average $M_*$ grows more rapidly, while the decreasing $\theta$ slows the growth of the SMBH resulting in a nearly constant BHAR at late times. $\theta(t)$ is strongly dependent on the disc fraction ($\theta(t) \propto f_d^{5/2}$) in the GTDA model and we expect a decrease in disc fraction at late times as massive galaxies transition from disc to early type morphology.

The final scenario assumes an increasing SGR for $z<1$. Subsequently, the stellar mass grows substantially (by about one dex) between $z=1$ and $z=0$. Even though the larger stellar mass also boosts the BHAR, the fast galaxy growth results in a slope that is only mildly super-linear and approaches unity at late times. Physically, an increasing SGR at low redshift may arise from the late assembly of massive galaxies via merging expected in a hierarchical Universe.

\subsubsection{Predictions of the two-zone model}\label{sec:2-Zone Model}


In contrast to the one-zone model, the two-zone model uses the stellar mass within $R_0$, and not within the galaxy, to calculate the BHAR.
Furthermore, the link between BHAR and stellar mass (as opposed to SFR) within the central region as given by equation \ref{eq:GTDA equation} has also observational support. \cite{Yang::etal::2019} find that the BHAR correlates better with $M_*$ rather than SFR in the central region in non-bulge dominated galaxies which applies to the majority of the galaxies in our sample, see \Cref{fig:fbulge for MAX}.

In the one-zone model, the slope of the $M_*-M_{\rm BH}$ relation depends mainly on the total SGR of a given galaxy. Here, however, the slope is affected both by the galaxy-wide SGR (which drives $M_*$) and by the central SGR (which drives the BHAR). We can thus investigate how the slope changes for various options of a constant or increasing SGR, either galaxy-wide or in the central region. The slope is provided in \Cref{tab:Slope values for 2-Zone model} in the limits of $M_{\rm seed}$ exceeding, is equal to, or smaller than $\int_0^t {\rm BHAR} \, dt'$.

\begin{table}
\centering
\begin{tabular}{|l|l|l|l|l|}

\hline
SGR & Case 1 & Case 2 & Case 3 & Case 4 \\
\hline
Inner & c & c & ct & ct \\
Total & c & ct & c & ct \\
\hline
$\mathrm{M_{seed}}$ vs $\mathrm{\int_0^t BHAR \, dt'}$ &  &  &  &  \\
\hline
Greater & 0 & 0 & 0 & 0 \\
Smaller & 2 & 1 & 3 & 3/2 \\
\hline

\end{tabular}
\caption{The slopes of trajectories in $M_*-M_{\rm BH}$ space for simple cases of the two-zone model. The top part of the table details the four the combinations of constant or linearly increasing SGRs in the central region of a galaxy or for galaxy-wide SGR. The bottom part of the table lists the slopes depending on whether the SMBH seed mass is greater than, equal to, or smaller than the current SMBH mass, $\int_0^t \rm BHAR \, dt'$.
Different cases stand for combination of constant and linear increasing SGR in the central region and the whole galaxy}. Depending on the specific SGR histories, $M_*-M_{\rm BH}$ trajectories can have a  shallow or steep slope.

\label{tab:Slope values for 2-Zone model}
\end{table}

\begin{figure*}
    
    \includegraphics[width=\textwidth]{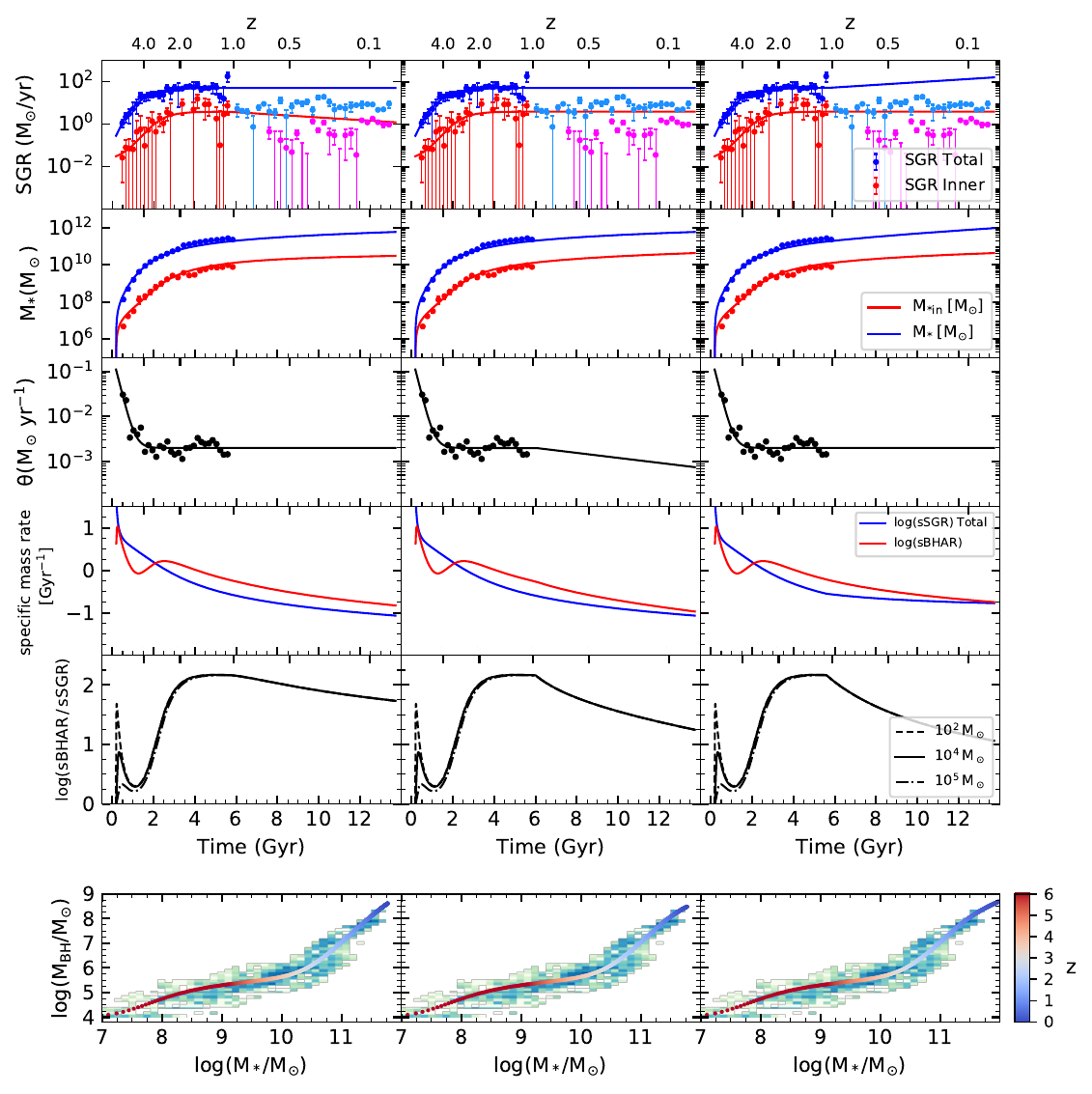}
    \caption{Results of the two-zone model, similar to the one-zone model. We consider three different scenarios (columns left or right) for the late time evolution of SGR and $\theta$. The first column considers a scenario in which central SGR decreases while the total SGR and $\theta(t)$ remains constant. In the second column, central and total SGRs are constant, and $\theta(t)$ decreases by half a dex after $\rm z=1$. The final column studies the impact of growing total SGR while central SGR and $\theta$ constant. The first row shows the average points of central (red) and total (blue) SGR of the A, B, and C-series of \textsc{MassiveFIRE} where $\log(M_{\rm halo}) = 12.5$ at $z=2$ and best fit lines to the average central and total SGR for $z\gtrsim1$. The light blue and light red colours represent the averaged data from A1 and A4 simulations with FIRE-1 physics that we do not include in the best fit. The solid red line in the second row shows the central stellar mass (integration of the red line in the first row) denoted as $\rm{M}_{*,in}$. The solid black line in the third row shows $\rm BHAR / \rm{M}_{*,in}$ in the \textsc{few mergers} MAX centring case. The solid blue line is the integration of the total SGR, blue line in the first row. In the fourth row, we show the specific growth rates of the galaxies ($\rm sSGR = SGR / M_*$) and SMBHs ($\rm sBHAR = BHAR / M_{BH}$). The fifth row consists of the slope of the scaling relation ($\rm sBHAR / sSGR$) as a solid black line. The black dashed and dotted dashed lines are the slope for different seed masses, $10^2$ and $10^5 M_\odot$. Further reduction in central SGR and $\theta(t)$ give flatter slopes, and a boost in the total stellar growth also has a similar effect on the slope of the $M_*-M_{\rm BH}$ scaling relation. The final panel at the bottom shows the prediction of our two-zone model for the $M_* - M_{\rm BH}$ scaling relation colour-coded by redshift. The shaded region shows the result from post-processing analysis using fiducial settings, see \Cref{fig:mstar vs mbh scaling relation}. Compared to the one-zone model, the two-zone model predicts a more noticeable shallow-to-steep transition of the slope of the $M_*-M_{BH}$ relation for $\log(M_*/M_\odot) = 10$ galaxies.}
    \label{fig:toy model sgr 2-Zone model}
\end{figure*}

Similar to \Cref{sec:1-Zone Model}, we can apply also the two-zone model to the SGR histories of \textsc{MassiveFIRE} galaxies. Given the lack of many galaxies at low $z$, we again explore three possibilities for the SGR evolution at $z<1$. We will show that while these scenarios differ, they predict similar results for the slope of $M_* - M_{\rm BH}$ scaling relation in the local Universe. The three scenarios are as follows: 

\begin{itemize}
    \item The central SGR decreases by half a dex since $z=1$ while $\theta(t)$ and the total SGR remain constant.
    \item $\theta(t)$ decreases with time at $z<1$ while the central and total SGR remain constant.
    \item The central SGR and $\theta(t)$ are constant while the total SGR \emph{increases} by half a dex since $z=1$.
\end{itemize}

A decrease in the central SGR, as speculated in the first scenario, could originate, e.g., in inside-out quenching due to the AGN feedback \citep{Tacchella::etal::2015, Ellison::etal::2018, Abdurrouf2018, Tacchella::etal::2018}. Strong stellar feedback \citep{Cox::etal::2006}, merger quenching \citep{Gabor::etal::2010}, or gravitational heating due to clumpy accretion \citep{Birnboim::etal::2007, DekelBirnboim2008, Dekel::etal::2009} could also reduce the central SGR by lowering the central star formation activity. The decrease in $\theta(t)$, proposed in the second scenario, is expected from a disc-to-early type morphological transformation, while the increase in the total SGR (but not central SGR) could arise from galaxy merging. 

As \Cref{fig:toy model sgr 2-Zone model} shows, all three scenarios result in a $M_* - M_{\rm BH}$ relation that is consistent with the aggregated \textsc{MassiveFIRE} results even at $z<1$. In particular, it shows the shallow-to-steep transition of the slope when galaxies approach $M_*\sim 10^{10}$ $M_\odot$. In addition, all three scenarios show a reduction in slope in massive galaxies at late times. Scenario 2 and 3 leads to final slopes of order unity, while the first scenario lowers the slope to about 1.5.

We expect that all three scenarios are partly at play in the real Universe. The $M_*-M_{\rm BH}$ relation at late times may thus be especially susceptible to the differential growth of galaxies in their central region and on galaxy-wide scales. In addition, similar to the one-zone model, galaxy merging or a change in galaxy morphology can strongly affect the $M_*-M_{\rm BH}$ slope.

\begin{figure*}
    \includegraphics[width=\textwidth]{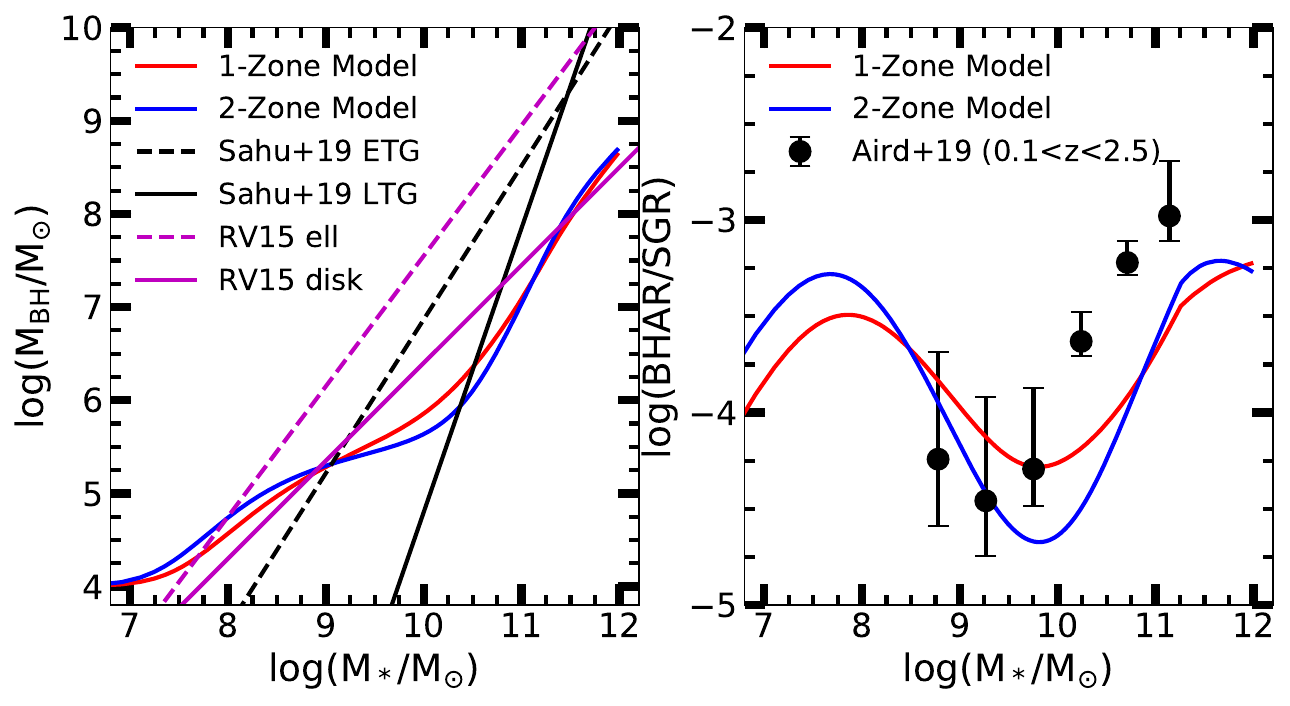}
    \caption{\textbf{Left Panel:} Comparison of the predictions of one- and two-zone models with the scaling relations from the literature. The solid and dashed black lines represent the best fit line for late and early-type galaxies in \protect\cite{Sahu::etal::2019b}, respectively. The solid and dashed purple lines show the best fit lines for the elliptical and disc galaxies in \protect\cite{ReinesVolonteri2015}. The red and blue solid lines are the predictions of our one-zone and two-zone models, where the BHAR is linked to the stellar growth history. The slopes predicted by the one- and two-zone models flatten at high masses $\log(M_* / M_\odot) \sim 11-11.5$ as in the transition from LTGs to ETGs in \protect\cite{Sahu::etal::2019b}. \textbf{Right Panel:} Predictions for how the BHAR-SGR ratio changes with the total stellar mass. This panel is the same as \Cref{fig:bhar vs sfr}, but compares the BHAR-SGR ratio from the toy models with the observational result of \protect\cite{Aird::etal::2019}, instead of being based on the post-processing data from \textsc{\small{MassiveFIRE}}. The solid red line indicates the BHAR-SGR ratio from the one-zone model and the solid blue line represents the ratio between the BHAR and the total SGR in two-zone model.}
    \label{fig:toy model mstar mbh sgr bhar}
\end{figure*}

\Cref{fig:toy model mstar mbh sgr bhar} compares the predictions of one-zone and two-zone models. A main difference is the somewhat more pronounced change in the slope of the $M_*-M_{\rm BH}$ relation at $t\sim 1-4$ Gyr in the two-zone model compared to the one-zone model. However, overall the predictions are rather similar. The right panel of \Cref{fig:toy model mstar mbh sgr bhar} shows the BHAR-SGR ratio for both the one-zone and the two-zone model. Again, both models make rather similar predictions for the  $M_*-M_{\rm BH}$ relation and for the stellar mass dependence of the ratio between the BHAR and the SGR. Overall, the two-zone model results in a slightly more accurate representation of the full post-processing analysis, see \Cref{fig:toy model sgr 1-Zone model} and \Cref{fig:toy model sgr 2-Zone model}, specifically in more pronounced changes from the shallow to the steep slope regime as well as from the steep to the approximately linear slope regime.

\subsubsection{Implications of the toy models}

As we saw in \Cref{fig:bhar vs sfr}, the SGR and the BHAR trace each other reasonably well both in observations and simulations albeit with significant scatter. In the context of the GTDA model, this link is facilitated via the amount of mass (stars and gas) in the central region of a galaxy which is directly driving the BHAR. In the previous sections, we discussed the consequences of simplified toy models based on this general idea. Specifically, in the one-zone model (two-zone model) the galaxy-wide stellar mass (the central stellar mass) is assumed to be proportional to the baryonic mass in the central region. Hence, the BHAR and thus SMBH growth can be calculated once the SGR history is known.

We have shown that there are in general (at least) 3 distinct epochs as galaxies move through $M_*-M_{\rm BH}$ space. First, galaxies grow quickly while the SMBH mass does not (shallow slope). This epoch lasts until the total stellar mass reaches about $10^{10} \, M_\odot$. This threshold is generally reached during cosmic noon ($z \sim 2-3$) for our \textsc{\small{MassiveFIRE}} sample. Secondly, the slope increases steadily marking efficient SMBH growth (steep slope). Finally, our analysis predicts that under certain assumptions the slope decreases again at late times (approximately linear slope). 

During early times, galaxies have relatively low masses, form stars at high rates \citep{Riechers::etal::2013, Finkelstein::etal::2013, Caitlin::etal::2014, Zavala::etal::2018, Bowler::etal::2018, Berta::etal::2020} and thus increase their stellar masses quickly. Specifically, the shallower potential wells in the centers of galaxies at earlier times may boost SN-driven mass ejections \citep{Dubois::etal::2015} which plays a crucial role in suppressing the accretion onto SMBH. Also, stellar feedback may drive buoyant outflows of high-entropy gas from the central regions \citep{Bower::etal::2016} which ceases to be effective in massive halos with hot gas coronas. In addition, a virialization of the circum-galactic medium down to the central galaxy (`inner CGM virialization', \citealt{Stern::etal::2020}), which can stabilize disks against feedback driven outflows, has typically not yet taken place. Consequently, galaxies grow much quicker than their SMBHs at those early times resulting in a shallow slope for the trajectory in $M_*-M_{\rm BH}$ space.

As a galaxy becomes moderately massive, $\sim 10^{10} \, M_\odot$, the escape speed of the galaxy exceeds the characteristic speed of SN-driven winds thus strongly reducing the effectiveness of galactic outflows \citep{Angles-Alcazar+17c}. In fact, a stellar compactness of $10^{10} M_\odot \, \rm kpc^{-1}$ in the central region would be enough to keep the SN-driven winds within the host galaxy centre \citep{Dubois::etal::2015}. During this time, SMBHs will be able to grow quickly resulting in a steep slope because (i) galaxies may contain significant reservoirs of gas in their centers and (ii) galaxies of this mass have often a major disc component allowing gravitational torques to operate efficiently \citep{Querejeta::etal::2016, Angles-Alcazar+17a, Blumenthal::etal::2018, Thomas::etal::2019}.

Finally, we find that when galaxies become very massive, $\sim 10^{11} \, M_\odot$, the masses of galaxies and of their SMBHs often grow at similar rates. The one- and two-zone models presented in the previous sections provide some insights into the origin of this behavior. Specifically, we pointed to two possibilities consistent with our analysis. First, the close to linear slope could originate in a reduction in the $\theta(t)$ term in the GTDA model. $\theta(t)$ depends super-linearly on the disc fraction, $\theta(t) \propto f_{\rm disc}^{5/2}$. Hence, a transition from disc to elliptical morphology in the central regions of galaxies reduces the SMBH accretion rate. Such a transition is expected given the change in overall Hubble type when galaxies grow in mass \citep{D'Onofrio::etal::2015, Tacchella::etal::2019, Cooke::etal::2019}. Supporting this scenario is also the observational finding that the slope of the $M_*-M_{\rm BH}$ relation depends on galaxy morphology \cite{Davis::etal::2019}. Secondly, the slope could flatten not because the SMBH accretion rate decreases, but because the SGR increases, e.g., due to a larger number of galaxy mergers in massive galaxies \citep{Marchesini::etal::2014, Bellstedt::etal::2016, BuchanShankar2016, Vulcani::etal::2016, Nipoti::etal::2018}.

\section{Discussion}\label{sec:Discussion}

The results of this study are subject to a few potential caveats.

First, there is the potential question whether the GTDA model is applicable to galaxies over a large range of masses and redshift. The GTDA model was developed for disc galaxies and estimates the accretion from circum-nuclear ($\sim 100$ pc) to sub-pc scales using the properties of the circum-nuclear region, including its stellar mass, baryonic disc fraction, and gas mass. However, especially at high redshift, the host galaxies may not always be disc galaxies \citep{Cowie::etal::1995, vandenBergh::etal::1996, Tacconi::etal::2010, Genzel::etal::2011, Guo::etal::2012, Zanella::etal::2015} even though massive disc galaxies have certainly been found even at $z=4$ and beyond \citep{Hodge::etal::2012, Neeleman::etal::2020}. Fortunately, the specific details of the GTDA model appear to be less important given that a very similar growth history for SMBHS can be obtained by using the much simpler dynamical accretion model (\Cref{eq:dyn accretion equation}) with $\gamma = 10^{-4}$. Hence, our predictionns are likely robust as long as the accretion model results in SMBHs accreting only a small fraction ($\sim 0.1\%$) of the available gas per free-fall time \citep{Angles-Alcazar+17b, Angles-Alcazar+17c}. Models that may contribute to a faster, more efficient gas accretion onto SMBHs, such as chaotic accretion of hot gas \citep{Thomas::etal::2019, Dave::etal::2019} or merger-triggered accretion \citep{CapeloDotti2017, Ricarte::etal::2018} will be left to future work.

A second potential concern is that none of the SMBHs in our post-processing analysis are as massive as the most luminous AGN observed at $z \gtrsim 6$ \citep{Mortlock+2011, Wu+2015, Mazzucchelli::etal::2017, Banados+2018, Yang::etal::2020, Wang::etal::2021}. While some of our haloes are sufficiently massive enough ($\gtrsim 10^{12} M_\odot$) to potentially host a very luminous AGN, simple number density and clustering arguments show that our simulation volume is likely too small to contain even a single luminous AGN. In particular, the number density of luminous AGNs at $z \sim 6-7$ is $\sim 1\,{\rm cGpc}^{-3}$ for $M_{1450} < -26$ \citep{Wang::etal::2019}. While some of our simulations are run in boxes with sizes of $\sim 0.8$ cGpc, we only simulate a small number of (massive) halos selected from those boxes via the zoom-in approach. Hence, the chance of selecting the halo of even a single luminous AGN is very small. The most luminous AGNs may have an atypical formation path which leads to a larger seed mass, e.g. direct collapse black holes \citep{BrommLoeb2003, Volonteri2010}. These high-mass SMBH seeds can grow quickly when the accretion rate is strongly dependent on the SMBH mass. Furthermore, the duty cycle of luminous AGN is close to the unity at $z \gtrsim 6$ \citep{Shankar::etal::2010a, Shankar::etal::2010b, Shankar::etal::2020} implying that analyzing snapshots at different times does not substantially increase the odds of reproducing a luminous AGN in our simulations.

Thirdly, a significant simplification of our model is the treatment of SMBH mergers. To robustly asses the impact of SMBH merging on our result we have considered two extreme scenarios in addition to our fiducial (\textsc{few mergers}) case. In the first of these extreme scenarios, no SMBH mergers take place. In the second extreme case, SMBH of galaxies merge as soon as their parent halos become sub-halos of each other. While the predictions for the low $z$ Universe are shown to be rather robust to the specifics of the SMBH merger model, the SMBH-galaxy scaling relation at early times are sensitive to details of SMBH merging, thus highlighting the importance of properly accounting for SMBH mergers especially in the young Universe \citep{Ma::etal::2021}.

Finally, most of the \textsc{\small{MassiveFIRE}} simulations used in this paper do not include black hole physics on-the-fly, especially AGN feedback. This is by design and allows us to study SMBH growth and scaling relations in the absence of AGN feedback. Thus, this work provides a basis for the future comparison with the FIRE simulations including black hole physics.

Importantly, as we showed in \Cref{fig:mstar vs mbh scaling relation}, accretion models that are weakly dependent on the SMBH mass can reproduce the $M_*-M_{\rm BH}$ local scaling relation without AGN feedback \citep{AnglesAlcazar::etal::2013}. We speculate that the negative effect of AGN feedback should result in overall slower growth of SMBHs. If true, the current analysis provides an upper limit on how fast SMBHs can grow in the context of the GTDA model. We leave the study of BH growth in simulations with AGN feedback for the future.

\section{Summary \& Outlook}\label{sec:Summary}

We have carried out a post-processing analysis of $34$ high resolution cosmological zoom-in simulations from the \textsc{\small{MassiveFIRE}} suite to study the growth of SMBHs and their host galaxies across cosmic history in the \emph{absence} of AGN feedback. In particular, we have analyzed the effect of SMBH placements, SMBH merger treatments, and specifics of the accretion models on the $M_* - M_{\rm BH}$ scaling relation, focusing in particular on the gravitational torque driven accretion model (GTDA) by \cite{HopkinsQuataert2011} and on the evolution at $z \gtrsim 2$. Our main findings are as follows.
    
\begin{itemize}
    
    \item The masses of galaxies and their central SMBHs co-evolve, even in the absence of AGN feedback, in the GTDA model approximately in line with the local $M_*-M_{\rm BH}$ scaling relation, see \Cref{fig:mstar vs mbh scaling relation}.
    
    \item While overall in line with the local $M_*-M_{\rm BH}$ scaling relation, we find clear evidence of a significant deviation from a simple power-law relationship (a "shallow-to-steep" transition of the slope) in low to moderately-massive galaxies. The strength of this deviation depends on the specific modeling assumptions. In particular, it is more pronounced if early SMBH growth is stunted by placing them on more typical (i.e., not the most gas rich) regions near the centers of galaxies and if SMBHs are not allowed to merge. SMBHs mergers and efficient early growth of SMBHs significantly reduces this deviation from the local $M_*-M_{\rm BH}$ scaling relation.
    
    \item Model assumptions, especially, about SMBH placements and mergers leave a clear imprint on the  $M_*-M_{\rm BH}$ scaling relation \emph{at high redshift} in the absence of AGN feedback (see \Cref{fig:scaling relations MDC}). Hence, we expect a link between the SMBH merger rate and their mass ratios with any deviations of the $M_*-M_{\rm BH}$ scaling relation at high $z$ from those of local galaxies.
    
    \item Different SMBH placement and merger models have no apparent efffect on the final SMBH mass at low redshift in the context of the GTDA model. The masses of SMBHs at late times are also largely independent of the BH seed mass.
    
    \item Aside from the GTDA model, we also study alternative accretion models. \Cref{fig:BH growth for different accretion models} shows that large-scale accretion models can be divided into two major groups depending on whether the accretion rate scales superlinear (e.g., Bondi-like models) or sub-linear (e.g., the GTDA model) with the SMBH mass. The first class of models results typically in over-massive SMBHs strongly indicating the need for AGN feedback. The second class of models, however, is able to reproduce the local scaling relations \emph{without} the inclusion of AGN feedback.
        
    \item Currently, none of the SMBHs in our post-processing analysis are as luminous as the billion solar mass SMBHs in the early Universe likely due to the limited volume probed by our simulations. However, we have considerable overlap in the IR luminosity. 
    
    \item This study predicts a large number of low luminosity AGN at high redshift which may be potentially observable with JWST. Mergers between these SMBHs may be detectable by gravitational wave experiments.
        
    \item The offset of the SMBH mass from the local $M_*-M_{\rm BH}$ scaling relation, $\Delta \log(M_{\rm BH})_{\rm obs}$, increases towards higher redshift. While our finding differs from observational data taken at face value, such a comparison does not account for observational selection biases. We thus predict that the discovery of dimmer AGNs at high redshift could decrease the slope of $z-\Delta \log(M_{\rm BH})_{\rm obs}$ relation.
   
    \item We develop two variants of an analytical model in \Cref{sec:Toy Model} that link the growth of SMBHs to the stellar growth history of the host galaxies within the frame of the GTDA model. These models capture the $M_*-M_{\rm BH}$ trajectory predicted by the full post-processing analysis remarkably well thus allowing us to understand the shape and normalization of the $M_*-M_{\rm BH}$ relation in terms of the stellar growth history of galaxies.
    
\end{itemize} 

A high merging efficiency results in a close to linear slope of the $M_* - M_{\rm BH}$ scaling relation for all stellar masses, see \Cref{fig:delta2 vs mstar}. In contrast, the $M_* - M_{\rm BH}$ shows a clear non-linear scaling if BH mergers are rare. Consequently, the slope of the $M_*-M_{\rm BH}$ relation and the merger rate of SMBHs appear intricately linked. This link may be explored observationally by constraining the mass distribution of SMBHs residing in moderately massive galaxies and by measuring BH merger rates via gravitational wave signals.

\section{Acknowledgements}

O\c{C} thanks Pedro R. Capelo, Alexander P. Hobbs and Mehmet Hakan Erkut for the discussion and their valuable feedback on the manuscript. O\c{C} also thanks his wife M\"{O}\c{C} and two cats L\c{C} and K\c{C} for their tireless efforts to create a motivating working environment during COVID-19 lockdown. RF acknowledges financial support from the Swiss National Science Foundation (grant no 157591 and 194814). DAA acknowledges support by NSF grant AST-2009687 and by the Flatiron Institute, which is supported by the Simons Foundation. CAFG was supported by NSF through grants AST-1517491, AST-1715216, and CAREER award AST-1652522; by NASA through grant 17-ATP17-0067; and by a Cottrell Scholar Award and Scialog Award $\#$26968 from the Research Corporation for Science Advancement.

Simulations were run with resources provided by the NASA High-End Computing (HEC) Program through the NASA Advanced Supercomputing (NAS) Division at Ames Research centre, proposal SMD-14-5492. Additional computing support was provided by HEC allocations SMD-14-5189, SMD-15-5950, by NSF XSEDE allocations AST120025, AST150045, AST160048, by allocations s697, s698 at the Swiss National Supercomputing Centre (CSCS), and by S3IT resources at the University of Zurich. Numerical calculations were run on the Quest computing cluster at Northwestern University; XSEDE allocation TG-AST140023; and NASA HEC allocation SMD-16-7561 and SMD-17-1204.

\section{Data Availability Statement}
Please contact the corresponding author if you have a sharing request for the data underlying this article.


\bibliographystyle{mnras}
\bibliography{mnras} 

\begin{thebibliography}{}
\makeatletter
\relax
\def\mn@urlcharsother{\let\do\@makeother \do\$\do\&\do\#\do\^\do\_\do\%\do\~}
\def\mn@doi{\begingroup\mn@urlcharsother \@ifnextchar [ {\mn@doi@}
  {\mn@doi@[]}}
\def\mn@doi@[#1]#2{\def\@tempa{#1}\ifx\@tempa\@empty \href
  {http://dx.doi.org/#2} {doi:#2}\else \href {http://dx.doi.org/#2} {#1}\fi
  \endgroup}
\def\mn@eprint#1#2{\mn@eprint@#1:#2::\@nil}
\def\mn@eprint@arXiv#1{\href {http://arxiv.org/abs/#1} {{\tt arXiv:#1}}}
\def\mn@eprint@dblp#1{\href {http://dblp.uni-trier.de/rec/bibtex/#1.xml}
  {dblp:#1}}
\def\mn@eprint@#1:#2:#3:#4\@nil{\def\@tempa {#1}\def\@tempb {#2}\def\@tempc
  {#3}\ifx \@tempc \@empty \let \@tempc \@tempb \let \@tempb \@tempa \fi \ifx
  \@tempb \@empty \def\@tempb {arXiv}\fi \@ifundefined
  {mn@eprint@\@tempb}{\@tempb:\@tempc}{\expandafter \expandafter \csname
  mn@eprint@\@tempb\endcsname \expandafter{\@tempc}}}

\bibitem[\protect\citeauthoryear{{Abdurro'uf}}{{Abdurro'uf}}{2018}]{Abdurrouf2018}
{Abdurro'uf} Akiyama M.,  2018, \mn@doi [\mnras] {10.1093/mnras/sty1771}, \href
  {https://ui.adsabs.harvard.edu/abs/2018MNRAS.479.5083A} {479, 5083}

\bibitem[\protect\citeauthoryear{{Aird}, {Coil}  \& {Georgakakis}}{{Aird}
  et~al.}{2019}]{Aird::etal::2019}
{Aird} J.,  {Coil} A.~L.,   {Georgakakis} A.,  2019, \mn@doi [\mnras]
  {10.1093/mnras/stz125}, \href
  {https://ui.adsabs.harvard.edu/abs/2019MNRAS.484.4360A} {484, 4360}

\bibitem[\protect\citeauthoryear{{Angl{\'e}s-Alc{\'a}zar}, {{\"O}zel}  \&
  {Dav{\'e}}}{{Angl{\'e}s-Alc{\'a}zar}
  et~al.}{2013}]{AnglesAlcazar::etal::2013}
{Angl{\'e}s-Alc{\'a}zar} D.,  {{\"O}zel} F.,   {Dav{\'e}} R.,  2013, \mn@doi
  [\apj] {10.1088/0004-637X/770/1/5}, \href
  {https://ui.adsabs.harvard.edu/abs/2013ApJ...770....5A} {770, 5}

\bibitem[\protect\citeauthoryear{{Angl{\'e}s-Alc{\'a}zar}, {Dav{\'e}},
  {{\"O}zel}  \& {Oppenheimer}}{{Angl{\'e}s-Alc{\'a}zar}
  et~al.}{2014}]{Angles-Alcazar+14}
{Angl{\'e}s-Alc{\'a}zar} D.,  {Dav{\'e}} R.,  {{\"O}zel} F.,   {Oppenheimer}
  B.~D.,  2014, \mn@doi [\apj] {10.1088/0004-637X/782/2/84}, \href
  {https://ui.adsabs.harvard.edu/#abs/2014ApJ...782...84A} {782, 84}

\bibitem[\protect\citeauthoryear{{Angl{\'e}s-Alc{\'a}zar}, {{\"O}zel},
  {Dav{\'e}}, {Katz}, {Kollmeier}  \& {Oppenheimer}}{{Angl{\'e}s-Alc{\'a}zar}
  et~al.}{2015}]{AnglesAlcazar::etal::2015}
{Angl{\'e}s-Alc{\'a}zar} D.,  {{\"O}zel} F.,  {Dav{\'e}} R.,  {Katz} N.,
  {Kollmeier} J.~A.,   {Oppenheimer} B.~D.,  2015, \mn@doi [\apj]
  {10.1088/0004-637X/800/2/127}, \href
  {https://ui.adsabs.harvard.edu/abs/2015ApJ...800..127A} {800, 127}

\bibitem[\protect\citeauthoryear{{Angl{\'e}s-Alc{\'a}zar}, {Dav{\'e}},
  {Faucher- Gigu{\`e}re}, {{\"O}zel}  \& {Hopkins}}{{Angl{\'e}s-Alc{\'a}zar}
  et~al.}{2017a}]{Angles-Alcazar+17a}
{Angl{\'e}s-Alc{\'a}zar} D.,  {Dav{\'e}} R.,  {Faucher- Gigu{\`e}re} C.-A.,
  {{\"O}zel} F.,   {Hopkins} P.~F.,  2017a, \mn@doi [\mnras]
  {10.1093/mnras/stw2565}, \href
  {https://ui.adsabs.harvard.edu/#abs/2017MNRAS.464.2840A} {464, 2840}

\bibitem[\protect\citeauthoryear{{Angl{\'e}s-Alc{\'a}zar},
  {Faucher-Gigu{\`e}re}, {Kere{\v{s}}}, {Hopkins}, {Quataert}  \&
  {Murray}}{{Angl{\'e}s-Alc{\'a}zar} et~al.}{2017b}]{Angles-Alcazar+17b}
{Angl{\'e}s-Alc{\'a}zar} D.,  {Faucher-Gigu{\`e}re} C.-A.,  {Kere{\v{s}}} D.,
  {Hopkins} P.~F.,  {Quataert} E.,   {Murray} N.,  2017b, \mn@doi [\mnras]
  {10.1093/mnras/stx1517}, \href
  {https://ui.adsabs.harvard.edu/abs/2017MNRAS.470.4698A} {470, 4698}

\bibitem[\protect\citeauthoryear{{Angl{\'e}s-Alc{\'a}zar},
  {Faucher-Gigu{\`e}re}, {Quataert}, {Hopkins}, {Feldmann}, {Torrey}, {Wetzel}
  \& {Kere{\v{s}}}}{{Angl{\'e}s-Alc{\'a}zar}
  et~al.}{2017c}]{Angles-Alcazar+17c}
{Angl{\'e}s-Alc{\'a}zar} D.,  {Faucher-Gigu{\`e}re} C.-A.,  {Quataert} E.,
  {Hopkins} P.~F.,  {Feldmann} R.,  {Torrey} P.,  {Wetzel} A.,   {Kere{\v{s}}}
  D.,  2017c, \mn@doi [\mnras] {10.1093/mnrasl/slx161}, \href
  {https://ui.adsabs.harvard.edu/#abs/2017MNRAS.472L.109A} {472, L109}

\bibitem[\protect\citeauthoryear{{Angles-Alcazar} et~al.,}{{Angles-Alcazar}
  et~al.}{2020}]{AnglesAlcazar::etal::2020}
{Angles-Alcazar} D.,  et~al., 2020, arXiv e-prints, \href
  {https://ui.adsabs.harvard.edu/abs/2020arXiv200812303A} {p. arXiv:2008.12303}

\bibitem[\protect\citeauthoryear{{Ba{\~n}ados} et~al.,}{{Ba{\~n}ados}
  et~al.}{2015}]{Banados::etal::2015}
{Ba{\~n}ados} E.,  et~al., 2015, \mn@doi [\apj] {10.1088/0004-637X/804/2/118},
  \href {https://ui.adsabs.harvard.edu/abs/2015ApJ...804..118B} {804, 118}

\bibitem[\protect\citeauthoryear{{Ba{\~n}ados} et~al.,}{{Ba{\~n}ados}
  et~al.}{2018}]{Banados+2018}
{Ba{\~n}ados} E.,  et~al., 2018, \mn@doi [\nat] {10.1038/nature25180}, \href
  {https://ui.adsabs.harvard.edu/#abs/2018Natur.553..473B} {553, 473}

\bibitem[\protect\citeauthoryear{{Beifiori}, {Courteau}, {Corsini}  \&
  {Zhu}}{{Beifiori} et~al.}{2012}]{Beifiori::etal::2012}
{Beifiori} A.,  {Courteau} S.,  {Corsini} E.~M.,   {Zhu} Y.,  2012, \mn@doi
  [\mnras] {10.1111/j.1365-2966.2011.19903.x}, \href
  {https://ui.adsabs.harvard.edu/abs/2012MNRAS.419.2497B} {419, 2497}

\bibitem[\protect\citeauthoryear{{Bellstedt} et~al.,}{{Bellstedt}
  et~al.}{2016}]{Bellstedt::etal::2016}
{Bellstedt} S.,  et~al., 2016, \mn@doi [\mnras] {10.1093/mnras/stw1184}, \href
  {https://ui.adsabs.harvard.edu/abs/2016MNRAS.460.2862B} {460, 2862}

\bibitem[\protect\citeauthoryear{{Bennert} et~al.,}{{Bennert}
  et~al.}{2015}]{Bennert::etal::2015}
{Bennert} V.~N.,  et~al., 2015, \mn@doi [\apj] {10.1088/0004-637X/809/1/20},
  \href {https://ui.adsabs.harvard.edu/abs/2015ApJ...809...20B} {809, 20}

\bibitem[\protect\citeauthoryear{{Bentz} \& {Manne-Nicholas}}{{Bentz} \&
  {Manne-Nicholas}}{2018}]{Bentz::etal::2018}
{Bentz} M.~C.,  {Manne-Nicholas} E.,  2018, \mn@doi [\apj]
  {10.3847/1538-4357/aad808}, \href
  {https://ui.adsabs.harvard.edu/abs/2018ApJ...864..146B} {864, 146}

\bibitem[\protect\citeauthoryear{{Berta} et~al.,}{{Berta}
  et~al.}{2020}]{Berta::etal::2020}
{Berta} S.,  et~al., 2020, arXiv e-prints, \href
  {https://ui.adsabs.harvard.edu/abs/2020arXiv201201448B} {p. arXiv:2012.01448}

\bibitem[\protect\citeauthoryear{{Best} \& {Heckman}}{{Best} \&
  {Heckman}}{2012}]{BestHeckman2012}
{Best} P.~N.,  {Heckman} T.~M.,  2012, \mn@doi [\mnras]
  {10.1111/j.1365-2966.2012.20414.x}, \href
  {https://ui.adsabs.harvard.edu/abs/2012MNRAS.421.1569B} {421, 1569}

\bibitem[\protect\citeauthoryear{{Binney} \& {Tremaine}}{{Binney} \&
  {Tremaine}}{1987}]{BinneyTremaine1987}
{Binney} J.,  {Tremaine} S.,  1987, {Galactic dynamics}

\bibitem[\protect\citeauthoryear{{Birnboim}, {Dekel}  \& {Neistein}}{{Birnboim}
  et~al.}{2007}]{Birnboim::etal::2007}
{Birnboim} Y.,  {Dekel} A.,   {Neistein} E.,  2007, \mn@doi [\mnras]
  {10.1111/j.1365-2966.2007.12074.x}, \href
  {https://ui.adsabs.harvard.edu/abs/2007MNRAS.380..339B} {380, 339}

\bibitem[\protect\citeauthoryear{{Bischetti} et~al.,}{{Bischetti}
  et~al.}{2016}]{Bischetti::etal::2016}
{Bischetti} M.,  et~al., 2016, in Active Galactic Nuclei 12: A Multi-Messenger
  Perspective (AGN12). p.~12, \mn@doi{10.5281/zenodo.163588}

\bibitem[\protect\citeauthoryear{{Bischetti} et~al.,}{{Bischetti}
  et~al.}{2018}]{Bischetti::etal::2018}
{Bischetti} M.,  et~al., 2018, \mn@doi [\aap] {10.1051/0004-6361/201833249},
  \href {https://ui.adsabs.harvard.edu/abs/2018A&A...617A..82B} {617, A82}

\bibitem[\protect\citeauthoryear{{Blumenthal} \& {Barnes}}{{Blumenthal} \&
  {Barnes}}{2018}]{Blumenthal::etal::2018}
{Blumenthal} K.~A.,  {Barnes} J.~E.,  2018, \mn@doi [\mnras]
  {10.1093/mnras/sty1605}, \href
  {https://ui.adsabs.harvard.edu/abs/2018MNRAS.479.3952B} {479, 3952}

\bibitem[\protect\citeauthoryear{{Bondi}}{{Bondi}}{1952}]{Bondi1952}
{Bondi} H.,  1952, \mn@doi [\mnras] {10.1093/mnras/112.2.195}, \href
  {https://ui.adsabs.harvard.edu/#abs/1952MNRAS.112..195B} {112, 195}

\bibitem[\protect\citeauthoryear{{Bondi} \& {Hoyle}}{{Bondi} \&
  {Hoyle}}{1944}]{BondiHoyle1944}
{Bondi} H.,  {Hoyle} F.,  1944, \mn@doi [\mnras] {10.1093/mnras/104.5.273},
  \href {https://ui.adsabs.harvard.edu/\#abs/1944MNRAS.104..273B} {104, 273}

\bibitem[\protect\citeauthoryear{{Bongiorno} et~al.,}{{Bongiorno}
  et~al.}{2014}]{Bongiorno::etal::2014}
{Bongiorno} A.,  et~al., 2014, \mn@doi [\mnras] {10.1093/mnras/stu1248}, \href
  {https://ui.adsabs.harvard.edu/abs/2014MNRAS.443.2077B} {443, 2077}

\bibitem[\protect\citeauthoryear{{Booth} \& {Schaye}}{{Booth} \&
  {Schaye}}{2009}]{BoothSchaye2009}
{Booth} C.~M.,  {Schaye} J.,  2009, \mn@doi [\mnras]
  {10.1111/j.1365-2966.2009.15043.x}, \href
  {https://ui.adsabs.harvard.edu/abs/2009MNRAS.398...53B} {398, 53}

\bibitem[\protect\citeauthoryear{{Bower}, {Schaye}, {Frenk}, {Theuns},
  {Schaller}, {Crain}  \& {McAlpine}}{{Bower}
  et~al.}{2017a}]{Bower::etal::2017}
{Bower} R.~G.,  {Schaye} J.,  {Frenk} C.~S.,  {Theuns} T.,  {Schaller} M.,
  {Crain} R.~A.,   {McAlpine} S.,  2017a, \mn@doi [\mnras]
  {10.1093/mnras/stw2735}, \href
  {https://ui.adsabs.harvard.edu/abs/2017MNRAS.465...32B} {465, 32}

\bibitem[\protect\citeauthoryear{{Bower}, {Schaye}, {Frenk}, {Theuns},
  {Schaller}, {Crain}  \& {McAlpine}}{{Bower}
  et~al.}{2017b}]{Bower::etal::2016}
{Bower} R.~G.,  {Schaye} J.,  {Frenk} C.~S.,  {Theuns} T.,  {Schaller} M.,
  {Crain} R.~A.,   {McAlpine} S.,  2017b, \mn@doi [\mnras]
  {10.1093/mnras/stw2735}, \href
  {https://ui.adsabs.harvard.edu/abs/2017MNRAS.465...32B} {465, 32}

\bibitem[\protect\citeauthoryear{{Bowler}, {Bourne}, {Dunlop}, {McLure}  \&
  {McLeod}}{{Bowler} et~al.}{2018}]{Bowler::etal::2018}
{Bowler} R.~A.~A.,  {Bourne} N.,  {Dunlop} J.~S.,  {McLure} R.~J.,   {McLeod}
  D.~J.,  2018, \mn@doi [\mnras] {10.1093/mnras/sty2368}, \href
  {https://ui.adsabs.harvard.edu/abs/2018MNRAS.481.1631B} {481, 1631}

\bibitem[\protect\citeauthoryear{{Bromm} \& {Loeb}}{{Bromm} \&
  {Loeb}}{2003}]{BrommLoeb2003}
{Bromm} V.,  {Loeb} A.,  2003, \mn@doi [\apj] {10.1086/377529}, \href
  {http://adsabs.harvard.edu/abs/2003ApJ...596...34B} {596, 34}

\bibitem[\protect\citeauthoryear{{Buchan} \& {Shankar}}{{Buchan} \&
  {Shankar}}{2016}]{BuchanShankar2016}
{Buchan} S.,  {Shankar} F.,  2016, \mn@doi [\mnras] {10.1093/mnras/stw1771},
  \href {https://ui.adsabs.harvard.edu/abs/2016MNRAS.462.2001B} {462, 2001}

\bibitem[\protect\citeauthoryear{{Calhau}, {Sobral}, {Stroe}, {Best}, {Smail},
  {Lehmer}, {Harrison}  \& {Thomson}}{{Calhau}
  et~al.}{2017}]{Calhau::etal::2017}
{Calhau} J.,  {Sobral} D.,  {Stroe} A.,  {Best} P.,  {Smail} I.,  {Lehmer} B.,
  {Harrison} C.,   {Thomson} A.,  2017, \mn@doi [\mnras]
  {10.1093/mnras/stw2295}, \href
  {https://ui.adsabs.harvard.edu/abs/2017MNRAS.464..303C} {464, 303}

\bibitem[\protect\citeauthoryear{{Capelo} \& {Dotti}}{{Capelo} \&
  {Dotti}}{2017}]{CapeloDotti2017}
{Capelo} P.~R.,  {Dotti} M.,  2017, \mn@doi [\mnras] {10.1093/mnras/stw2872},
  \href {https://ui.adsabs.harvard.edu/abs/2017MNRAS.465.2643C} {465, 2643}

\bibitem[\protect\citeauthoryear{{Casey}, {Narayanan}  \& {Cooray}}{{Casey}
  et~al.}{2014}]{Caitlin::etal::2014}
{Casey} C.~M.,  {Narayanan} D.,   {Cooray} A.,  2014, \mn@doi [\physrep]
  {10.1016/j.physrep.2014.02.009}, \href
  {https://ui.adsabs.harvard.edu/abs/2014PhR...541...45C} {541, 45}

\bibitem[\protect\citeauthoryear{{Chabrier}}{{Chabrier}}{2003}]{Chabrier2003}
{Chabrier} G.,  2003, \mn@doi [\pasp] {10.1086/376392}, \href
  {https://ui.adsabs.harvard.edu/abs/2003PASP..115..763C} {115, 763}

\bibitem[\protect\citeauthoryear{{Cisternas} et~al.,}{{Cisternas}
  et~al.}{2011}]{Cisternas::etal::2011}
{Cisternas} M.,  et~al., 2011, \mn@doi [\apj] {10.1088/2041-8205/741/1/L11},
  \href {https://ui.adsabs.harvard.edu/abs/2011ApJ...741L..11C} {741, L11}

\bibitem[\protect\citeauthoryear{{Cochrane} et~al.,}{{Cochrane}
  et~al.}{2019}]{Cochrane::etal::2019}
{Cochrane} R.~K.,  et~al., 2019, \mn@doi [\mnras] {10.1093/mnras/stz1736},
  \href {https://ui.adsabs.harvard.edu/abs/2019MNRAS.488.1779C} {488, 1779}

\bibitem[\protect\citeauthoryear{{Cooke}, {Kartaltepe}, {Tyler}, {Darvish},
  {Casey}, {Le F{\`e}vre}, {Salvato}  \& {Scoville}}{{Cooke}
  et~al.}{2019}]{Cooke::etal::2019}
{Cooke} K.~C.,  {Kartaltepe} J.~S.,  {Tyler} K.~D.,  {Darvish} B.,  {Casey}
  C.~M.,  {Le F{\`e}vre} O.,  {Salvato} M.,   {Scoville} N.,  2019, \mn@doi
  [\apj] {10.3847/1538-4357/ab30c9}, \href
  {https://ui.adsabs.harvard.edu/abs/2019ApJ...881..150C} {881, 150}

\bibitem[\protect\citeauthoryear{{Cowie}, {Hu}  \& {Songaila}}{{Cowie}
  et~al.}{1995}]{Cowie::etal::1995}
{Cowie} L.~L.,  {Hu} E.~M.,   {Songaila} A.,  1995, \mn@doi [\aj]
  {10.1086/117631}, \href
  {https://ui.adsabs.harvard.edu/abs/1995AJ....110.1576C} {110, 1576}

\bibitem[\protect\citeauthoryear{{Cox}, {Jonsson}, {Primack}  \&
  {Somerville}}{{Cox} et~al.}{2006}]{Cox::etal::2006}
{Cox} T.~J.,  {Jonsson} P.,  {Primack} J.~R.,   {Somerville} R.~S.,  2006,
  \mn@doi [\mnras] {10.1111/j.1365-2966.2006.11107.x}, \href
  {https://ui.adsabs.harvard.edu/abs/2006MNRAS.373.1013C} {373, 1013}

\bibitem[\protect\citeauthoryear{{Cresci} et~al.,}{{Cresci}
  et~al.}{2015}]{Cresci::etal::2015}
{Cresci} G.,  et~al., 2015, \mn@doi [\apj] {10.1088/0004-637X/799/1/82}, \href
  {https://ui.adsabs.harvard.edu/abs/2015ApJ...799...82C} {799, 82}

\bibitem[\protect\citeauthoryear{{D'Onofrio}, {Marziani}  \&
  {Buson}}{{D'Onofrio} et~al.}{2015}]{D'Onofrio::etal::2015}
{D'Onofrio} M.,  {Marziani} P.,   {Buson} L.,  2015, \mn@doi [Frontiers in
  Astronomy and Space Sciences] {10.3389/fspas.2015.00004}, \href
  {https://ui.adsabs.harvard.edu/abs/2015FrASS...2....4D} {2, 4}

\bibitem[\protect\citeauthoryear{{Dai}, {Wilkes}, {Bergeron}, {Kuraszkiewicz},
  {Omont}, {Atanas}  \& {Teplitz}}{{Dai} et~al.}{2018a}]{Dai::etal::2018}
{Dai} Y.~S.,  {Wilkes} B.~J.,  {Bergeron} J.,  {Kuraszkiewicz} J.,  {Omont} A.,
   {Atanas} A.,   {Teplitz} H.~I.,  2018a, \mn@doi [\mnras]
  {10.1093/mnras/sty1341}, \href
  {https://ui.adsabs.harvard.edu/abs/2018MNRAS.478.4238D} {478, 4238}

\bibitem[\protect\citeauthoryear{{Dai}, {Wilkes}, {Bergeron}, {Kuraszkiewicz},
  {Omont}, {Atanas}  \& {Teplitz}}{{Dai} et~al.}{2018b}]{Dai::etal::2015}
{Dai} Y.~S.,  {Wilkes} B.~J.,  {Bergeron} J.,  {Kuraszkiewicz} J.,  {Omont} A.,
   {Atanas} A.,   {Teplitz} H.~I.,  2018b, \mn@doi [\mnras]
  {10.1093/mnras/sty1341}, \href
  {https://ui.adsabs.harvard.edu/abs/2018MNRAS.478.4238D} {478, 4238}

\bibitem[\protect\citeauthoryear{{Dav{\'e}}, {Angl{\'e}s-Alc{\'a}zar},
  {Narayanan}, {Li}, {Rafieferantsoa}  \& {Appleby}}{{Dav{\'e}}
  et~al.}{2019}]{Dave::etal::2019}
{Dav{\'e}} R.,  {Angl{\'e}s-Alc{\'a}zar} D.,  {Narayanan} D.,  {Li} Q.,
  {Rafieferantsoa} M.~H.,   {Appleby} S.,  2019, \mn@doi [\mnras]
  {10.1093/mnras/stz937}, \href
  {https://ui.adsabs.harvard.edu/abs/2019MNRAS.486.2827D} {486, 2827}

\bibitem[\protect\citeauthoryear{{Davis}, {Graham}  \& {Cameron}}{{Davis}
  et~al.}{2019}]{Davis::etal::2019}
{Davis} B.~L.,  {Graham} A.~W.,   {Cameron} E.,  2019, \mn@doi [The
  Astrophysical Journal] {10.3847/1538-4357/aaf3b8}, \href
  {https://ui.adsabs.harvard.edu/abs/2019ApJ...873...85D} {873, 85}

\bibitem[\protect\citeauthoryear{{Decarli} et~al.,}{{Decarli}
  et~al.}{2018}]{Decarli::etal::2018}
{Decarli} R.,  et~al., 2018, \mn@doi [\apj] {10.3847/1538-4357/aaa5aa}, \href
  {https://ui.adsabs.harvard.edu/abs/2018ApJ...854...97D} {854, 97}

\bibitem[\protect\citeauthoryear{{Dehnen} \& {Aly}}{{Dehnen} \&
  {Aly}}{2012}]{DehnenAly2012}
{Dehnen} W.,  {Aly} H.,  2012, \mn@doi [\mnras]
  {10.1111/j.1365-2966.2012.21439.x}, \href
  {https://ui.adsabs.harvard.edu/abs/2012MNRAS.425.1068D} {425, 1068}

\bibitem[\protect\citeauthoryear{{Dekel} \& {Birnboim}}{{Dekel} \&
  {Birnboim}}{2008}]{DekelBirnboim2008}
{Dekel} A.,  {Birnboim} Y.,  2008, \mn@doi [\mnras]
  {10.1111/j.1365-2966.2007.12569.x}, \href
  {https://ui.adsabs.harvard.edu/abs/2008MNRAS.383..119D} {383, 119}

\bibitem[\protect\citeauthoryear{{Dekel}, {Sari}  \& {Ceverino}}{{Dekel}
  et~al.}{2009}]{Dekel::etal::2009}
{Dekel} A.,  {Sari} R.,   {Ceverino} D.,  2009, \mn@doi [\apj]
  {10.1088/0004-637X/703/1/785}, \href
  {https://ui.adsabs.harvard.edu/abs/2009ApJ...703..785D} {703, 785}

\bibitem[\protect\citeauthoryear{{Delvecchio} et~al.,}{{Delvecchio}
  et~al.}{2019}]{Delvecchio::etal::2019}
{Delvecchio} I.,  et~al., 2019, arXiv e-prints, \href
  {https://ui.adsabs.harvard.edu/abs/2019arXiv191008114D} {p. arXiv:1910.08114}

\bibitem[\protect\citeauthoryear{{Devriendt}, {Slyz}, {Powell}, {Pichon}  \&
  {Teyssier}}{{Devriendt} et~al.}{2010}]{Devriendt::etal::2010}
{Devriendt} J.,  {Slyz} A.,  {Powell} L.,  {Pichon} C.,   {Teyssier} R.,  2010,
  \mn@doi [] {10.1017/S1743921310002863}, \href
  {https://ui.adsabs.harvard.edu/abs/2010IAUS..262..248D} {262, 248}

\bibitem[\protect\citeauthoryear{{Di Matteo}, {Springel}  \& {Hernquist}}{{Di
  Matteo} et~al.}{2005}]{DiMatteo::etal::2005}
{Di Matteo} T.,  {Springel} V.,   {Hernquist} L.,  2005, \mn@doi [\nat]
  {10.1038/nature03335}, \href
  {https://ui.adsabs.harvard.edu/abs/2005Natur.433..604D} {433, 604}

\bibitem[\protect\citeauthoryear{{D{\'\i}az-Santos} et~al.,}{{D{\'\i}az-Santos}
  et~al.}{2018}]{Diaz-Santos::etal::2018}
{D{\'\i}az-Santos} T.,  et~al., 2018, \mn@doi [Science]
  {10.1126/science.aap7605}, \href
  {https://ui.adsabs.harvard.edu/abs/2018Sci...362.1034D} {362, 1034}

\bibitem[\protect\citeauthoryear{{Ding} et~al.,}{{Ding}
  et~al.}{2017}]{Ding::etal::2017}
{Ding} X.,  et~al., 2017, \mn@doi [\mnras] {10.1093/mnras/stx1972}, \href
  {https://ui.adsabs.harvard.edu/abs/2017MNRAS.472...90D} {472, 90}

\bibitem[\protect\citeauthoryear{{Ding} et~al.,}{{Ding}
  et~al.}{2020}]{Ding::etal::2020}
{Ding} X.,  et~al., 2020, \mn@doi [\apj] {10.3847/1538-4357/ab5b90}, \href
  {https://ui.adsabs.harvard.edu/abs/2020ApJ...888...37D} {888, 37}

\bibitem[\protect\citeauthoryear{{Dubois} et~al.,}{{Dubois}
  et~al.}{2014}]{Dubois::etal::2014}
{Dubois} Y.,  et~al., 2014, \mn@doi [\mnras] {10.1093/mnras/stu1227}, \href
  {https://ui.adsabs.harvard.edu/abs/2014MNRAS.444.1453D} {444, 1453}

\bibitem[\protect\citeauthoryear{{Dubois}, {Volonteri}, {Silk}, {Devriendt},
  {Slyz}  \& {Teyssier}}{{Dubois} et~al.}{2015}]{Dubois::etal::2015}
{Dubois} Y.,  {Volonteri} M.,  {Silk} J.,  {Devriendt} J.,  {Slyz} A.,
  {Teyssier} R.,  2015, \mn@doi [\mnras] {10.1093/mnras/stv1416}, \href
  {https://ui.adsabs.harvard.edu/abs/2015MNRAS.452.1502D} {452, 1502}

\bibitem[\protect\citeauthoryear{{Duncan} et~al.,}{{Duncan}
  et~al.}{2019}]{Duncan::etal::2019}
{Duncan} K.,  et~al., 2019, \mn@doi [\apj] {10.3847/1538-4357/ab148a}, \href
  {https://ui.adsabs.harvard.edu/abs/2019ApJ...876..110D} {876, 110}

\bibitem[\protect\citeauthoryear{{Duras} et~al.,}{{Duras}
  et~al.}{2017}]{Duras::etal::2017}
{Duras} F.,  et~al., 2017, \mn@doi [\aap] {10.1051/0004-6361/201731052}, \href
  {https://ui.adsabs.harvard.edu/abs/2017A&A...604A..67D} {604, A67}

\bibitem[\protect\citeauthoryear{{Eilers}, {Hennawi}  \& {Davies}}{{Eilers}
  et~al.}{2018}]{Eilers::etal::2018}
{Eilers} A.-C.,  {Hennawi} J.~F.,   {Davies} F.~B.,  2018, \mn@doi [\apj]
  {10.3847/1538-4357/aae081}, \href
  {https://ui.adsabs.harvard.edu/abs/2018ApJ...867...30E} {867, 30}

\bibitem[\protect\citeauthoryear{{Ellison}, {S{\'a}nchez}, {Ibarra-Medel},
  {Antonio}, {Mendel}  \& {Barrera-Ballesteros}}{{Ellison}
  et~al.}{2018}]{Ellison::etal::2018}
{Ellison} S.~L.,  {S{\'a}nchez} S.~F.,  {Ibarra-Medel} H.,  {Antonio} B.,
  {Mendel} J.~T.,   {Barrera-Ballesteros} J.,  2018, \mn@doi [\mnras]
  {10.1093/mnras/stx2882}, \href
  {https://ui.adsabs.harvard.edu/abs/2018MNRAS.474.2039E} {474, 2039}

\bibitem[\protect\citeauthoryear{{Fan} et~al.,}{{Fan}
  et~al.}{2016}]{Fan::etal::2016}
{Fan} L.,  et~al., 2016, \mn@doi [\apjl] {10.3847/2041-8205/822/2/L32}, \href
  {https://ui.adsabs.harvard.edu/abs/2016ApJ...822L..32F} {822, L32}

\bibitem[\protect\citeauthoryear{{Faucher-Gigu{\`e}re}}{{Faucher-Gigu{\`e}re}}{2018}]{Faucher-Giguere2018}
{Faucher-Gigu{\`e}re} C.-A.,  2018, \mn@doi [\mnras] {10.1093/mnras/stx2595},
  \href {https://ui.adsabs.harvard.edu/abs/2018MNRAS.473.3717F} {473, 3717}

\bibitem[\protect\citeauthoryear{{Faucher-Gigu{\`e}re}, {Lidz}, {Zaldarriaga}
  \& {Hernquist}}{{Faucher-Gigu{\`e}re}
  et~al.}{2009}]{Faucher-Giguere::etal::2009}
{Faucher-Gigu{\`e}re} C.-A.,  {Lidz} A.,  {Zaldarriaga} M.,   {Hernquist} L.,
  2009, \mn@doi [\apj] {10.1088/0004-637X/703/2/1416}, \href
  {https://ui.adsabs.harvard.edu/abs/2009ApJ...703.1416F} {703, 1416}

\bibitem[\protect\citeauthoryear{{Feldmann}}{{Feldmann}}{2017}]{Feldmann17}
{Feldmann} R.,  2017, \mn@doi [\mnras] {10.1093/mnrasl/slx073}, \href
  {https://ui.adsabs.harvard.edu/#abs/2017MNRAS.470L..59F} {470, L59}

\bibitem[\protect\citeauthoryear{{Feldmann}, {Hopkins}, {Quataert},
  {Faucher-Gigu{\`e}re}  \& {Kere{\v{s}}}}{{Feldmann}
  et~al.}{2016}]{Feldmann+16}
{Feldmann} R.,  {Hopkins} P.~F.,  {Quataert} E.,  {Faucher-Gigu{\`e}re} C.-A.,
   {Kere{\v{s}}} D.,  2016, \mn@doi [\mnras] {10.1093/mnrasl/slw014}, \href
  {https://ui.adsabs.harvard.edu/#abs/2016MNRAS.458L..14F} {458, L14}

\bibitem[\protect\citeauthoryear{{Feldmann}, {Quataert}, {Hopkins},
  {Faucher-Gigu{\`e}re}  \& {Kere{\v{s}}}}{{Feldmann}
  et~al.}{2017}]{Feldmann+17}
{Feldmann} R.,  {Quataert} E.,  {Hopkins} P.~F.,  {Faucher-Gigu{\`e}re} C.-A.,
   {Kere{\v{s}}} D.,  2017, \mn@doi [\mnras] {10.1093/mnras/stx1120}, \href
  {https://ui.adsabs.harvard.edu/#abs/2017MNRAS.470.1050F} {470, 1050}

\bibitem[\protect\citeauthoryear{{Feng}, {Di-Matteo}, {Croft}, {Bird},
  {Battaglia}  \& {Wilkins}}{{Feng} et~al.}{2016}]{Feng::etal::2016}
{Feng} Y.,  {Di-Matteo} T.,  {Croft} R.~A.,  {Bird} S.,  {Battaglia} N.,
  {Wilkins} S.,  2016, \mn@doi [\mnras] {10.1093/mnras/stv2484}, \href
  {https://ui.adsabs.harvard.edu/abs/2016MNRAS.455.2778F} {455, 2778}

\bibitem[\protect\citeauthoryear{{Ferrarese} \& {Merritt}}{{Ferrarese} \&
  {Merritt}}{2000}]{FerrareseMerritt2000}
{Ferrarese} L.,  {Merritt} D.,  2000, \mn@doi [\apj] {10.1086/312838}, \href
  {https://ui.adsabs.harvard.edu/abs/2000ApJ...539L...9F} {539, L9}

\bibitem[\protect\citeauthoryear{{Ferreras} et~al.,}{{Ferreras}
  et~al.}{2014}]{Ferreras::etal::2014}
{Ferreras} I.,  et~al., 2014, \mn@doi [\mnras] {10.1093/mnras/stu1425}, \href
  {https://ui.adsabs.harvard.edu/abs/2014MNRAS.444..906F} {444, 906}

\bibitem[\protect\citeauthoryear{{Ferreras}, {Trujillo},
  {M{\'a}rmol-Queralt{\'o}}  \& {P{\'e}rez-Gonz{\'a}lez}}{{Ferreras}
  et~al.}{2016}]{Ferreras::etal::2016}
{Ferreras} I.,  {Trujillo} I.,  {M{\'a}rmol-Queralt{\'o}} E.,
  {P{\'e}rez-Gonz{\'a}lez} P.,  2016, \mn@doi [] {10.1017/S174392131501087X},
  \href {https://ui.adsabs.harvard.edu/abs/2016IAUS..319..114F} {319, 114}

\bibitem[\protect\citeauthoryear{{Feruglio} et~al.,}{{Feruglio}
  et~al.}{2018}]{Feruglio::etal::2018}
{Feruglio} C.,  et~al., 2018, \mn@doi [\aap] {10.1051/0004-6361/201833174},
  \href {https://ui.adsabs.harvard.edu/abs/2018A&A...619A..39F} {619, A39}

\bibitem[\protect\citeauthoryear{{Finkelstein} et~al.,}{{Finkelstein}
  et~al.}{2013}]{Finkelstein::etal::2013}
{Finkelstein} S.~L.,  et~al., 2013, \mn@doi [\nat] {10.1038/nature12657}, \href
  {https://ui.adsabs.harvard.edu/abs/2013Natur.502..524F} {502, 524}

\bibitem[\protect\citeauthoryear{{Gabor}, {Dav{\'e}}, {Finlator}  \&
  {Oppenheimer}}{{Gabor} et~al.}{2010}]{Gabor::etal::2010}
{Gabor} J.~M.,  {Dav{\'e}} R.,  {Finlator} K.,   {Oppenheimer} B.~D.,  2010,
  \mn@doi [\mnras] {10.1111/j.1365-2966.2010.16961.x}, \href
  {https://ui.adsabs.harvard.edu/abs/2010MNRAS.407..749G} {407, 749}

\bibitem[\protect\citeauthoryear{{Gebhardt} et~al.,}{{Gebhardt}
  et~al.}{2000}]{Gebhardt::etal::2000}
{Gebhardt} K.,  et~al., 2000, \mn@doi [\apj] {10.1086/312840}, \href
  {https://ui.adsabs.harvard.edu/abs/2000ApJ...539L..13G} {539, L13}

\bibitem[\protect\citeauthoryear{{Genel} et~al.,}{{Genel}
  et~al.}{2014}]{Genel::etal::2014}
{Genel} S.,  et~al., 2014, \mn@doi [\mnras] {10.1093/mnras/stu1654}, \href
  {https://ui.adsabs.harvard.edu/abs/2014MNRAS.445..175G} {445, 175}

\bibitem[\protect\citeauthoryear{{Genzel} et~al.,}{{Genzel}
  et~al.}{2011}]{Genzel::etal::2011}
{Genzel} R.,  et~al., 2011, \mn@doi [\apj] {10.1088/0004-637X/733/2/101}, \href
  {https://ui.adsabs.harvard.edu/abs/2011ApJ...733..101G} {733, 101}

\bibitem[\protect\citeauthoryear{{Graham} \& {Scott}}{{Graham} \&
  {Scott}}{2015}]{GrahamScott2015}
{Graham} A.~W.,  {Scott} N.,  2015, \mn@doi [\apj]
  {10.1088/0004-637X/798/1/54}, \href
  {https://ui.adsabs.harvard.edu/abs/2015ApJ...798...54G} {798, 54}

\bibitem[\protect\citeauthoryear{{Gruppioni} et~al.,}{{Gruppioni}
  et~al.}{2016}]{Gruppioni::etal::2016}
{Gruppioni} C.,  et~al., 2016, \mn@doi [\mnras] {10.1093/mnras/stw577}, \href
  {https://ui.adsabs.harvard.edu/abs/2016MNRAS.458.4297G} {458, 4297}

\bibitem[\protect\citeauthoryear{{G{\"u}ltekin} et~al.,}{{G{\"u}ltekin}
  et~al.}{2009}]{Gultekin+2009}
{G{\"u}ltekin} K.,  et~al., 2009, \mn@doi [\apj] {10.1088/0004-637X/698/1/198},
  \href {https://ui.adsabs.harvard.edu/#abs/2009ApJ...698..198G} {698, 198}

\bibitem[\protect\citeauthoryear{{Guo}, {Giavalisco}, {Ferguson}, {Cassata}  \&
  {Koekemoer}}{{Guo} et~al.}{2012}]{Guo::etal::2012}
{Guo} Y.,  {Giavalisco} M.,  {Ferguson} H.~C.,  {Cassata} P.,   {Koekemoer}
  A.~M.,  2012, \mn@doi [\apj] {10.1088/0004-637X/757/2/120}, \href
  {https://ui.adsabs.harvard.edu/abs/2012ApJ...757..120G} {757, 120}

\bibitem[\protect\citeauthoryear{{Habouzit}, {Volonteri}  \&
  {Dubois}}{{Habouzit} et~al.}{2017}]{Habouzit::etal::2017}
{Habouzit} M.,  {Volonteri} M.,   {Dubois} Y.,  2017, \mn@doi [\mnras]
  {10.1093/mnras/stx666}, \href
  {https://ui.adsabs.harvard.edu/abs/2017MNRAS.468.3935H} {468, 3935}

\bibitem[\protect\citeauthoryear{{Hahn} \& {Abel}}{{Hahn} \&
  {Abel}}{2011}]{Hahn&Abel2011}
{Hahn} O.,  {Abel} T.,  2011, \mn@doi [\mnras]
  {10.1111/j.1365-2966.2011.18820.x}, \href
  {https://ui.adsabs.harvard.edu/abs/2011MNRAS.415.2101H} {415, 2101}

\bibitem[\protect\citeauthoryear{{H{\"a}ring} \& {Rix}}{{H{\"a}ring} \&
  {Rix}}{2004}]{HaringRix2004}
{H{\"a}ring} N.,  {Rix} H.-W.,  2004, \mn@doi [\apj] {10.1086/383567}, \href
  {https://ui.adsabs.harvard.edu/#abs/2004ApJ...604L..89H} {604, L89}

\bibitem[\protect\citeauthoryear{{Hobbs}, {Power}, {Nayakshin}  \&
  {King}}{{Hobbs} et~al.}{2012}]{Hobbs:etal:2012}
{Hobbs} A.,  {Power} C.,  {Nayakshin} S.,   {King} A.~R.,  2012, \mn@doi
  [\mnras] {10.1111/j.1365-2966.2012.20563.x}, \href
  {https://ui.adsabs.harvard.edu/\#abs/2012MNRAS.421.3443H} {421, 3443}

\bibitem[\protect\citeauthoryear{{Hodge}, {Carilli}, {Walter}, {de Blok},
  {Riechers}, {Daddi}  \& {Lentati}}{{Hodge} et~al.}{2012}]{Hodge::etal::2012}
{Hodge} J.~A.,  {Carilli} C.~L.,  {Walter} F.,  {de Blok} W.~J.~G.,  {Riechers}
  D.,  {Daddi} E.,   {Lentati} L.,  2012, \mn@doi [\apj]
  {10.1088/0004-637X/760/1/11}, \href
  {https://ui.adsabs.harvard.edu/abs/2012ApJ...760...11H} {760, 11}

\bibitem[\protect\citeauthoryear{{Hopkins} \& {Quataert}}{{Hopkins} \&
  {Quataert}}{2010}]{HopkinsQuataert2010}
{Hopkins} P.~F.,  {Quataert} E.,  2010, \mn@doi [\mnras]
  {10.1111/j.1365-2966.2010.17064.x}, \href
  {https://ui.adsabs.harvard.edu/abs/2010MNRAS.407.1529H} {407, 1529}

\bibitem[\protect\citeauthoryear{{Hopkins} \& {Quataert}}{{Hopkins} \&
  {Quataert}}{2011}]{HopkinsQuataert2011}
{Hopkins} P.~F.,  {Quataert} E.,  2011, \mn@doi [\mnras]
  {10.1111/j.1365-2966.2011.18542.x}, \href
  {https://ui.adsabs.harvard.edu/#abs/2011MNRAS.415.1027H} {415, 1027}

\bibitem[\protect\citeauthoryear{{Hopkins}, {Kere{\v{s}}}, {O{\~n}orbe},
  {Faucher-Gigu{\`e}re}, {Quataert}, {Murray}  \& {Bullock}}{{Hopkins}
  et~al.}{2014}]{Hopkins:etal:2014}
{Hopkins} P.~F.,  {Kere{\v{s}}} D.,  {O{\~n}orbe} J.,  {Faucher-Gigu{\`e}re}
  C.-A.,  {Quataert} E.,  {Murray} N.,   {Bullock} J.~S.,  2014, \mn@doi
  [\mnras] {10.1093/mnras/stu1738}, \href
  {https://ui.adsabs.harvard.edu/\#abs/2014MNRAS.445..581H} {445, 581}

\bibitem[\protect\citeauthoryear{{Hopkins} et~al.,}{{Hopkins}
  et~al.}{2018}]{Hopkins+18}
{Hopkins} P.~F.,  et~al., 2018, \mn@doi [\mnras] {10.1093/mnras/sty1690}, \href
  {https://ui.adsabs.harvard.edu/#abs/2018MNRAS.480..800H} {480, 800}

\bibitem[\protect\citeauthoryear{{Hoyle} \& {Lyttleton}}{{Hoyle} \&
  {Lyttleton}}{1939}]{HoyleLyttleton1939}
{Hoyle} F.,  {Lyttleton} R.~A.,  1939, \mn@doi [PCPS]
  {10.1017/S0305004100021150}, \href
  {https://ui.adsabs.harvard.edu/\#abs/1939PCPS...35..405H} {35, 405}

\bibitem[\protect\citeauthoryear{{Huang}, {Di Matteo}, {Bhowmick}, {Feng}  \&
  {Ma}}{{Huang} et~al.}{2018}]{Huang::etal::2018}
{Huang} K.-W.,  {Di Matteo} T.,  {Bhowmick} A.~K.,  {Feng} Y.,   {Ma} C.-P.,
  2018, \mn@doi [\mnras] {10.1093/mnras/sty1329}, \href
  {https://ui.adsabs.harvard.edu/abs/2018MNRAS.478.5063H} {478, 5063}

\bibitem[\protect\citeauthoryear{{Ivison} et~al.,}{{Ivison}
  et~al.}{2012}]{Ivison::etal::2012}
{Ivison} R.~J.,  et~al., 2012, \mn@doi [\mnras]
  {10.1111/j.1365-2966.2012.21544.x}, \href
  {https://ui.adsabs.harvard.edu/abs/2012MNRAS.425.1320I} {425, 1320}

\bibitem[\protect\citeauthoryear{{Izumi} et~al.,}{{Izumi}
  et~al.}{2018}]{Izumi::etal::2018}
{Izumi} T.,  et~al., 2018, \mn@doi [\pasj] {10.1093/pasj/psy026}, \href
  {https://ui.adsabs.harvard.edu/abs/2018PASJ...70...36I} {70, 36}

\bibitem[\protect\citeauthoryear{{Jahnke} et~al.,}{{Jahnke}
  et~al.}{2009}]{Jahnke::etal::2009}
{Jahnke} K.,  et~al., 2009, \mn@doi [\apj] {10.1088/0004-637X/706/2/L215},
  \href {https://ui.adsabs.harvard.edu/abs/2009ApJ...706L.215J} {706, L215}

\bibitem[\protect\citeauthoryear{{Kauffmann} et~al.,}{{Kauffmann}
  et~al.}{2003}]{Kauffmann::etal::2003}
{Kauffmann} G.,  et~al., 2003, \mn@doi [\mnras]
  {10.1046/j.1365-8711.2003.06292.x}, \href
  {https://ui.adsabs.harvard.edu/abs/2003MNRAS.341...54K} {341, 54}

\bibitem[\protect\citeauthoryear{{Kaviraj} et~al.,}{{Kaviraj}
  et~al.}{2017}]{Kaviraj::etal::2017}
{Kaviraj} S.,  et~al., 2017, \mn@doi [\mnras] {10.1093/mnras/stx126}, \href
  {https://ui.adsabs.harvard.edu/abs/2017MNRAS.467.4739K} {467, 4739}

\bibitem[\protect\citeauthoryear{{Kennicutt}}{{Kennicutt}}{1998}]{Kennicutt1998}
{Kennicutt} Robert~C. J.,  1998, \mn@doi [Annual Review of Astronomy and
  Astrophysics] {10.1146/annurev.astro.36.1.189}, \href
  {https://ui.adsabs.harvard.edu/abs/1998ARA&A..36..189K} {36, 189}

\bibitem[\protect\citeauthoryear{{Khandai}, {Di Matteo}, {Croft}, {Wilkins},
  {Feng}, {Tucker}, {DeGraf}  \& {Liu}}{{Khandai}
  et~al.}{2015}]{Khandai::etal::2015}
{Khandai} N.,  {Di Matteo} T.,  {Croft} R.,  {Wilkins} S.,  {Feng} Y.,
  {Tucker} E.,  {DeGraf} C.,   {Liu} M.-S.,  2015, \mn@doi [\mnras]
  {10.1093/mnras/stv627}, \href
  {https://ui.adsabs.harvard.edu/abs/2015MNRAS.450.1349K} {450, 1349}

\bibitem[\protect\citeauthoryear{{Kimball}, {Lacy}, {Lonsdale}  \&
  {Macquart}}{{Kimball} et~al.}{2015}]{Kimball::etal::2015}
{Kimball} A.~E.,  {Lacy} M.,  {Lonsdale} C.~J.,   {Macquart} J.~P.,  2015,
  \mn@doi [\mnras] {10.1093/mnras/stv1160}, \href
  {https://ui.adsabs.harvard.edu/abs/2015MNRAS.452...88K} {452, 88}

\bibitem[\protect\citeauthoryear{{Knollmann} \& {Knebe}}{{Knollmann} \&
  {Knebe}}{2009}]{KnollmannKnebe2009}
{Knollmann} S.~R.,  {Knebe} A.,  2009, \mn@doi [The Astrophysical Journal
  Supplement Series] {10.1088/0067-0049/182/2/608}, \href
  {https://ui.adsabs.harvard.edu/#abs/2009ApJS..182..608K} {182, 608}

\bibitem[\protect\citeauthoryear{{Kormendy} \& {Ho}}{{Kormendy} \&
  {Ho}}{2013}]{KormendyHo2013}
{Kormendy} J.,  {Ho} L.~C.,  2013, \mn@doi [Annual Review of Astronomy and
  Astrophysics] {10.1146/annurev-astro-082708-101811}, \href
  {https://ui.adsabs.harvard.edu/#abs/2013ARA&A..51..511K} {51, 511}

\bibitem[\protect\citeauthoryear{{Kormendy} \& {Richstone}}{{Kormendy} \&
  {Richstone}}{1995}]{KormendyRichstone1995}
{Kormendy} J.,  {Richstone} D.,  1995, \mn@doi [\araa]
  {10.1146/annurev.aa.33.090195.003053}, \href
  {https://ui.adsabs.harvard.edu/abs/1995ARA&A..33..581K} {33, 581}

\bibitem[\protect\citeauthoryear{{Lynden-Bell}}{{Lynden-Bell}}{1969}]{Lynden-Bell1969}
{Lynden-Bell} D.,  1969, \mn@doi [\nat] {10.1038/223690a0}, \href
  {https://ui.adsabs.harvard.edu/abs/1969Natur.223..690L} {223, 690}

\bibitem[\protect\citeauthoryear{{Lynden-Bell} \& {Rees}}{{Lynden-Bell} \&
  {Rees}}{1971}]{Lynden-BellRees1971}
{Lynden-Bell} D.,  {Rees} M.~J.,  1971, \mn@doi [\mnras]
  {10.1093/mnras/152.4.461}, \href
  {https://ui.adsabs.harvard.edu/abs/1971MNRAS.152..461L} {152, 461}

\bibitem[\protect\citeauthoryear{{Ma}, {Hopkins}, {Ma},
  {Angl{\'e}s-Alc{\'a}zar}, {Faucher-Gigu{\`e}re}  \& {Kelley}}{{Ma}
  et~al.}{2021}]{Ma::etal::2021}
{Ma} L.,  {Hopkins} P.~F.,  {Ma} X.,  {Angl{\'e}s-Alc{\'a}zar} D.,
  {Faucher-Gigu{\`e}re} C.-A.,   {Kelley} L.~Z.,  2021, arXiv e-prints, \href
  {https://ui.adsabs.harvard.edu/abs/2021arXiv210102727M} {p. arXiv:2101.02727}

\bibitem[\protect\citeauthoryear{{Magorrian} et~al.,}{{Magorrian}
  et~al.}{1998}]{Magorrian+1998}
{Magorrian} J.,  et~al., 1998, \mn@doi [\aj] {10.1086/300353}, \href
  {https://ui.adsabs.harvard.edu/#abs/1998AJ....115.2285M} {115, 2285}

\bibitem[\protect\citeauthoryear{{Maiolino} et~al.,}{{Maiolino}
  et~al.}{2005}]{Maiolino::etal::2005}
{Maiolino} R.,  et~al., 2005, \mn@doi [\aap] {10.1051/0004-6361:200500165},
  \href {https://ui.adsabs.harvard.edu/abs/2005A&A...440L..51M} {440, L51}

\bibitem[\protect\citeauthoryear{{Maiolino} et~al.,}{{Maiolino}
  et~al.}{2017}]{Maiolino::etal::2017}
{Maiolino} R.,  et~al., 2017, \mn@doi [\nat] {10.1038/nature21677}, \href
  {https://ui.adsabs.harvard.edu/abs/2017Natur.544..202M} {544, 202}

\bibitem[\protect\citeauthoryear{{Marchesini} et~al.,}{{Marchesini}
  et~al.}{2014}]{Marchesini::etal::2014}
{Marchesini} D.,  et~al., 2014, \mn@doi [\apj] {10.1088/0004-637X/794/1/65},
  \href {https://ui.adsabs.harvard.edu/abs/2014ApJ...794...65M} {794, 65}

\bibitem[\protect\citeauthoryear{{Marconi} \& {Hunt}}{{Marconi} \&
  {Hunt}}{2003}]{MarconiHunt2003}
{Marconi} A.,  {Hunt} L.~K.,  2003, \mn@doi [\apj] {10.1086/375804}, \href
  {https://ui.adsabs.harvard.edu/abs/2003ApJ...589L..21M} {589, L21}

\bibitem[\protect\citeauthoryear{{Mart{\'\i}n-Navarro} \&
  {Mezcua}}{{Mart{\'\i}n-Navarro} \& {Mezcua}}{2018}]{Martin-NavarroMezcua2018}
{Mart{\'\i}n-Navarro} I.,  {Mezcua} M.,  2018, \mn@doi [The Astrophysical
  Journal] {10.3847/2041-8213/aab103}, \href
  {https://ui.adsabs.harvard.edu/abs/2018ApJ...855L..20M} {855, L20}

\bibitem[\protect\citeauthoryear{{Matsuoka} et~al.,}{{Matsuoka}
  et~al.}{2016}]{Matsuoka::etal::2016}
{Matsuoka} Y.,  et~al., 2016, \mn@doi [\apj] {10.3847/0004-637X/828/1/26},
  \href {https://ui.adsabs.harvard.edu/abs/2016ApJ...828...26M} {828, 26}

\bibitem[\protect\citeauthoryear{{Mazzucchelli} et~al.,}{{Mazzucchelli}
  et~al.}{2017}]{Mazzucchelli::etal::2017}
{Mazzucchelli} C.,  et~al., 2017, \mn@doi [\apj] {10.3847/1538-4357/aa9185},
  \href {https://ui.adsabs.harvard.edu/abs/2017ApJ...849...91M} {849, 91}

\bibitem[\protect\citeauthoryear{{McAlpine}, {Bower}, {Rosario}, {Crain},
  {Schaye}  \& {Theuns}}{{McAlpine} et~al.}{2018}]{McAlpine::etal::2018}
{McAlpine} S.,  {Bower} R.~G.,  {Rosario} D.~J.,  {Crain} R.~A.,  {Schaye} J.,
   {Theuns} T.,  2018, \mn@doi [\mnras] {10.1093/mnras/sty2489}, \href
  {https://ui.adsabs.harvard.edu/abs/2018MNRAS.481.3118M} {481, 3118}

\bibitem[\protect\citeauthoryear{{McConnell} \& {Ma}}{{McConnell} \&
  {Ma}}{2013}]{McConnellMa2013}
{McConnell} N.~J.,  {Ma} C.-P.,  2013, \mn@doi [\apj]
  {10.1088/0004-637X/764/2/184}, \href
  {https://ui.adsabs.harvard.edu/#abs/2013ApJ...764..184M} {764, 184}

\bibitem[\protect\citeauthoryear{{Merloni} et~al.,}{{Merloni}
  et~al.}{2010}]{Merloni::etal::2010}
{Merloni} A.,  et~al., 2010, \mn@doi [\apj] {10.1088/0004-637X/708/1/137},
  \href {https://ui.adsabs.harvard.edu/abs/2010ApJ...708..137M} {708, 137}

\bibitem[\protect\citeauthoryear{{Moore}, {Governato}, {Quinn}, {Stadel}  \&
  {Lake}}{{Moore} et~al.}{1998}]{Moore::etal::1998}
{Moore} B.,  {Governato} F.,  {Quinn} T.,  {Stadel} J.,   {Lake} G.,  1998,
  \mn@doi [\apjl] {10.1086/311333}, \href
  {https://ui.adsabs.harvard.edu/abs/1998ApJ...499L...5M} {499, L5}

\bibitem[\protect\citeauthoryear{{Morris}}{{Morris}}{1996}]{Morris1996}
{Morris} J.~P.,  1996, \pasa, \href
  {https://ui.adsabs.harvard.edu/abs/1996PASA...13...97M} {13, 97}

\bibitem[\protect\citeauthoryear{{Mortlock} et~al.,}{{Mortlock}
  et~al.}{2011}]{Mortlock+2011}
{Mortlock} D.~J.,  et~al., 2011, \mn@doi [\nat] {10.1038/nature10159}, \href
  {https://ui.adsabs.harvard.edu/#abs/2011Natur.474..616M} {474, 616}

\bibitem[\protect\citeauthoryear{{Mullaney} et~al.,}{{Mullaney}
  et~al.}{2012}]{Mullaney::etal::2012b}
{Mullaney} J.~R.,  et~al., 2012, \mn@doi [\apjl] {10.1088/2041-8205/753/2/L30},
  \href {https://ui.adsabs.harvard.edu/abs/2012ApJ...753L..30M} {753, L30}

\bibitem[\protect\citeauthoryear{{Muratov}, {Kere{\v{s}}},
  {Faucher-Gigu{\`e}re}, {Hopkins}, {Quataert}  \& {Murray}}{{Muratov}
  et~al.}{2015}]{Muratov::etal::2015}
{Muratov} A.~L.,  {Kere{\v{s}}} D.,  {Faucher-Gigu{\`e}re} C.-A.,  {Hopkins}
  P.~F.,  {Quataert} E.,   {Murray} N.,  2015, \mn@doi [\mnras]
  {10.1093/mnras/stv2126}, \href
  {https://ui.adsabs.harvard.edu/abs/2015MNRAS.454.2691M} {454, 2691}

\bibitem[\protect\citeauthoryear{{Neeleman}, {Prochaska}, {Kanekar}  \&
  {Rafelski}}{{Neeleman} et~al.}{2020}]{Neeleman::etal::2020}
{Neeleman} M.,  {Prochaska} J.~X.,  {Kanekar} N.,   {Rafelski} M.,  2020,
  \mn@doi [\nat] {10.1038/s41586-020-2276-y}, \href
  {https://ui.adsabs.harvard.edu/abs/2020Natur.581..269N} {581, 269}

\bibitem[\protect\citeauthoryear{{Netzer} \& {Trakhtenbrot}}{{Netzer} \&
  {Trakhtenbrot}}{2014}]{NetzerBenny2014}
{Netzer} H.,  {Trakhtenbrot} B.,  2014, \mn@doi [\mnras]
  {10.1093/mnras/stt2238}, \href
  {https://ui.adsabs.harvard.edu/\#abs/2014MNRAS.438..672N} {438, 672}

\bibitem[\protect\citeauthoryear{{Netzer}, {Lani}, {Nordon}, {Trakhtenbrot},
  {Lira}  \& {Shemmer}}{{Netzer} et~al.}{2016}]{Netzer::etal::2016}
{Netzer} H.,  {Lani} C.,  {Nordon} R.,  {Trakhtenbrot} B.,  {Lira} P.,
  {Shemmer} O.,  2016, \mn@doi [\apj] {10.3847/0004-637X/819/2/123}, \href
  {https://ui.adsabs.harvard.edu/abs/2016ApJ...819..123N} {819, 123}

\bibitem[\protect\citeauthoryear{{Nipoti}, {Giocoli}  \& {Despali}}{{Nipoti}
  et~al.}{2018}]{Nipoti::etal::2018}
{Nipoti} C.,  {Giocoli} C.,   {Despali} G.,  2018, \mn@doi [\mnras]
  {10.1093/mnras/sty278}, \href
  {https://ui.adsabs.harvard.edu/abs/2018MNRAS.476..705N} {476, 705}

\bibitem[\protect\citeauthoryear{{Norris}, {Lenc}, {Roy}  \& {Spoon}}{{Norris}
  et~al.}{2012}]{Norris::etal::2012}
{Norris} R.~P.,  {Lenc} E.,  {Roy} A.~L.,   {Spoon} H.,  2012, \mn@doi [\mnras]
  {10.1111/j.1365-2966.2012.20717.x}, \href
  {https://ui.adsabs.harvard.edu/abs/2012MNRAS.422.1453N} {422, 1453}

\bibitem[\protect\citeauthoryear{{Omont}, {Beelen}, {Bertoldi}, {Cox},
  {Carilli}, {Priddey}, {McMahon}  \& {Isaak}}{{Omont}
  et~al.}{2003}]{Omont::etal::2003}
{Omont} A.,  {Beelen} A.,  {Bertoldi} F.,  {Cox} P.,  {Carilli} C.~L.,
  {Priddey} R.~S.,  {McMahon} R.~G.,   {Isaak} K.~G.,  2003, \mn@doi [\aap]
  {10.1051/0004-6361:20021652}, \href
  {https://ui.adsabs.harvard.edu/abs/2003A&A...398..857O} {398, 857}

\bibitem[\protect\citeauthoryear{{Park}, {Ricotti}, {Natarajan},
  {Bogdanovi{\'c}}  \& {Wise}}{{Park} et~al.}{2016}]{Park::etal::2016}
{Park} K.,  {Ricotti} M.,  {Natarajan} P.,  {Bogdanovi{\'c}} T.,   {Wise}
  J.~H.,  2016, \mn@doi [\apj] {10.3847/0004-637X/818/2/184}, \href
  {https://ui.adsabs.harvard.edu/abs/2016ApJ...818..184P} {818, 184}

\bibitem[\protect\citeauthoryear{{Parsotan}, {Cochrane}, {Hayward},
  {Angl{\'e}s-Alc{\'a}zar}, {Feldmann}, {Faucher-Gigu{\`e}re}, {Wellons}  \&
  {Hopkins}}{{Parsotan} et~al.}{2021}]{Parsotan::etal::2021}
{Parsotan} T.,  {Cochrane} R.~K.,  {Hayward} C.~C.,  {Angl{\'e}s-Alc{\'a}zar}
  D.,  {Feldmann} R.,  {Faucher-Gigu{\`e}re} C.~A.,  {Wellons} S.,   {Hopkins}
  P.~F.,  2021, \mn@doi [\mnras] {10.1093/mnras/staa3765}, \href
  {https://ui.adsabs.harvard.edu/abs/2021MNRAS.501.1591P} {501, 1591}

\bibitem[\protect\citeauthoryear{{Price}, {Kriek}, {Feldmann}, {Quataert},
  {Hopkins}, {Faucher-Gigu{\`e}re}, {Kere{\v{s}}}  \& {Barro}}{{Price}
  et~al.}{2017}]{Price::etal::2017}
{Price} S.~H.,  {Kriek} M.,  {Feldmann} R.,  {Quataert} E.,  {Hopkins} P.~F.,
  {Faucher-Gigu{\`e}re} C.-A.,  {Kere{\v{s}}} D.,   {Barro} G.,  2017, \mn@doi
  [\apjl] {10.3847/2041-8213/aa7d4b}, \href
  {https://ui.adsabs.harvard.edu/abs/2017ApJ...844L...6P} {844, L6}

\bibitem[\protect\citeauthoryear{{Priddey}, {Isaak}, {McMahon}  \&
  {Omont}}{{Priddey} et~al.}{2003}]{Priddey::etal::2003}
{Priddey} R.~S.,  {Isaak} K.~G.,  {McMahon} R.~G.,   {Omont} A.,  2003, \mn@doi
  [\mnras] {10.1046/j.1365-8711.2003.06279.x}, \href
  {https://ui.adsabs.harvard.edu/abs/2003MNRAS.339.1183P} {339, 1183}

\bibitem[\protect\citeauthoryear{{Querejeta} et~al.,}{{Querejeta}
  et~al.}{2016}]{Querejeta::etal::2016}
{Querejeta} M.,  et~al., 2016, \mn@doi [\aap] {10.1051/0004-6361/201527536},
  \href {https://ui.adsabs.harvard.edu/abs/2016A&A...588A..33Q} {588, A33}

\bibitem[\protect\citeauthoryear{{Rees}}{{Rees}}{1984}]{Rees1984}
{Rees} M.~J.,  1984, \mn@doi [Annual Review of Astronomy and Astrophysics]
  {10.1146/annurev.aa.22.090184.002351}, \href
  {https://ui.adsabs.harvard.edu/abs/1984ARA&A..22..471R} {22, 471}

\bibitem[\protect\citeauthoryear{{Reines} \& {Volonteri}}{{Reines} \&
  {Volonteri}}{2015}]{ReinesVolonteri2015}
{Reines} A.~E.,  {Volonteri} M.,  2015, \mn@doi [\apj]
  {10.1088/0004-637X/813/2/82}, \href
  {https://ui.adsabs.harvard.edu/#abs/2015ApJ...813...82R} {813, 82}

\bibitem[\protect\citeauthoryear{{Ricarte} \& {Natarajan}}{{Ricarte} \&
  {Natarajan}}{2018}]{Ricarte::etal::2018}
{Ricarte} A.,  {Natarajan} P.,  2018, \mn@doi [\mnras] {10.1093/mnras/stx2851},
  \href {https://ui.adsabs.harvard.edu/abs/2018MNRAS.474.1995R} {474, 1995}

\bibitem[\protect\citeauthoryear{{Riechers} et~al.,}{{Riechers}
  et~al.}{2009}]{Riechers::etal::2009}
{Riechers} D.~A.,  et~al., 2009, \mn@doi [\apj] {10.1088/0004-637X/703/2/1338},
  \href {https://ui.adsabs.harvard.edu/abs/2009ApJ...703.1338R} {703, 1338}

\bibitem[\protect\citeauthoryear{{Riechers} et~al.,}{{Riechers}
  et~al.}{2013}]{Riechers::etal::2013}
{Riechers} D.~A.,  et~al., 2013, \mn@doi [\nat] {10.1038/nature12050}, \href
  {https://ui.adsabs.harvard.edu/abs/2013Natur.496..329R} {496, 329}

\bibitem[\protect\citeauthoryear{{Sahu}, {Graham}  \& {Davis}}{{Sahu}
  et~al.}{2019}]{Sahu::etal::2019b}
{Sahu} N.,  {Graham} A.~W.,   {Davis} B.~L.,  2019, \mn@doi [\apj]
  {10.3847/1538-4357/ab50b7}, \href
  {https://ui.adsabs.harvard.edu/abs/2019ApJ...887...10S} {887, 10}

\bibitem[\protect\citeauthoryear{{Salucci}, {Szuszkiewicz}, {Monaco}  \&
  {Danese}}{{Salucci} et~al.}{1999}]{Salucci::etal::1999}
{Salucci} P.,  {Szuszkiewicz} E.,  {Monaco} P.,   {Danese} L.,  1999, \mn@doi
  [\mnras] {10.1046/j.1365-8711.1999.02659.x}, \href
  {https://ui.adsabs.harvard.edu/abs/1999MNRAS.307..637S} {307, 637}

\bibitem[\protect\citeauthoryear{{Savorgnan}, {Graham}, {Marconi}  \&
  {Sani}}{{Savorgnan} et~al.}{2016}]{Savorgnan::etal::2016}
{Savorgnan} G. A.~D.,  {Graham} A.~W.,  {Marconi} A.~r.,   {Sani} E.,  2016,
  \mn@doi [\apj] {10.3847/0004-637X/817/1/21}, \href
  {https://ui.adsabs.harvard.edu/abs/2016ApJ...817...21S} {817, 21}

\bibitem[\protect\citeauthoryear{{Schaye} et~al.,}{{Schaye}
  et~al.}{2015}]{Schaye::etal::2015}
{Schaye} J.,  et~al., 2015, \mn@doi [\mnras] {10.1093/mnras/stu2058}, \href
  {https://ui.adsabs.harvard.edu/abs/2015MNRAS.446..521S} {446, 521}

\bibitem[\protect\citeauthoryear{{Shankar}, {Salucci}, {Granato}, {De Zotti}
  \& {Danese}}{{Shankar} et~al.}{2004}]{Shankar::etal::2004}
{Shankar} F.,  {Salucci} P.,  {Granato} G.~L.,  {De Zotti} G.,   {Danese} L.,
  2004, \mn@doi [\mnras] {10.1111/j.1365-2966.2004.08261.x}, \href
  {https://ui.adsabs.harvard.edu/abs/2004MNRAS.354.1020S} {354, 1020}

\bibitem[\protect\citeauthoryear{{Shankar}, {Weinberg}  \& {Shen}}{{Shankar}
  et~al.}{2010a}]{Shankar::etal::2010b}
{Shankar} F.,  {Weinberg} D.~H.,   {Shen} Y.,  2010a, \mn@doi [\mnras]
  {10.1111/j.1365-2966.2010.16801.x}, \href
  {https://ui.adsabs.harvard.edu/abs/2010MNRAS.406.1959S} {406, 1959}

\bibitem[\protect\citeauthoryear{{Shankar}, {Crocce}, {Miralda-Escud{\'e}},
  {Fosalba}  \& {Weinberg}}{{Shankar} et~al.}{2010b}]{Shankar::etal::2010a}
{Shankar} F.,  {Crocce} M.,  {Miralda-Escud{\'e}} J.,  {Fosalba} P.,
  {Weinberg} D.~H.,  2010b, \mn@doi [\apj] {10.1088/0004-637X/718/1/231}, \href
  {https://ui.adsabs.harvard.edu/abs/2010ApJ...718..231S} {718, 231}

\bibitem[\protect\citeauthoryear{{Shankar} et~al.,}{{Shankar}
  et~al.}{2019}]{Shankar::etal::2020}
{Shankar} F.,  et~al., 2019, \mn@doi [Nature Astronomy]
  {10.1038/s41550-019-0949-y}, \href
  {https://ui.adsabs.harvard.edu/abs/2020NatAs...4..282S} {4, 282}

\bibitem[\protect\citeauthoryear{{Shields}, {Gebhardt}, {Salviander}, {Wills},
  {Xie}, {Brotherton}, {Yuan}  \& {Dietrich}}{{Shields}
  et~al.}{2003}]{Shields::etal::2003}
{Shields} G.~A.,  {Gebhardt} K.,  {Salviander} S.,  {Wills} B.~J.,  {Xie} B.,
  {Brotherton} M.~S.,  {Yuan} J.,   {Dietrich} M.,  2003, \mn@doi [\apj]
  {10.1086/345348}, \href
  {https://ui.adsabs.harvard.edu/abs/2003ApJ...583..124S} {583, 124}

\bibitem[\protect\citeauthoryear{{Shirakata} et~al.,}{{Shirakata}
  et~al.}{2016}]{Shirakata::etal::2016}
{Shirakata} H.,  et~al., 2016, \mn@doi [\mnras] {10.1093/mnras/stw1798}, \href
  {https://ui.adsabs.harvard.edu/abs/2016MNRAS.461.4389S} {461, 4389}

\bibitem[\protect\citeauthoryear{{Sijacki}, {Springel}, {Di Matteo}  \&
  {Hernquist}}{{Sijacki} et~al.}{2007}]{Sijacki::etal::2007}
{Sijacki} D.,  {Springel} V.,  {Di Matteo} T.,   {Hernquist} L.,  2007, \mn@doi
  [\mnras] {10.1111/j.1365-2966.2007.12153.x}, \href
  {https://ui.adsabs.harvard.edu/abs/2007MNRAS.380..877S} {380, 877}

\bibitem[\protect\citeauthoryear{{Silva}, {Marchesini}, {Silverman}, {Martis},
  {Iono}, {Espada}  \& {Skelton}}{{Silva} et~al.}{2021}]{Silva::etal::2021}
{Silva} A.,  {Marchesini} D.,  {Silverman} J.~D.,  {Martis} N.,  {Iono} D.,
  {Espada} D.,   {Skelton} R.,  2021, arXiv e-prints, \href
  {https://ui.adsabs.harvard.edu/abs/2021arXiv210105000S} {p. arXiv:2101.05000}

\bibitem[\protect\citeauthoryear{{Soltan}}{{Soltan}}{1982}]{Soltan1982}
{Soltan} A.,  1982, \mn@doi [\mnras] {10.1093/mnras/200.1.115}, \href
  {https://ui.adsabs.harvard.edu/abs/1982MNRAS.200..115S} {200, 115}

\bibitem[\protect\citeauthoryear{{Sparre}, {Hayward}, {Feldmann},
  {Faucher-Gigu{\`e}re}, {Muratov}, {Kere{\v{s}}}  \& {Hopkins}}{{Sparre}
  et~al.}{2017}]{Sparre::etal::2017}
{Sparre} M.,  {Hayward} C.~C.,  {Feldmann} R.,  {Faucher-Gigu{\`e}re} C.-A.,
  {Muratov} A.~L.,  {Kere{\v{s}}} D.,   {Hopkins} P.~F.,  2017, \mn@doi
  [\mnras] {10.1093/mnras/stw3011}, \href
  {https://ui.adsabs.harvard.edu/abs/2017MNRAS.466...88S} {466, 88}

\bibitem[\protect\citeauthoryear{{Springel}, {Di Matteo}  \&
  {Hernquist}}{{Springel} et~al.}{2005}]{Springel::etal::2005}
{Springel} V.,  {Di Matteo} T.,   {Hernquist} L.,  2005, \mn@doi [\mnras]
  {10.1111/j.1365-2966.2005.09238.x}, \href
  {https://ui.adsabs.harvard.edu/abs/2005MNRAS.361..776S} {361, 776}

\bibitem[\protect\citeauthoryear{{Springel} et~al.,}{{Springel}
  et~al.}{2018}]{Springel::etal::2018}
{Springel} V.,  et~al., 2018, \mn@doi [\mnras] {10.1093/mnras/stx3304}, \href
  {https://ui.adsabs.harvard.edu/abs/2018MNRAS.475..676S} {475, 676}

\bibitem[\protect\citeauthoryear{{Stern} et~al.,}{{Stern}
  et~al.}{2020}]{Stern::etal::2020}
{Stern} J.,  et~al., 2020, arXiv e-prints, \href
  {https://ui.adsabs.harvard.edu/abs/2020arXiv200613976S} {p. arXiv:2006.13976}

\bibitem[\protect\citeauthoryear{{Tacchella} et~al.,}{{Tacchella}
  et~al.}{2015}]{Tacchella::etal::2015}
{Tacchella} S.,  et~al., 2015, \mn@doi [Science] {10.1126/science.1261094},
  \href {https://ui.adsabs.harvard.edu/abs/2015Sci...348..314T} {348, 314}

\bibitem[\protect\citeauthoryear{{Tacchella} et~al.,}{{Tacchella}
  et~al.}{2018}]{Tacchella::etal::2018}
{Tacchella} S.,  et~al., 2018, \mn@doi [\apj] {10.3847/1538-4357/aabf8b}, \href
  {https://ui.adsabs.harvard.edu/abs/2018ApJ...859...56T} {859, 56}

\bibitem[\protect\citeauthoryear{{Tacchella} et~al.,}{{Tacchella}
  et~al.}{2019}]{Tacchella::etal::2019}
{Tacchella} S.,  et~al., 2019, \mn@doi [\mnras] {10.1093/mnras/stz1657}, \href
  {https://ui.adsabs.harvard.edu/abs/2019MNRAS.487.5416T} {487, 5416}

\bibitem[\protect\citeauthoryear{{Tacconi} et~al.,}{{Tacconi}
  et~al.}{2010}]{Tacconi::etal::2010}
{Tacconi} L.~J.,  et~al., 2010, \mn@doi [\nat] {10.1038/nature08773}, \href
  {https://ui.adsabs.harvard.edu/abs/2010Natur.463..781T} {463, 781}

\bibitem[\protect\citeauthoryear{{Targett}, {Dunlop}  \& {McLure}}{{Targett}
  et~al.}{2012}]{Targett::etal::2012}
{Targett} T.~A.,  {Dunlop} J.~S.,   {McLure} R.~J.,  2012, \mn@doi [\mnras]
  {10.1111/j.1365-2966.2011.20286.x}, \href
  {https://ui.adsabs.harvard.edu/abs/2012MNRAS.420.3621T} {420, 3621}

\bibitem[\protect\citeauthoryear{{Taylor} \& {Kobayashi}}{{Taylor} \&
  {Kobayashi}}{2014}]{Taylor::etal::2014}
{Taylor} P.,  {Kobayashi} C.,  2014, \mn@doi [\mnras] {10.1093/mnras/stu983},
  \href {https://ui.adsabs.harvard.edu/abs/2014MNRAS.442.2751T} {442, 2751}

\bibitem[\protect\citeauthoryear{{Taylor}, {Federrath}  \&
  {Kobayashi}}{{Taylor} et~al.}{2017}]{Taylor::etal::2017}
{Taylor} P.,  {Federrath} C.,   {Kobayashi} C.,  2017, \mn@doi [\mnras]
  {10.1093/mnras/stx1128}, \href
  {https://ui.adsabs.harvard.edu/abs/2017MNRAS.469.4249T} {469, 4249}

\bibitem[\protect\citeauthoryear{{Thomas}, {Dav{\'e}}, {Angl{\'e}s-Alc{\'a}zar}
   \& {Jarvis}}{{Thomas} et~al.}{2019}]{Thomas::etal::2019}
{Thomas} N.,  {Dav{\'e}} R.,  {Angl{\'e}s-Alc{\'a}zar} D.,   {Jarvis} M.,
  2019, arXiv e-prints, \href
  {https://ui.adsabs.harvard.edu/abs/2019arXiv190502741T} {p. arXiv:1905.02741}

\bibitem[\protect\citeauthoryear{{Torrey}, {Hopkins}, {Faucher-Gigu{\`e}re},
  {Vogelsberger}, {Quataert}, {Kere{\v{s}}}  \& {Murray}}{{Torrey}
  et~al.}{2017}]{Torrey::etal::2017}
{Torrey} P.,  {Hopkins} P.~F.,  {Faucher-Gigu{\`e}re} C.-A.,  {Vogelsberger}
  M.,  {Quataert} E.,  {Kere{\v{s}}} D.,   {Murray} N.,  2017, \mn@doi [\mnras]
  {10.1093/mnras/stx254}, \href
  {https://ui.adsabs.harvard.edu/abs/2017MNRAS.467.2301T} {467, 2301}

\bibitem[\protect\citeauthoryear{{Trakhtenbrot}, {Lira}, {Netzer}, {Cicone},
  {Maiolino}  \& {Shemmer}}{{Trakhtenbrot}
  et~al.}{2017}]{Trakhtenbrot::etal::2017}
{Trakhtenbrot} B.,  {Lira} P.,  {Netzer} H.,  {Cicone} C.,  {Maiolino} R.,
  {Shemmer} O.,  2017, \mn@doi [\apj] {10.3847/1538-4357/836/1/8}, \href
  {https://ui.adsabs.harvard.edu/abs/2017ApJ...836....8T} {836, 8}

\bibitem[\protect\citeauthoryear{{Tremmel}, {Karcher}, {Governato},
  {Volonteri}, {Quinn}, {Pontzen}, {Anderson}  \& {Bellovary}}{{Tremmel}
  et~al.}{2017}]{Tremmel+2017}
{Tremmel} M.,  {Karcher} M.,  {Governato} F.,  {Volonteri} M.,  {Quinn} T.~R.,
  {Pontzen} A.,  {Anderson} L.,   {Bellovary} J.,  2017, \mn@doi [\mnras]
  {10.1093/mnras/stx1160}, \href
  {https://ui.adsabs.harvard.edu/#abs/2017MNRAS.470.1121T} {470, 1121}

\bibitem[\protect\citeauthoryear{{Treu}, {Malkan}  \& {Blandford}}{{Treu}
  et~al.}{2004}]{Treu::etal::2004}
{Treu} T.,  {Malkan} M.~A.,   {Blandford} R.~D.,  2004, \mn@doi [\apj]
  {10.1086/426437}, \href
  {https://ui.adsabs.harvard.edu/abs/2004ApJ...615L..97T} {615, L97}

\bibitem[\protect\citeauthoryear{{Tsai} et~al.,}{{Tsai}
  et~al.}{2018}]{Tsai::etal::2018}
{Tsai} C.-W.,  et~al., 2018, \mn@doi [\apj] {10.3847/1538-4357/aae698}, \href
  {https://ui.adsabs.harvard.edu/abs/2018ApJ...868...15T} {868, 15}

\bibitem[\protect\citeauthoryear{{Venemans} et~al.,}{{Venemans}
  et~al.}{2012}]{Venemans::etal::2012}
{Venemans} B.~P.,  et~al., 2012, \mn@doi [\apjl] {10.1088/2041-8205/751/2/L25},
  \href {https://ui.adsabs.harvard.edu/abs/2012ApJ...751L..25V} {751, L25}

\bibitem[\protect\citeauthoryear{{Venemans} et~al.,}{{Venemans}
  et~al.}{2013}]{Venemans::etal::2013}
{Venemans} B.~P.,  et~al., 2013, \mn@doi [\apj] {10.1088/0004-637X/779/1/24},
  \href {https://ui.adsabs.harvard.edu/abs/2013ApJ...779...24V} {779, 24}

\bibitem[\protect\citeauthoryear{{Venemans}, {Walter}, {Zschaechner},
  {Decarli}, {De Rosa}, {Findlay}, {McMahon}  \& {Sutherland}}{{Venemans}
  et~al.}{2016}]{Venemans::etal::2016}
{Venemans} B.~P.,  {Walter} F.,  {Zschaechner} L.,  {Decarli} R.,  {De Rosa}
  G.,  {Findlay} J.~R.,  {McMahon} R.~G.,   {Sutherland} W.~J.,  2016, \mn@doi
  [\apj] {10.3847/0004-637X/816/1/37}, \href
  {https://ui.adsabs.harvard.edu/abs/2016ApJ...816...37V} {816, 37}

\bibitem[\protect\citeauthoryear{{Vogelsberger} et~al.,}{{Vogelsberger}
  et~al.}{2014}]{Vogelsberger::etal::2014}
{Vogelsberger} M.,  et~al., 2014, \mn@doi [\mnras] {10.1093/mnras/stu1536},
  \href {https://ui.adsabs.harvard.edu/abs/2014MNRAS.444.1518V} {444, 1518}

\bibitem[\protect\citeauthoryear{{Volonteri}}{{Volonteri}}{2010}]{Volonteri2010}
{Volonteri} M.,  2010, \mn@doi [\aapr] {10.1007/s00159-010-0029-x}, \href
  {https://ui.adsabs.harvard.edu/abs/2010A&ARv..18..279V} {18, 279}

\bibitem[\protect\citeauthoryear{{Volonteri}}{{Volonteri}}{2012}]{Volonteri2012}
{Volonteri} M.,  2012, \mn@doi [Science] {10.1126/science.1220843}, \href
  {https://ui.adsabs.harvard.edu/abs/2012Sci...337..544V} {337, 544}

\bibitem[\protect\citeauthoryear{{Volonteri}, {Capelo}, {Netzer}, {Bellovary},
  {Dotti}  \& {Governato}}{{Volonteri} et~al.}{2015}]{Volonteri::etal::2015b}
{Volonteri} M.,  {Capelo} P.~R.,  {Netzer} H.,  {Bellovary} J.,  {Dotti} M.,
  {Governato} F.,  2015, \mn@doi [\mnras] {10.1093/mnrasl/slv078}, \href
  {https://ui.adsabs.harvard.edu/abs/2015MNRAS.452L...6V} {452, L6}

\bibitem[\protect\citeauthoryear{{Volonteri}, {Dubois}, {Pichon}  \&
  {Devriendt}}{{Volonteri} et~al.}{2016}]{Volonteri::etal::2016}
{Volonteri} M.,  {Dubois} Y.,  {Pichon} C.,   {Devriendt} J.,  2016, \mn@doi
  [\mnras] {10.1093/mnras/stw1123}, \href
  {https://ui.adsabs.harvard.edu/abs/2016MNRAS.460.2979V} {460, 2979}

\bibitem[\protect\citeauthoryear{{Vulcani} et~al.,}{{Vulcani}
  et~al.}{2016}]{Vulcani::etal::2016}
{Vulcani} B.,  et~al., 2016, \mn@doi [\apj] {10.3847/0004-637X/816/2/86}, \href
  {https://ui.adsabs.harvard.edu/abs/2016ApJ...816...86V} {816, 86}

\bibitem[\protect\citeauthoryear{{Walter}, {Carilli}, {Bertoldi}, {Menten},
  {Cox}, {Lo}, {Fan}  \& {Strauss}}{{Walter} et~al.}{2004}]{Walter::etal::2004}
{Walter} F.,  {Carilli} C.,  {Bertoldi} F.,  {Menten} K.,  {Cox} P.,  {Lo}
  K.~Y.,  {Fan} X.,   {Strauss} M.~A.,  2004, \mn@doi [\apj] {10.1086/426017},
  \href {https://ui.adsabs.harvard.edu/abs/2004ApJ...615L..17W} {615, L17}

\bibitem[\protect\citeauthoryear{{Wang} et~al.,}{{Wang}
  et~al.}{2011}]{Wang::etal::2011}
{Wang} R.,  et~al., 2011, \mn@doi [\aj] {10.1088/0004-6256/142/4/101}, \href
  {https://ui.adsabs.harvard.edu/abs/2011AJ....142..101W} {142, 101}

\bibitem[\protect\citeauthoryear{{Wang} et~al.,}{{Wang}
  et~al.}{2013}]{Wang::etal::2013}
{Wang} R.,  et~al., 2013, \mn@doi [\apj] {10.1088/0004-637X/773/1/44}, \href
  {https://ui.adsabs.harvard.edu/abs/2013ApJ...773...44W} {773, 44}

\bibitem[\protect\citeauthoryear{{Wang} et~al.,}{{Wang}
  et~al.}{2016}]{Wang::etal::2016}
{Wang} R.,  et~al., 2016, \mn@doi [\apj] {10.3847/0004-637X/830/1/53}, \href
  {https://ui.adsabs.harvard.edu/abs/2016ApJ...830...53W} {830, 53}

\bibitem[\protect\citeauthoryear{{Wang} et~al.,}{{Wang}
  et~al.}{2019}]{Wang::etal::2019}
{Wang} F.,  et~al., 2019, \mn@doi [\apj] {10.3847/1538-4357/ab2be5}, \href
  {https://ui.adsabs.harvard.edu/abs/2019ApJ...884...30W} {884, 30}

\bibitem[\protect\citeauthoryear{{Wang} et~al.,}{{Wang}
  et~al.}{2021}]{Wang::etal::2021}
{Wang} F.,  et~al., 2021, \mn@doi [\apjl] {10.3847/2041-8213/abd8c6}, \href
  {https://ui.adsabs.harvard.edu/abs/2021ApJ...907L...1W} {907, L1}

\bibitem[\protect\citeauthoryear{{Wellons}, {Faucher-Gigu{\`e}re},
  {Angl{\'e}s-Alc{\'a}zar}, {Hayward}, {Feldmann}, {Hopkins}  \&
  {Kere{\v{s}}}}{{Wellons} et~al.}{2020}]{Wellons::etal::2020}
{Wellons} S.,  {Faucher-Gigu{\`e}re} C.-A.,  {Angl{\'e}s-Alc{\'a}zar} D.,
  {Hayward} C.~C.,  {Feldmann} R.,  {Hopkins} P.~F.,   {Kere{\v{s}}} D.,  2020,
  \mn@doi [\mnras] {10.1093/mnras/staa2229}, \href
  {https://ui.adsabs.harvard.edu/abs/2020MNRAS.497.4051W} {497, 4051}

\bibitem[\protect\citeauthoryear{{Willott}, {Bergeron}  \& {Omont}}{{Willott}
  et~al.}{2015}]{Willott::etal::2015}
{Willott} C.~J.,  {Bergeron} J.,   {Omont} A.,  2015, \mn@doi [\apj]
  {10.1088/0004-637X/801/2/123}, \href
  {https://ui.adsabs.harvard.edu/abs/2015ApJ...801..123W} {801, 123}

\bibitem[\protect\citeauthoryear{{Wu} et~al.,}{{Wu} et~al.}{2015}]{Wu+2015}
{Wu} X.-B.,  et~al., 2015, \mn@doi [\nat] {10.1038/nature14241}, \href
  {https://ui.adsabs.harvard.edu/#abs/2015Natur.518..512W} {518, 512}

\bibitem[\protect\citeauthoryear{{Xu}, {Rieke}, {Egami}, {Haines}, {Pereira}
  \& {Smith}}{{Xu} et~al.}{2015}]{Lei::etal::2015}
{Xu} L.,  {Rieke} G.~H.,  {Egami} E.,  {Haines} C.~P.,  {Pereira} M.~J.,
  {Smith} G.~P.,  2015, \mn@doi [\apj] {10.1088/0004-637X/808/2/159}, \href
  {https://ui.adsabs.harvard.edu/abs/2015ApJ...808..159X} {808, 159}

\bibitem[\protect\citeauthoryear{{Yang} et~al.,}{{Yang}
  et~al.}{2017}]{Yang::etal::2017}
{Yang} G.,  et~al., 2017, \mn@doi [\apj] {10.3847/1538-4357/aa7564}, \href
  {https://ui.adsabs.harvard.edu/abs/2017ApJ...842...72Y} {842, 72}

\bibitem[\protect\citeauthoryear{{Yang}, {Brandt}, {Alexander}, {Chen}, {Ni},
  {Vito}  \& {Zhu}}{{Yang} et~al.}{2019}]{Yang::etal::2019}
{Yang} G.,  {Brandt} W.~N.,  {Alexander} D.~M.,  {Chen} C. T.~J.,  {Ni} Q.,
  {Vito} F.,   {Zhu} F.~F.,  2019, \mn@doi [\mnras] {10.1093/mnras/stz611},
  \href {https://ui.adsabs.harvard.edu/abs/2019MNRAS.485.3721Y} {485, 3721}

\bibitem[\protect\citeauthoryear{{Yang} et~al.,}{{Yang}
  et~al.}{2020}]{Yang::etal::2020}
{Yang} J.,  et~al., 2020, \mn@doi [\apjl] {10.3847/2041-8213/ab9c26}, \href
  {https://ui.adsabs.harvard.edu/abs/2020ApJ...897L..14Y} {897, L14}

\bibitem[\protect\citeauthoryear{{Zahid}, {Geller}, {Damjanov}  \&
  {Sohn}}{{Zahid} et~al.}{2019}]{Zahid::etal::2019}
{Zahid} H.~J.,  {Geller} M.~J.,  {Damjanov} I.,   {Sohn} J.,  2019, \mn@doi
  [\apj] {10.3847/1538-4357/ab21b9}, \href
  {https://ui.adsabs.harvard.edu/abs/2019ApJ...878..158Z} {878, 158}

\bibitem[\protect\citeauthoryear{{Zanella} et~al.,}{{Zanella}
  et~al.}{2015}]{Zanella::etal::2015}
{Zanella} A.,  et~al., 2015, \mn@doi [\nat] {10.1038/nature14409}, \href
  {https://ui.adsabs.harvard.edu/abs/2015Natur.521...54Z} {521, 54}

\bibitem[\protect\citeauthoryear{{Zavala} et~al.,}{{Zavala}
  et~al.}{2018}]{Zavala::etal::2018}
{Zavala} J.~A.,  et~al., 2018, \mn@doi [Nature Astronomy]
  {10.1038/s41550-017-0297-8}, \href
  {https://ui.adsabs.harvard.edu/abs/2018NatAs...2...56Z} {2, 56}

\bibitem[\protect\citeauthoryear{{van den Bergh}, {Abraham}, {Ellis}, {Tanvir},
  {Santiago}  \& {Glazebrook}}{{van den Bergh}
  et~al.}{1996}]{vandenBergh::etal::1996}
{van den Bergh} S.,  {Abraham} R.~G.,  {Ellis} R.~S.,  {Tanvir} N.~R.,
  {Santiago} B.~X.,   {Glazebrook} K.~G.,  1996, \mn@doi [\aj]
  {10.1086/118020}, \href
  {https://ui.adsabs.harvard.edu/abs/1996AJ....112..359V} {112, 359}

\makeatother
\end{thebibliography}


\appendix
\section{Additional Analysis} \label{Appendix:Additional analysis}

\begin{figure*}
    \includegraphics[width=\linewidth]{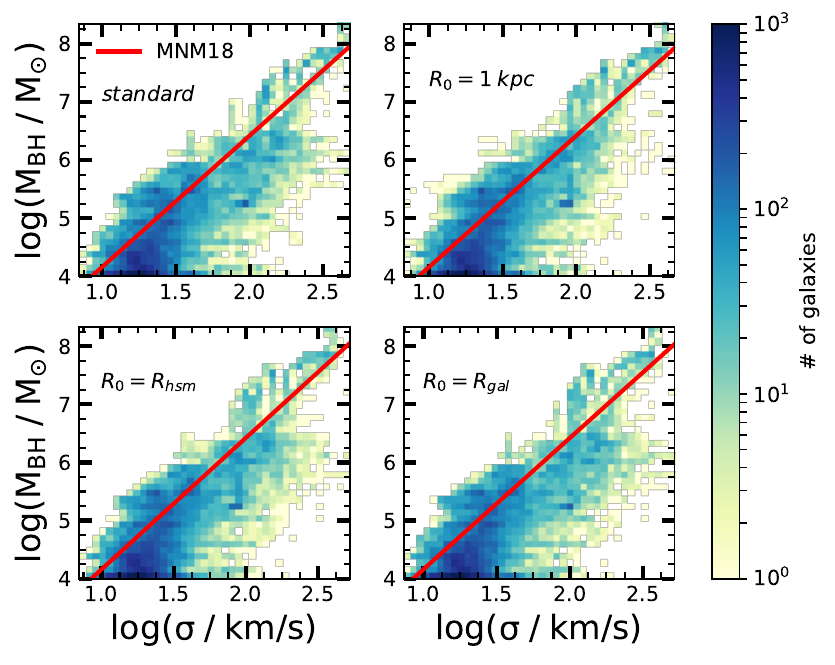}
    \caption{Comparison of velocity dispersion estimation for non-rotating bulges versus velocity dispersion of all star particles within a sphere for different radii. We use the fiducial settings on top of the "\textsc{few mergers}" model. The red line shows the fit of \protect\cite{McConnellMa2013}. The bulge-disc decomposition of \protect\cite{Angles-Alcazar+14} yields similar results with the velocity dispersion of all stars regardless the radii.}
    \label{fig:vel disp comparison}
\end{figure*}

We perform a bulge-disc decomposition following \citep{Angles-Alcazar+14}, which is explained in \Cref{sec:Parameter Study} in detail. In \Cref{fig:vel disp comparison}, we compare what we found based on the recipe of \cite{Angles-Alcazar+14} to the velocity dispersion of all star particles within different radii; 1 kpc, the half stellar mass radius, and galactic radius $R_{\rm gal}$. The black hole masses are calculated using our fiducial settings and "\textsc{few mergers}" model. The different radii we used to estimate the velocity dispersion of star particles gives indifferent results from the bulge-disc decomposition method we used.

\begin{figure}
    \centering
        \includegraphics[height=0.9\textwidth]{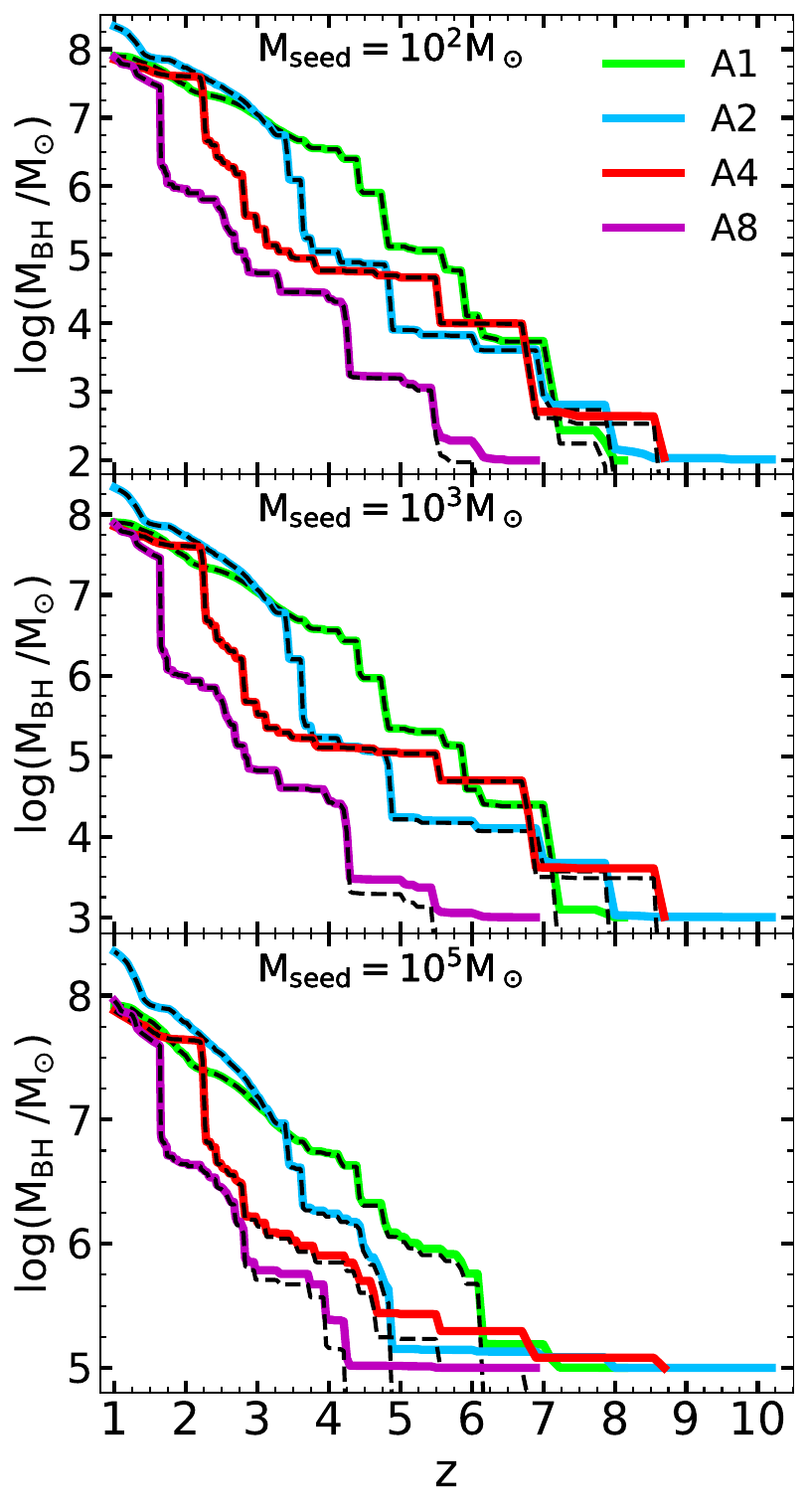}
    
    \caption{Accretion contribution to the total SMBH mass for different seed masses for A-series simulaitons run with FIRE-2 physics. The solid lines show the total SMBH mass while the dashed lines shows the contribution from SMBH seed mass where it becomes important towards heavier seed masses.}
    \label{fig:Accretion contribution for different seed masses}
\end{figure}

In \Cref{fig:BH growth for different accretion contribution}, we show the contribution of both accretion and seed mass to the total SMBH mass for a seed of $10^4 M_\odot$. \Cref{fig:Accretion contribution for different seed masses} searches for the effect of different seed masses on the total SMBH mass for FIRE-2 A series simulations. As the seed mass increases, the offset between total SMBH mass and in-situ accretion contribution increases, especially at high redshifts. The effect of seed mass choice is more visible at $z \gtrsim  2$ for seed masses equal to or above $10^4 M_\odot$, which points out that the effect of small seeds on SMBH mass is not observable.

\begin{figure}
    \centering
    \includegraphics[width=\linewidth]{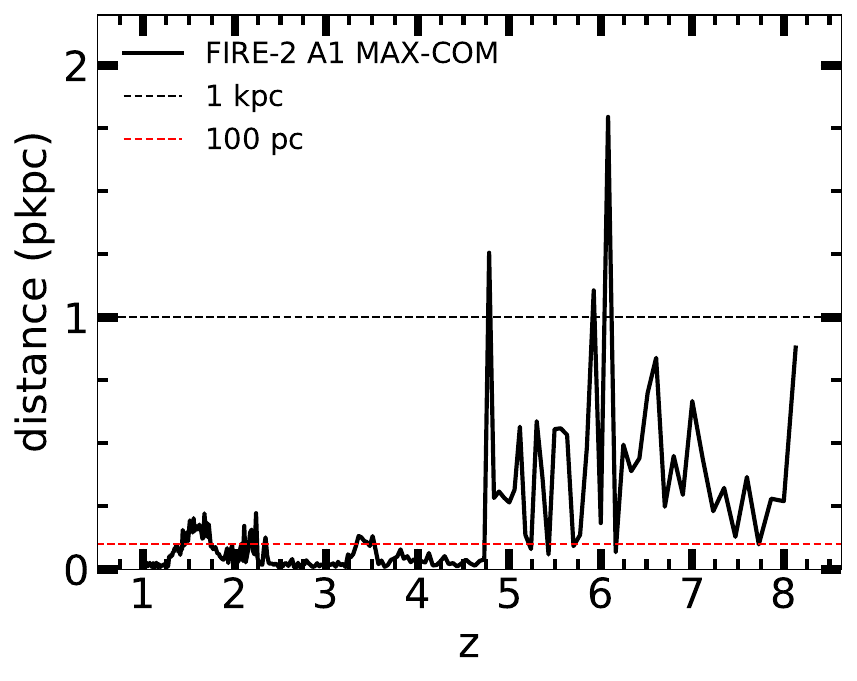}
    \caption{The distance between two different centering methods, MAX and COM, for FIRE-2 A1 simulation. The black dashed line marks the $1$ kpc distance while red dashed line shows the $100$ pc distance. The difference between the centres found with different methods becomes important at high redshift, where the structures are thought to be not very well settled.}
    \label{fig:halo center diff for MAX and COM}
\end{figure}

\Cref{fig:halo center diff for MAX and COM} shows the difference between the central coordinates of two different centring methods available in AHF. The difference between MAX and COM centring methods are generally above half of a kpc in physical units for $z\gtrsim4.5$ for a selected simulation. The coordinates of maximum density centre and centre-of-mass of the host halo are roughly the same for the rest of the analysis. This finding suggests that the central galaxy settles at the host halo centre for the sample simulation around $z\sim4.5$

\begin{figure}
    \centering
    \includegraphics[width=\linewidth]{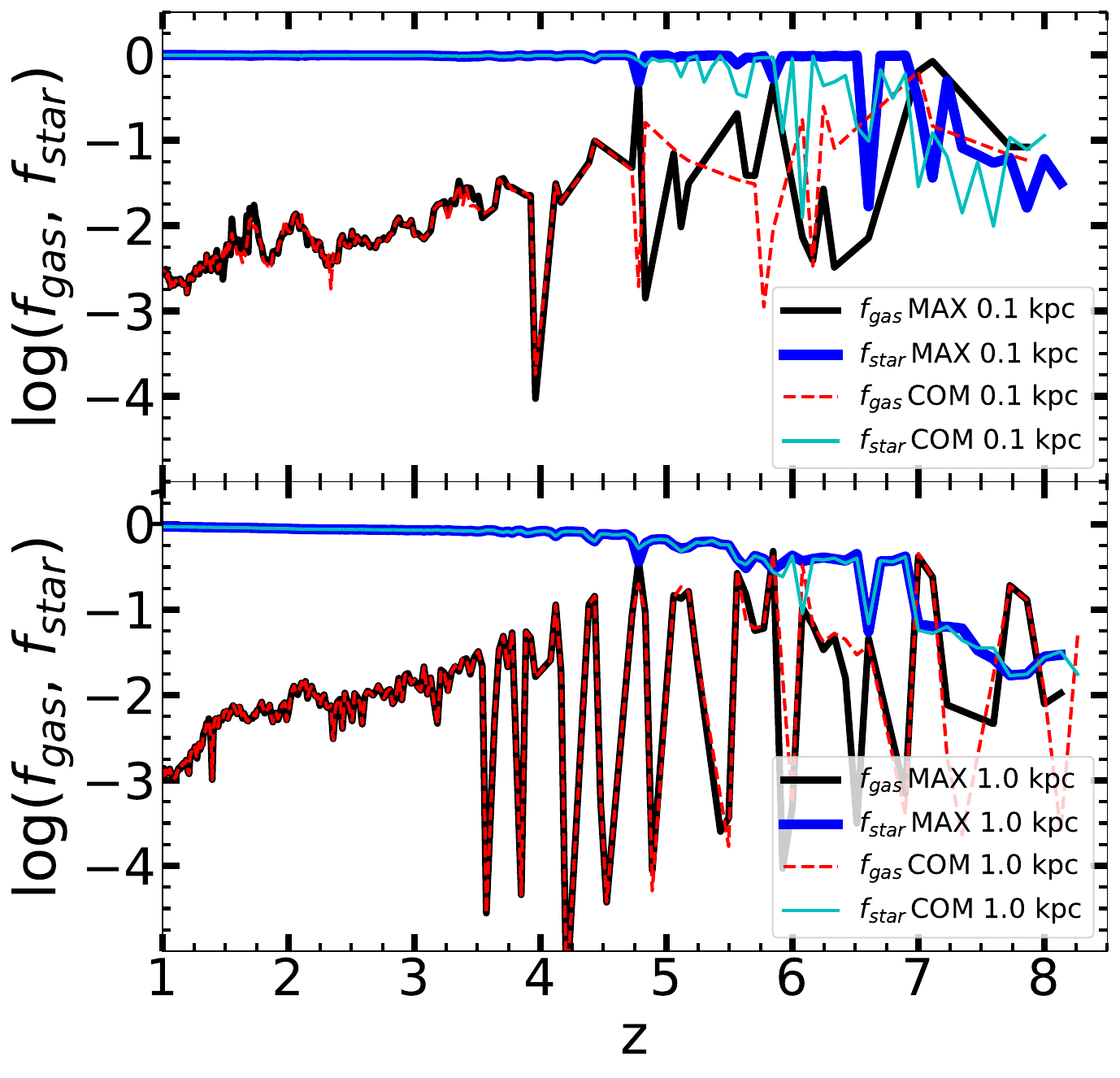}
    \caption{Gas-to-total ($f_{\rm gas}$) and stellar-to-total ($f_{\rm star}$) ratios for different halo centring methods within $100$ pc and $1$ kpc central regions for FIRE-2 A1 simulation. The solid black line shows the gas fraction for the densest central region, and the red dashed line shows the same quantities for the average density central region. The blue and cyan lines stand for stellar-to-total ratio for MAX and COM centring methods, respectively.}
    \label{fig:fgas for MAX and COM}
\end{figure}

The gas-to-total ($f_{\rm gas}$) and stellar-to-total ($f_{\rm star}$) mass ratios for different halo centring methods within $R_0$ are shown in \Cref{fig:fgas for MAX and COM}. Here, the total mass is the summation of the gas, stellar and dark matter mass of the galaxy. The $f_{\rm gas}$ and $f_{\rm star}$ are the ratios of gas mass and stellar mass to the total mass, respectively. The gas and stellar fractions within $100$ pc track each other for different centring methods until $z \sim 4.5$. The difference between MAX and COM within $1$ kpc track each more consistently compared to $100$ pc case. For GTDA model, it is conceivable to assume that the total stellar mass within $R_0$ is the dictating term in \Cref{eq:GTDA equation} to determine the BHAR since the ratio of stellar and total mass within $R_0$ is close to unity for $z \lesssim 6-7$. Moreover, a high gas fraction at high redshift is the one of the caveats discussed in \Cref{sec:Discussion} for the modelling of SMBH growth using GTDA model in the early Universe.

\begin{figure*}
      \centering
      \subfloat[$M_{\rm seed}=10^2 M_\odot$]{\includegraphics[width=0.5\textwidth]{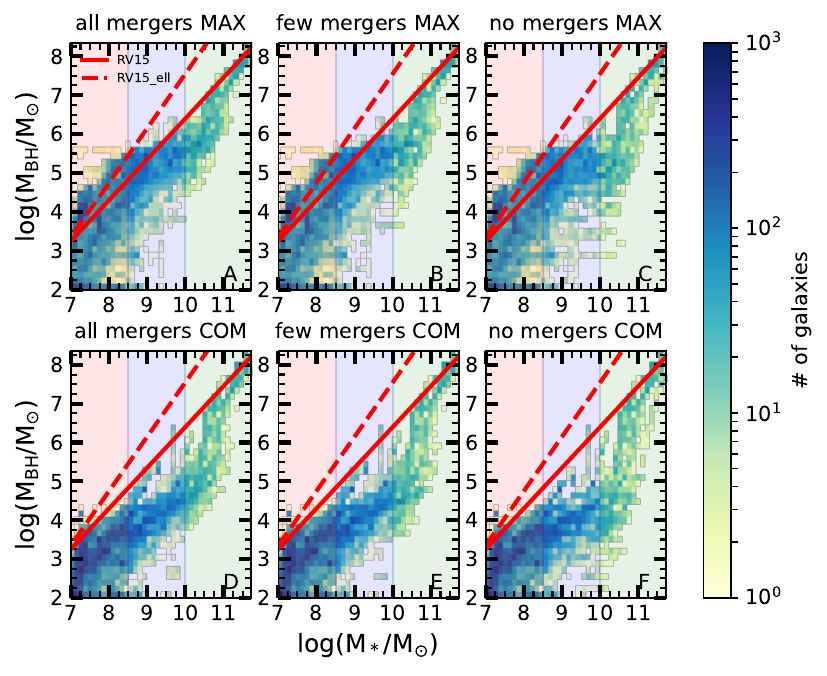}} 
      \subfloat[$M_{\rm seed}=10^3 M_\odot$]{\includegraphics[width=0.5\textwidth]{figures/mstar_mbh_GTDA_seed_1e2.pdf}}
    
      \subfloat[$M_{\rm seed}=10^4 M_\odot$]{\includegraphics[width=0.5\textwidth]{figures/mstar_mbh_GTDA_seed_1e4.pdf}}
      \subfloat[$M_{\rm seed}=10^5 M_\odot$]{\includegraphics[width=0.5\textwidth]{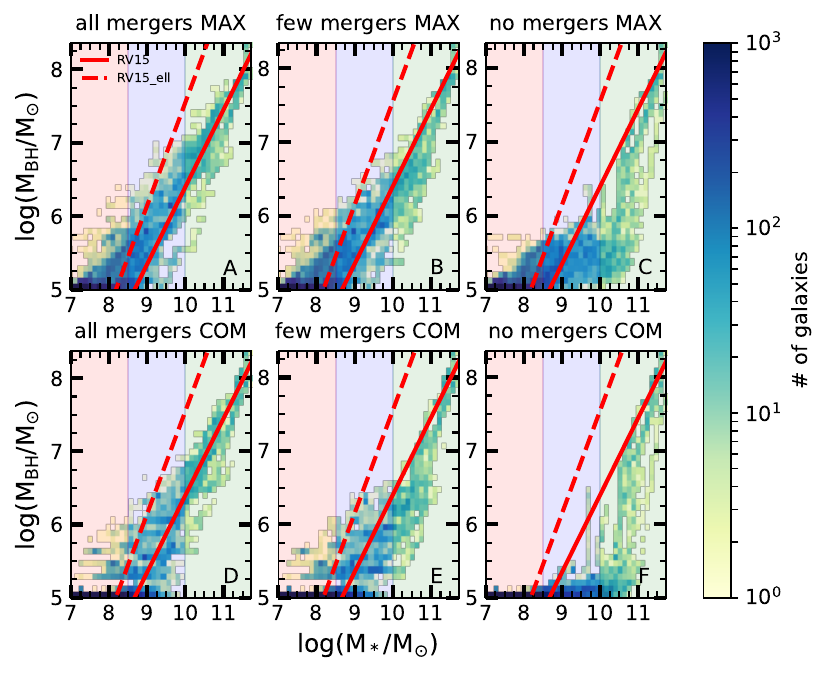}}

    \caption{$M_* - M_{\rm BH}$ scaling relation for different SMBH seed mass choices, $10^2 M_\odot$ (top left), $10^3 M_\odot$ (top right), $10^4 M_\odot$ (bottom left) and $10^5 M_\odot$ (bottom right). The redshift range is same as in \Cref{fig:mstar vs mbh scaling relation}. The red solid line shows the best fit line for the spiral galaxies in \protect\cite{ReinesVolonteri2015}. The effect of gas-rich early region on the early evolution of SMBH growth decreases as seed mass increases, especially for seeds heavier than $10^4 M_\odot$.}
    \label{fig:mstar vs mbh scaling relation for different seed masses}
\end{figure*}

\Cref{fig:mstar vs mbh scaling relation for different seed masses} produces the $M_* - M_{\rm BH}$ scaling relation for different seed masses, halo centring methods, and merger treatments. The effect of full merger treatment ("\textsc{all mergers}") is distinguishable again for seed masses that are equal to or above $10^4 M_\odot$. The increase in seed mass boosts the importance of mergers on the $M_*-M_{\rm BH}$ scaling relation. The smaller seed masses that were born in gas-rich environments follow local scaling relation since the growth of the black hole is dominated by in-situ accretion.

\begin{figure*}
    \centering
    \begin{tabular}{cc}
        \includegraphics[width=0.5\textwidth]{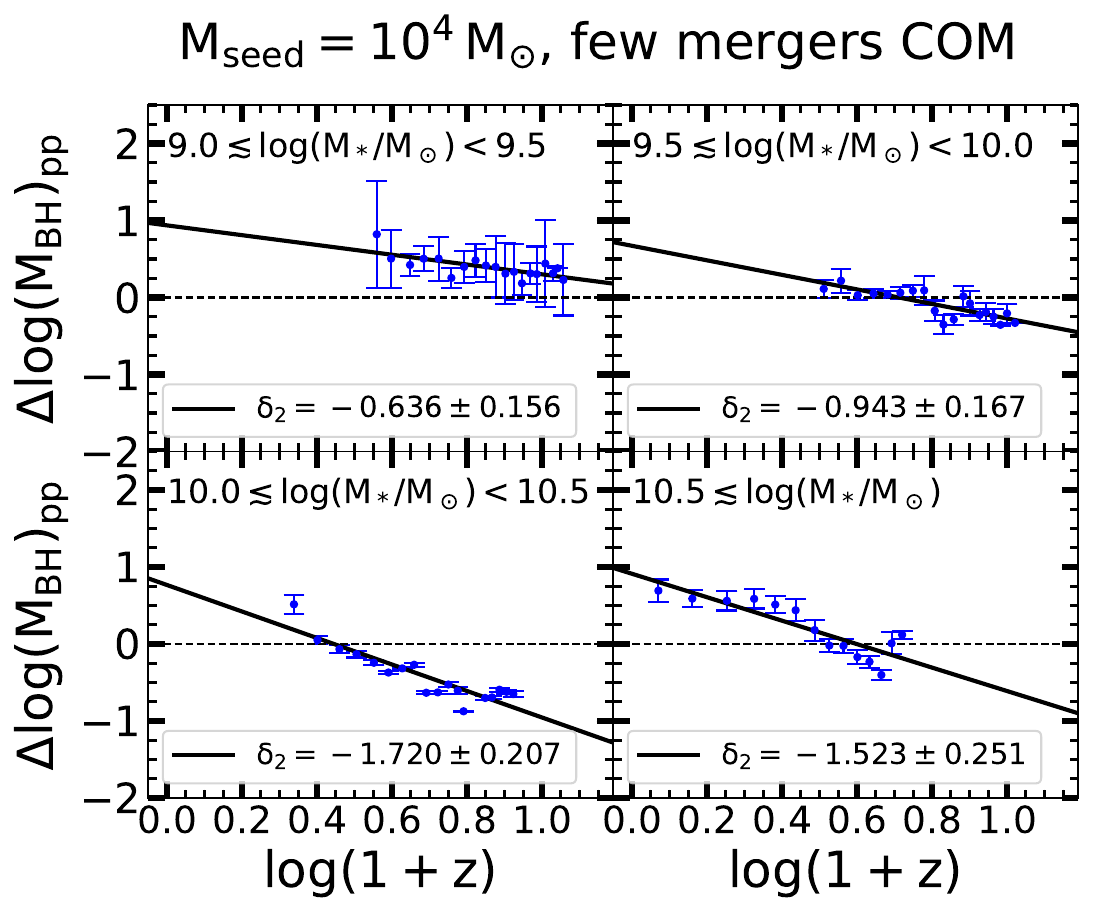}
        &
        \includegraphics[width=0.5\textwidth]{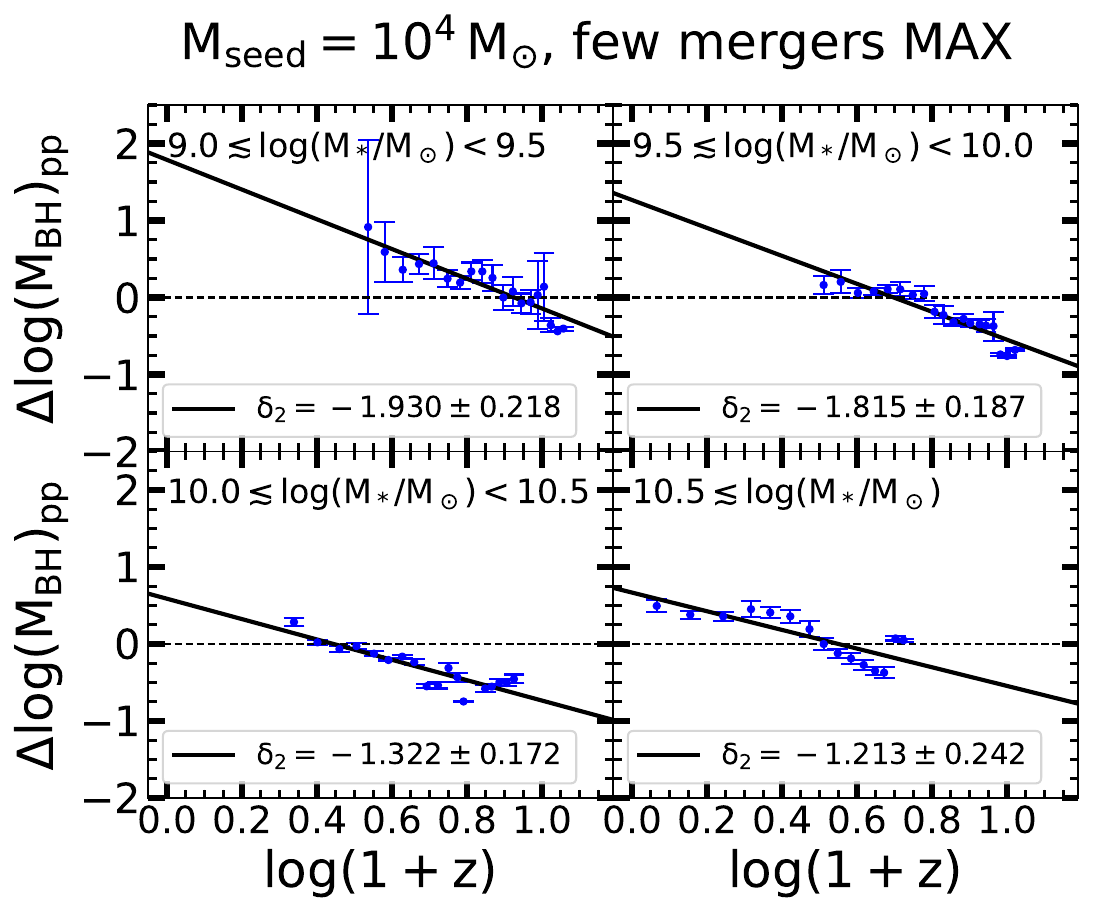}  
    \end{tabular}
    
    \begin{tabular}{cc}
        \includegraphics[width=0.5\textwidth]{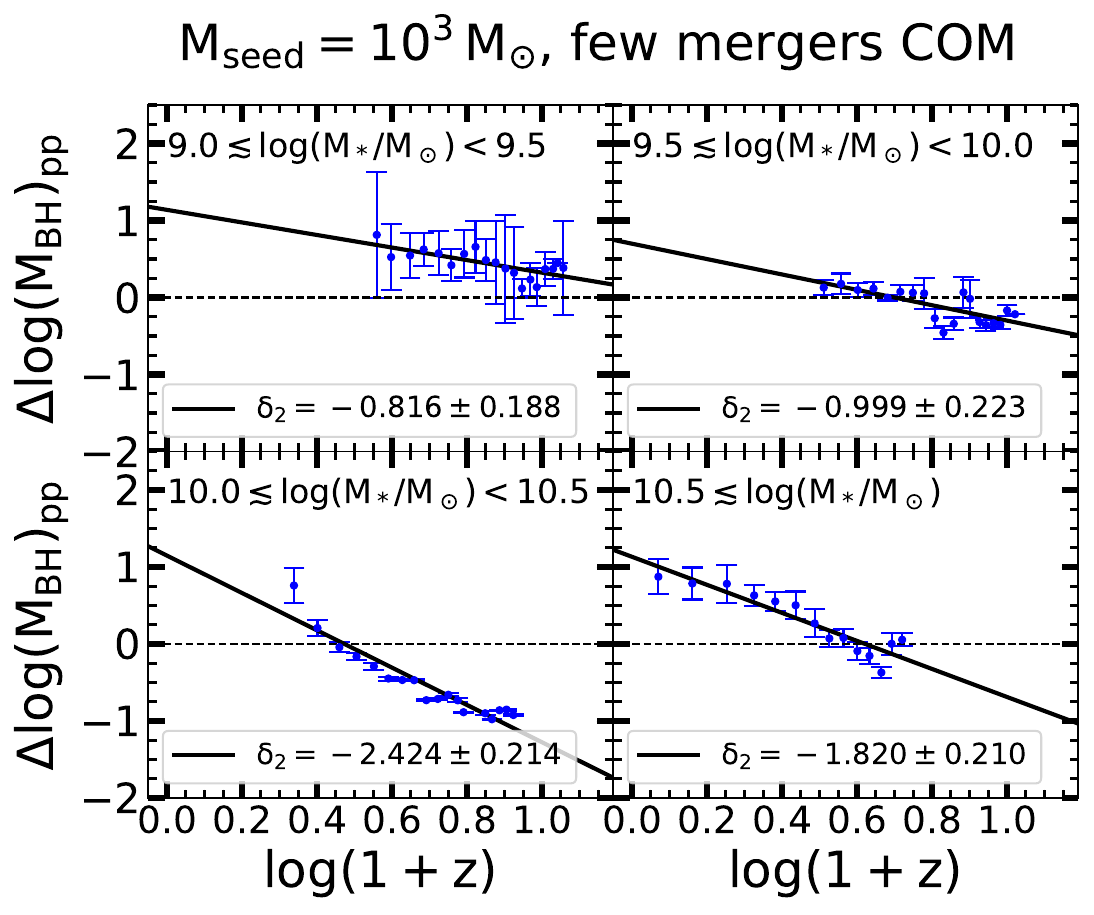}
        &
        \includegraphics[width=0.5\textwidth]{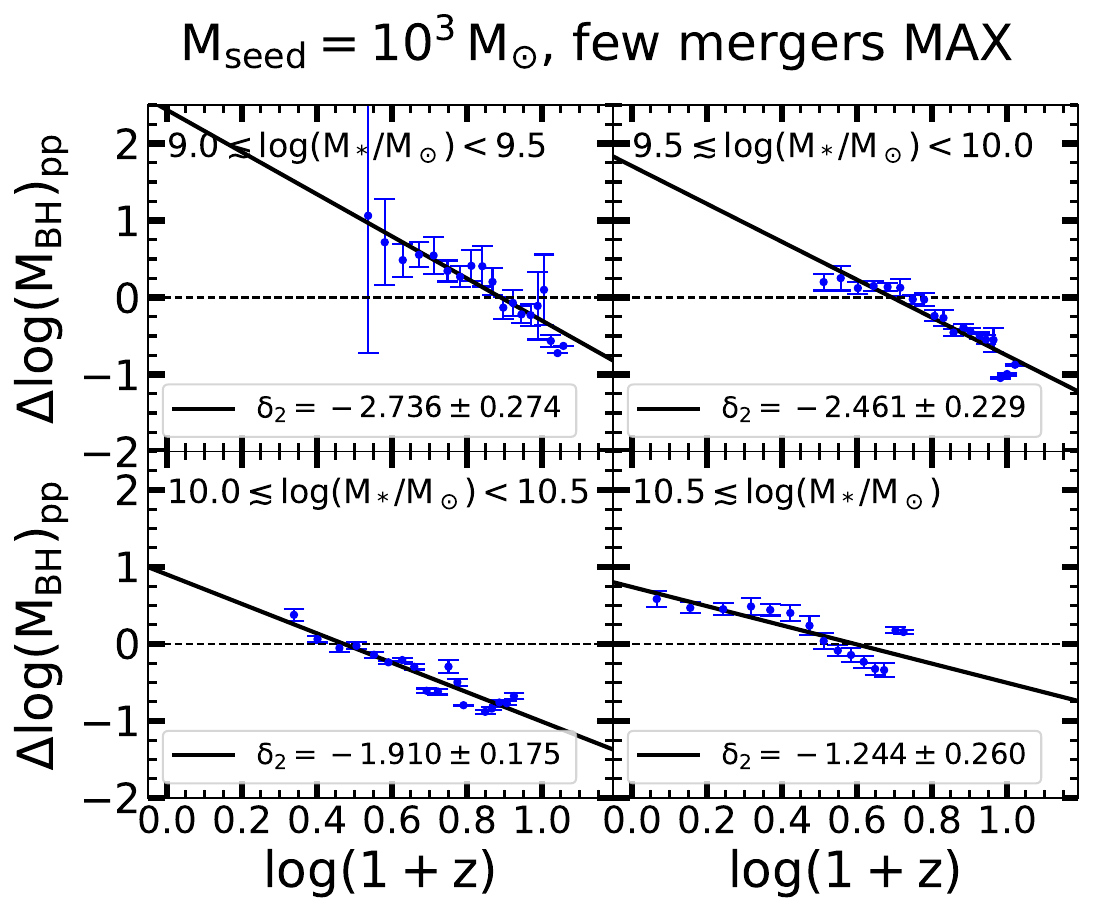}  
    \end{tabular}
    
    \caption{Redshift evolution of the offset of the $M_* - M_{\rm BH}$ scaling relation from our best fit line for different seed mass choices, centring methods, and the \textsc{few mergers} model in different stellar mass bins. We follow the same method as in \Cref{fig:redshift evolution of mstar vs mbh scaling relation} and define the offset as the difference between the best fit line to the whole sample and the post-processing data. There is a strong negative correlation between the offset from our best fit line at high stellar masses. Seed mass and the slope are inversely proportional. The maximum density centring results in a stepper negative slope for $\log(M_* / M_\odot) < 10.0$. The opposite is true for stellar masses greater than $10^{10} \, M_\odot$. Finally, the slope tends to flatten above $M_* = 10^{10.5} \, M_\odot$.} 
    \label{fig:redshift evolution of scaling relation}
\end{figure*}

\Cref{fig:redshift evolution of scaling relation} shows the different slopes following the same analysis method in \Cref{fig:redshift evolution of mstar vs mbh scaling relation} for \textsc{\small{MassiveFIRE}} data in post-processing analysis using fiducial settings. Each panel in \Cref{fig:redshift evolution of scaling relation} represents the blue and the red data points in the right panel of \Cref{fig:delta2 vs mstar}. The larger seed mass and COM centring choices flattens the slope.

\begin{figure*}
    \centering
    \begin{tabular}{cc}
        \includegraphics[width=0.5\linewidth]{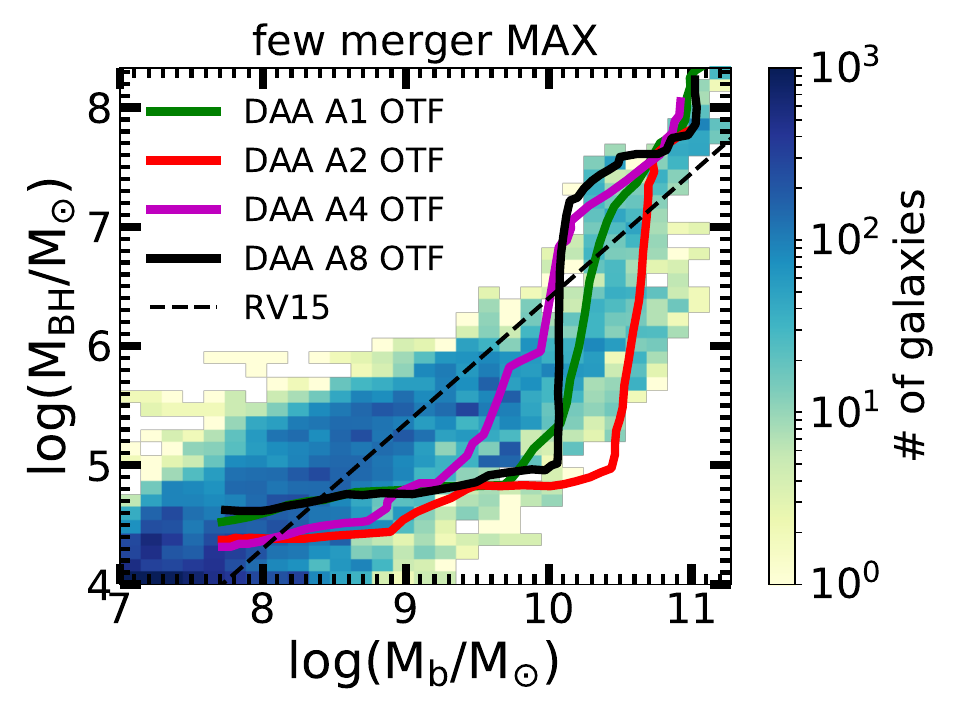}
        &
        \includegraphics[width=0.5\linewidth]{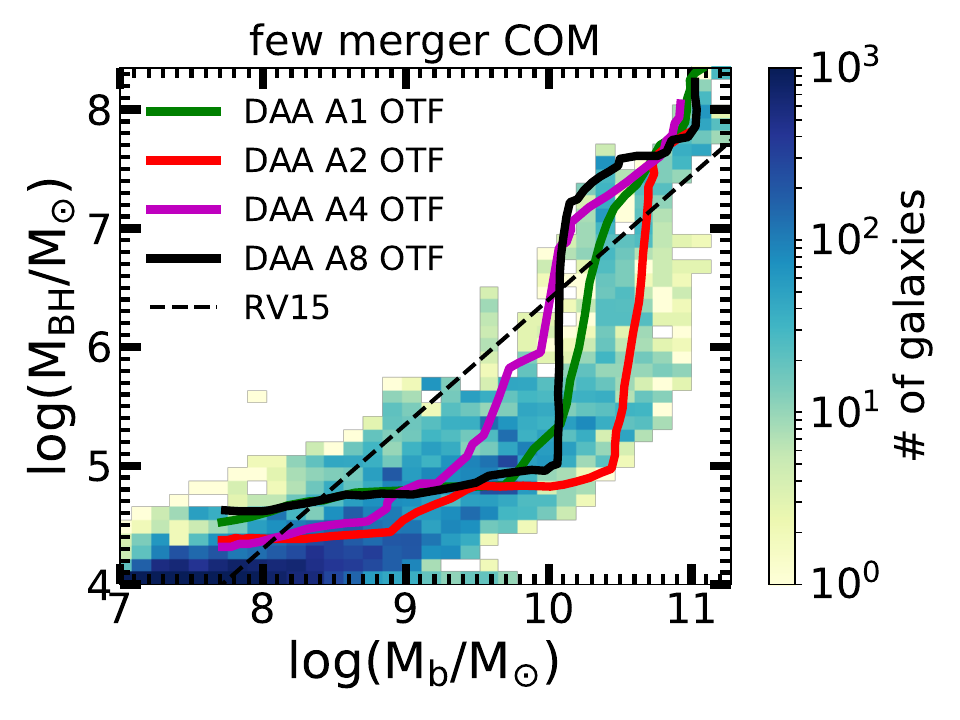}  
    \end{tabular}
    
    \begin{tabular}{cc}
        \includegraphics[width=0.5\linewidth]{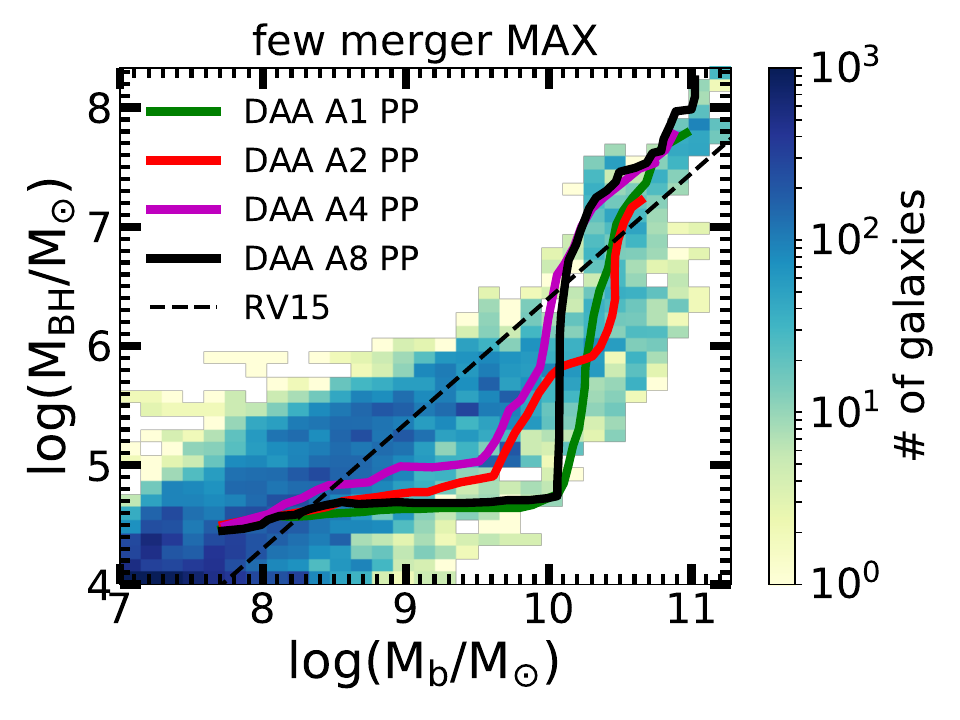}
        &
        \includegraphics[width=0.5\linewidth]{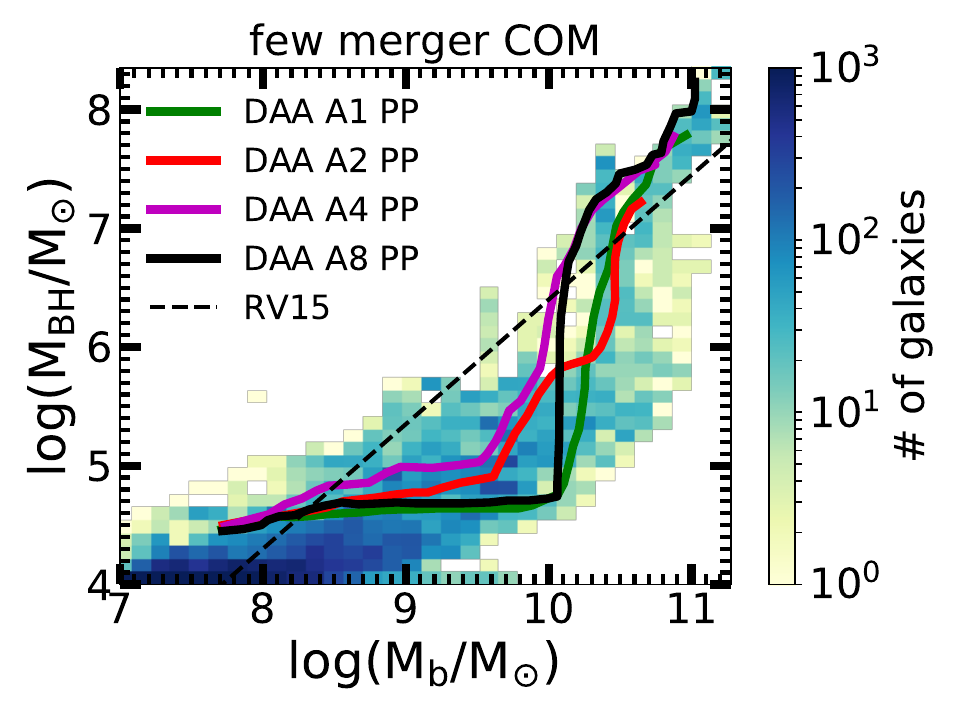}  
    \end{tabular}

    \caption{Comparison of the $M_{\rm b} - M_{\rm BH}$ scaling relation for our post-processing analysis with fiducial settings except the AHF centring method to the post-processing red(PP) and on-the-fly red (OTF) calculations of \protect\cite{Angles-Alcazar+17c} for simulations A$1$, A$2$, A$4$, and A$8$ run with FIRE-2 physics for a redshift range of $12< \rm z < 1$. (Top Panel) The dashed line shows the local scaling relation from \protect\cite{ReinesVolonteri2015}, data in the background is for the post-processing analysis of this paper and coloured solid lines represent the post-processing results of \protect\cite{Angles-Alcazar+17c}. (Bottom Panel) Everything is the same as in the left panel except we compare our results to the on-the-fly results of \protect\cite{Angles-Alcazar+17c} for the $M_{\rm b} - M_{\rm BH}$ scaling relation. Seed mass choices have small differences as we adopt a SMBH seed mass of $10^4 M_\odot$, while \protect\cite{Angles-Alcazar+17c} uses $M_{\rm seed} = 10^4 M_\odot/\rm h$. However, this small difference does not have any impact on the final SMBH mass.}
    \label{fig:DAA vs our PP}
\end{figure*}

Our post-processing analysis of 34 high-resolution \textsc{\small{MassiveFIRE}} simulations gives similar results with post-processing (PP) and on-the-fly (OTF) calculations of \cite{Angles-Alcazar+17c} (\Cref{fig:DAA vs our PP}). In this paper, we choose $10^4 M_\odot$ seed mass that is slightly smaller than what is chosen in \cite{Angles-Alcazar+17c}; $10^4 M_\odot/\rm h$.

\begin{figure*}
    \centering
    \includegraphics[width=\textwidth]{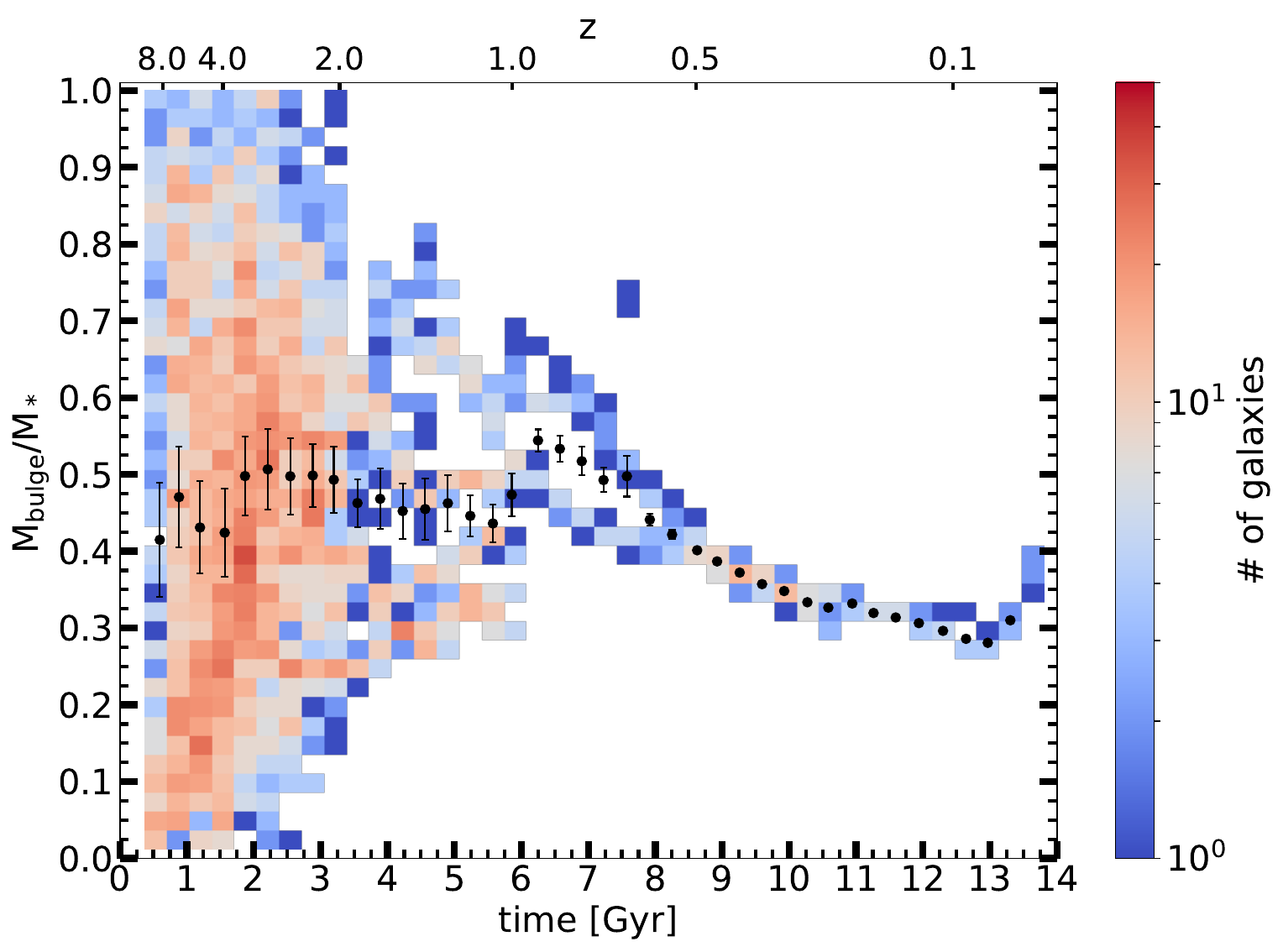}
    \caption{Bulge to total stellar mass of the galaxies in our sample in \Cref{fig:mstar vs mbh scaling relation} with ``\textsc{few mergers}'' model, SMBH seed mass of $10^4\, M_\odot$ and MAX centres. Colorbar shows the number of galaxies and black dots with error bars represent the mean value of each bin. Most of the galaxies in the sample reach $z=2$ and some reach $z=1$ while only galaxies from two simulations reach $z=0$. Galaxies in our sample are marginally disk dominated (with a bulge to total ratio of 0.3-0.5).}
    \label{fig:fbulge for MAX}
\end{figure*}

\Cref{fig:fbulge for MAX} shows the bulge mass to total stellar mass ratio of the sample regarding the \Cref{fig:mstar vs mbh scaling relation}. In general, the galaxies in our sample are marginally disk dominated, even at high redshift. This finding makes the GTDA model suitable for modelling the SMBH growth in \textsc{\small{MassiveFIRE}} galaxies.

\bsp	
\label{lastpage}
\end{document}